\documentstyle[12pt]{article}

\renewcommand{\theequation}{\thesection.\arabic{equation}}

\newcommand \beq{\begin{eqnarray}}
\newcommand \eeq{\end{eqnarray}}
\newcommand{\bea}{\begin{eqnarray*}}
\newcommand{\eea}{\end{eqnarray*}}
\newcommand{\be}{\begin{equation}}
\newcommand{\ee}{\end{equation}}
\newcommand{\ba}{\begin{array}}
\newcommand{\ea}{\end{array}}
\newcommand{\bc}{\begin{center}}
\newcommand{\ec}{\end{center}}

\begin{document}

\title{
Variational approximations for correlation functions in quantum field
theories 
}
\author{
C\'ecile Martin 
\\
\\
{Groupe de Physique Th\'eorique, 
}\\
{  Institut de Physique Nucl\'eaire,} \\
{ F-91406 , Orsay Cedex, France \footnote{E-mail address :
    martinc@ipno.in2p3.fr} }
}

\date{}

\maketitle 
\vspace*{1cm}
\begin{abstract}
Applying the time-dependent variational principle of Balian and
V\'en\'eroni, we derive variational approximations for multi-time
correlation functions in $\Phi^4$ field theory. We assume first that the
initial state is given and characterized by a density operator equal to a
Gaussian density matrix. Then, we study the more realistic situation where
only a few expectation values are given at the initial time and we perform
an optimization with respect to the initial state. We calculate explicitly
the two-time correlation functions with two and four field operators at
equilibrium in the symmetric phase. 
\end{abstract}




\newpage 

\begin{center}
Variational Correlation Functions 
\end{center}

\newpage

{\Large \bf Contents}

\vspace*{0.5cm}

{\bf 1-Introduction}

\vspace*{0.5cm}

 {\bf 2-Variational approximation for two-time correlation functions :
   optimization of the dynamics } 
 
 \begin{itemize}
 \item 2-1 Definition of the generating functional for correlation functions 
 
 \item 2-2 Variational evaluation of the generating functional 
 
 \item 2-3 Choice of the variational spaces 
 
 \item 2-4 Dynamical equations with sources
   
 \item 2-5 Expansion in powers of the sources
   \begin{itemize}
   \item 2-5a- TDHB equations for the expectation values  
    \item 2-5b-Two-time correlation functions  
 \end{itemize} 
    \end{itemize}

 {\bf 3-Optimization of the initial state }
 
\begin{itemize}
\item 3-1 Introduction

\item 3-2 Variational spaces 

\item 3-3 Dynamical equations 
  
\item 3-4 Expansion in powers of the sources 

\item 3-5 Approximation for the two-time and two-point correlation function
  in the symmetric phase. The case in equilibrium. The
  fluctuation-dissipation theorem.  

\item 3-6 Two-time and four-point correlation function : \\
Relation between the response function 
$\Pi_R({\bf q}^2,t',t'')$ and the four-point correlation function 
$\Sigma^{\Phi \Phi}$. The case of the symmetric phase. 
A perturbative resolution of the dynamical equations. 
The variational expression for the response function  
$\Pi_R({\bf q}^2,t',t'')$ at
the lowest order and at the first order in the loop term. 

\end{itemize}

{\bf 4-Conclusion}

\vspace*{0.5cm}
 
 {\bf Appendices}
 
 \begin{itemize}
 
\item  Appendix A : Characterization of a Gaussian state 
 
\item  Appendix B : Expression of the von Neuman entropy for bosons in the
 Hartree-Bogoliubov approximation
 
 \item Appendix C : Parametrization of the product of two Gaussians

\item Appendix D : 
  Dynamical equations for the expectation values  $\alpha_d, \Xi_d,
 \alpha_a, \Xi_a$

\item  Appendix E : Useful integrals for the resolution of the dynamical
 equations of section 3
 
 \item Appendix F : 
   Solutions of the backward dynamical equations to first order
 
\end{itemize}

\newpage 

\section{ Introduction}

The description of the time evolution of a system with initial
conditions which are out of equilibrium 
is one of the main challenges in quantum field
theory at finite temperature. This problem appears in different contexts in
cosmology or in particle physics. 
In inflationary models for the primordial universe, the problem is to describe
the evolution of the inflaton scalar field in order to be abble to predict
the duration of the inflationary phase. Another important problem is the
prediction of the density fluctuations and its relation 
to the distribution of galaxies \cite {INFLATION}.  
In particle physics, 
the experiments at the  Brookhaven Relativistic Heavy Ion Collider 
(RHIC) or soon at the CERN Large Hadron Collider (LHC)
may show a phase transition
between a state where the quarks and the gluons are deconfined and  chiral 
symmetry restored and the state of ordinary matter. The state of the
matter formed after the collision is out of equilibrium and the study of the
dynamics near the phase transition requires a nonperturbative method. The
chiral phase transition can be studied in a model of scalar fields with an 
O(N) symmetry. The relaxation of inhomogeneous  pions configurations formed
during the collision (called disoriented chiral condensates or DCC) could be
the signal of the chiral phase transition through a pions distribution with
strong correlations  \cite{DCC}. 
In the Standard model, one is interested in the dynamical properties of the
electroweak phase transition in order to describe the baryogenesis. The
violation of the baryonic number, which is related to the desintegration of
non-trivial topological field configurations, the sphalerons, is evaluated
from  a correlation function calculated at high temperature in a classical
approximation  \cite{BARYONS}. 

 The scale for the energy density involved in the primordial universe or the
 length scale which is important for the collective modes in Quantum
 Chromodynamics require non-perturbative approaches. Variational methods are
 therefore particularly adapted. 

In  the mean-field Hartree-Fock (HF) approximation for  fermions
or the Hartree-Bogoliubov (HB) approximation for  bosons, one 
minimizes the free energy or, at zero temperature, the energy with a
Gaussian variational ansatz for the density matrix or the pure state. For a
time-dependent problem, the corresponding approximations (TDHF or TDHB) are
obtained by minimizing an action also with a Gaussian variational ansatz.
(At zero temperature, a time-dependent variational principle was already
introduced by  Dirac \cite{DIRAC1}. It was used by the authors of reference 
\cite{JACKIWKER} in the context of quantum field theories .) 
These methods are well adapted to the calculation of expectation values. 
However, they are
not suitable for  multi-time correlation functions. In order to obtain
approximations for these multi-time correlation functions, a variational
principle was recently introduced in the context of non-relativistic fermions 
\cite{1}. The action to minimize has for stationary value the generating 
functional for the correlation functions. This functional is a functional of
the time-dependent sources coupled to the operators of interest. The action
involves two types of variational objects, one  
 ${\cal A}(t)$  related to the observables of interest and the other 
  ${\cal D}(t)$ akin to a density matrix. 
The temporal evolution of 
${\cal A}(t)$ is described by a backward equation while that of 
 ${\cal D}(t)$ involves a time running forward. Let us notice that this
 doubling of the number of degrees of freedom with two temporal arrows
 appears already in the formalism of Schwinger and Keldysh for 
 time-dependent problems at finite temperature \cite{SCHWINGER}. 
(For a review of the different formalisms used at finite temperature, 
 see  reference  \cite{LANDSMANN}). 
By restricting the variational spaces for  ${\cal A}(t)$ and  ${\cal D}(t)$ 
to Gaussian operators, one obtains evolution equations which are coupled.
The expansion at  first order in powers of the sources of these equations
(or at the second order for the generating functional ) provides
approximations for the two-time correlation functions. In spite of the
Gaussian character of the variational ans\"atze, the approximations obtained
in this way go beyond the usual mean-field approximations and include
correlations between particles. This comes from the fact that the variational
principle of Balian and V\'en\'eroni 
 \cite{1} has been  precisely constructed to provide directly the quantity
 of interest, namely the generating functional. 

Our aim is to generalize this formalism to quantum field theories. For
simplicity, we will consider the case of a scalar field described by a 
 $\Phi^4$ interaction in a  Minkowski space. We will choose sources coupled
 to local operators   such as $\Phi({\bf x})$ or $\Pi({\bf x})$ 
(written here in the Schr\"odinger representation) and to composite operators
such as 
 $\Phi({\bf x}) \Phi({\bf y})$ or $\Phi({\bf x}) \Pi({\bf y})$. The second 
 derivative of the generating functional with respect to the sources 
will thus give two-time correlation functions which have up to four field
operators. 

In the first part of the paper, we will consider that the initial state is
given and characterized by a density operator 
  $D(t_0)$ equal to a Gaussian density matrix. We will write the action
  functional to be minimized  (eqs. (\ref{2.4}) and 
  (\ref{2.5})). Then we will restrict our variational spaces to operators
  which are exponentials of quadratic and linear forms in the field operators 
 $\Phi({\bf x})$ and  $\Pi({\bf x})$ (called Gaussian operators). An
 adequate parametrization of the variational objects allows us to write the
 coupled dynamical equations in a compact form. We stress again the fact
 that the variational method we use introduces a doubling of the degrees of
 freedom. We obtain four coupled equations (two for the expectation values
 of one-field operator and two for the expectation values of two-field
 operators). Therefore we have twice as many dynamical equations as 
  the usual mean-field approach. The expansion of these equations 
 to lowest zero order in powers of the sources leads to a decoupling of these
 equations whose a  half is identical to the time-dependent mean-field
 equations for the  $\Phi^4$ theory  written in  reference  \cite{1a}. 
We have showed in a previous paper that these dynamical equations 
can be put into a classical noncanonical Hamiltonian framework with a 
Poisson structure  which is a generalization of the standard Poisson bracket
\cite{POISSON}.  
At the lowest order we thus obtain an approximation 
for the single-time expectation values of one
or two field operators. The expansion of the coupled equations 
to first order in powers of the
sources    provides an approximation for the two-time correlation functions 
with two, three or four field operators. 
 Their evaluation proceeds in the
following way :  first one   
solves the mean-field TDHB equations, then one  calculates the kernel which
corresponds to small deviations arround the mean-field solution (which is
analogous to the random phase approximation (RPA) 
kernel of reference \cite{1}) and finally one solves a
backward dynamical equation  which involves the kernel calculated for the
mean-field solution. These results have been obtained for the generating
functional of correlation functions with time-ordered product of the field
operators. We also give  the corresponding expressions for the retarded Green
functions. Indeed, these retarded Green functions appear when one studies
the response of the system to a small external perturbation. A less complete
version of this first part of the paper has been published in reference 
\cite{CECILE}.

In section 2, we have supposed that the initial state 
 $D(t_0)$ was known. In practice, this is not the case. Only the expectation
 values of a few observables are known. In section 3, we use the form of
 the variational principle introduced by  Balian and V\'en\'eroni which
 optimizes also the initial state  \cite{1}. The approximations obtained in
 this way for the two-time correlation functions include not only the
 correlations between particles originating from the dynamics but also
 those present in the initial state. One introduces again two variational
 objects which now evolve  in imaginary time. 
 We study the symmetric phase. We will consider the two-time function
 with two field operators  $\Phi$ and the two-time function with four field
 operators   $\Phi$. We will relate it  to the  standard
 polarization function  $\Pi_R$. At equilibrium, in the symmetric phase, 
 the case of the correlation function with two field operators is special 
 since the approximations obtained with or without optimization with respect
 to the initial state co\"\i ncide. For the correlation function with four
 field operators, we will be obliged to combine the variational
 approximation with a perturbative approximation valid for  small values of
 the coupling constant, by neglecting or taking into account at  first
 order a loop term in the dynamical equations. However, the solutions remain
 nonperturbative  since they depend on the self-consistent HB solution. At
 the lowest order, by neglecting the loop term, the two-time function with
 four field operators is simply related to the two-time function with two
 field operators according to the Wick theorem. At equilibrium, in the
 symmetric phase, we give to first order the expression for the two-time
 polarization functions with four field operators : 
 the retarded Green function $\Pi_R({\bf q}^2, t',t'')$  (eq. {4.133a}) and the
 non-retarded Green function 
 $\Pi({\bf q}^2, t',t'')$ (eq. (\ref{4.129})). 
We verify explicitly that these   functions depend only on the time difference 
$t'-t''$ and not on the initial time  $t_0$. This is not the case for the
non-retarded Green function if, like in the section 2, we optimize only
the dynamics and not the initial state. We check also that our approximations
for the two-time correlation functions satisfy the fluctuation-dissipation
theorem. The last section contains some conclusions and perspectives for
future work.

\section{Variational approximation for the two-time correlation functions :
  optimization of the dynamics }  

\setcounter{equation}{0}

\subsection{Definition of the generating functional 
  for correlation functions }

We define the operator  $A(J,t)$ which is a functional of the sources   $J$
and a function of the time $t$ in the following way : 
\be \ba{lll}\label{2.1}
A(J,t) & {\displaystyle = T \exp i \int_t^{\infty} dt' \: \left\{
\int d^3x \: \: J^{\Phi}({\bf x},t') \Phi^H({\bf x},t',t) + 
J^{\Pi}({\bf x},t') \Pi^H({\bf x},t',t) \right. } \\
& {\displaystyle  \left. 
+ \int d^3x \: d^3y \: \: J^{\Phi \Pi}({\bf x},{\bf y},t') \: 
\left( \Phi^H({\bf x},t',t) \Pi^H({\bf y},t',t) + \Pi^H({\bf y},t',t) 
\Phi^H({\bf x},t',t) \right) \right. } \\ 
& {\displaystyle \left. 
+J^{\Phi \Phi}({\bf x}, {\bf y}, t') \:  
\Phi^H({\bf x}, t',t) \Phi^H({\bf y},t',t) 
+J^{\Pi \Pi}({\bf x}, {\bf y},t') \: \Pi^H({\bf  x},t',t) \Pi^H({\bf y},t',t) 
\right\} } 
\ea  \ , \ee
where $T$ is the time-ordered product of operators and 
$\Phi^H({\bf x},t',t)$ and  $\Pi^H({\bf x},t',t)$ are respectively the field
and momentum operators defined in the Heisenberg representation. They
satisfy  the boundary condition : 
$\Phi^H({\bf x},t,t)=\Phi({\bf x})$ and $\Pi^H({\bf x},t,t)=\Pi({\bf x})$, 
$\Phi({\bf x})$ and $\Pi({\bf x})$ being the operators in the  Schr\"odinger
representation.  We use the short hand notation 
\be \label{2.1a}
A(J,t)=T \exp \left(
i \int_t^{\infty} dt' \: J_j(t') \: {\cal O}_j(t',t) \right) \ee
The sources  $J_j(t')$ (which are local or depend on two space points) are
turned on between  $t'=t$ and 
$t'=+\infty$. 

The functional $A(J,t)$ satisfies the equation 
\be \label{2.1b} 
\frac{d}{dt}A(J,t)=i[A(J,t),H] -i \: A(J,t) \sum_j J_j(t) {\cal O}_j \ee
with the boundary condition $A(t=+\infty)=1$. 
We will consider the case of a self-interacting scalar field in a Minkowski
metric described by the Hamiltonian  $H=\int d^3x \: {\cal
H}({\bf x}) $ with : 
\be \label{2.1c}
{\cal H}({\bf x})=\frac{1}{2}\Pi^2({\bf x})+\frac{1}{2}
[\vec \nabla \Phi({\bf x})]^2+\frac{m_0^2}{2}\Phi^2({\bf
x})+\frac{b}{4!}\Phi^4({\bf x}) \ . \ee
We work in three spatial dimensions. The constants $m_0$ and $b$ are
respectively the bare mass and the bare coupling constant.  

The generating functional for the causal Green functions writes 
\be \label{2.2}
Z(J,t_0)  = Tr \left(D(t_0) \: A(J,t_0) \right) \ , \ee
$D(t_0)$ being the initial state. The statistical operator  $D(t_0)$ may
represent a thermal equilibrium or a nonequilibrium state. At zero
temperature, it reduces to a projection operator. In addition to the sources 
$J$, the generating functional  $ Z$ depends on the initial time  $t_0$. 
This is a difference with the generating functional usually considered in
quantum field theories \cite{ZINN-JUSTIN}. We want to evaluate the
functional derivative of 
\be \label{2.3}
W(J,t_0) = -i \: \ln Z(J,t_0) \ . \ee
$W(J,t_0)$ is the generating functional for the connected Green functions.
Let us note that for hermitic  operators $A(t)$ and $D(t)$, 
$W(J,t)$ is real. The expansion in powers of the sources of 
$W(J,t)$ writes : 
\be \ba{lllll} \label{2.3a} 
& {\displaystyle W(J
, t_0)= -i \: \ln( n_0) + 
 \int_{t_0}^{+ \infty} dt' \: \{ 
\int d^3x \: J_{\Phi}({\bf x},t') \: \varphi({\bf x},t') + 
\int d^3x \: d^3y 
\: J_{\Phi \Phi}({\bf x},{\bf y},t') \: G({\bf x}, {\bf y},t') 
+ 
... \}
} \\ 
& {\displaystyle + \frac{i}{2} \: \int \int_{t_0}^{+ \infty} dt' \: dt'' \: 
\{ 
\int d^3x_1 \: d^3x_2 \: J_{\Phi}({\bf x}_1,t') \: J_{\Phi}({\bf x}_2,t'') \: 
C^2_{\Phi \Phi}({\bf x}_1,{\bf x}_2,t',t'') + ...  \}  } \\   
& {\displaystyle + ... } \ea \ ,  \ee 
where $n_0=Tr D(t_0)$ ( we choose not to normalize  the operator    
 $D(t_0)$). We introduce the following notations for the expectation values
 of one and two field operators : 
\be \label{2.3b} 
\varphi({\bf x},t) = \frac{1}{n_0} \: Tr \left( \Phi^{H}({\bf x},t,t_0) 
\: D(t_0) \right) \ , \ee 
\be \label{2.3c} 
G({\bf x}, {\bf y},t)=\frac{1}{n_0} \: Tr \left( \Phi^{H}({\bf x},t,t_0) 
\: \Phi^{H}({\bf y},t,t_0) \: D(t_0) \right) - \varphi({\bf x},t) \: 
\varphi({\bf y},t) \ . \ee 
The two-time causal functions with two field operators are defined according
to : 
\be \label{2.3d} 
C^2_{\Phi \Phi}({\bf x},{\bf y},t',t'')= \frac{1}{n_0} \: Tr \left( 
T \: \Phi^{H}({\bf x},t',t_0) \: \Phi^{H}({\bf y},t'',t_0) \: D(t_0) \right) 
- \varphi({\bf x},t') \: \varphi({\bf y},t'') \ . \ee 
Similarly, we define the functions 
$C_{\Phi \Pi}^2$, $C_{\Pi \Pi}^2$. 
From the expansion  (\ref{2.3a}), we have : 
\be \label{2.3dd}
\frac{1}{i} \: \frac{\delta ^2 W}{\delta J_{\Phi}({\bf x},t') \delta 
J_{\Phi}({\bf y},t'')} \bigg \vert_{J=0}=
C^2_{\Phi \Phi}({\bf x}, {\bf y},t',t'') 
\ . \ee 
We define also the three-point and four point two-time correlation function 
$ C^3$ et $C^4$  : 
\be \label{2.3e} \ba{ll} 
\frac{1}{i} & {\displaystyle 
 \frac{\delta^2 W}{\delta J_{\Phi}({\bf x},t') 
\delta J_{\Phi \Phi}({\bf y}, {\bf z},t")} \bigg\vert_{J=0} \equiv 
C^3({\bf x}, {\bf y}, {\bf  z}, t', t")  
} \\ 
& {\displaystyle 
+ \varphi({\bf y},t") \: \left( C^2({\bf x},{\bf z},t',t") + 
\varphi({\bf x},t') 
\: \varphi({\bf z}, t") \right) 
+ \varphi({\bf z},t") \: \left( C^2({\bf  x},{\bf y},t',t") + 
\varphi({\bf x},t') 
\: \varphi({\bf y}, t") \right) 
} \ , \ea \ee
\be \label{2.3f} \ba{lll} 
\frac{1}{i} & {\displaystyle 
\frac{\delta^2 W}{\delta J_{\Phi \Phi}({\bf x},{\bf y},t') 
\delta J_{\Phi \Phi}({\bf z}, {\bf u},t")} \bigg\vert_{J=0} \equiv 
C^4({\bf x}, {\bf y}, {\bf z}, {\bf u},t', t")  
} \\ 
& {\displaystyle 
+ \left( C^2({\bf x}, {\bf  z}, t',t") + \varphi({\bf x},t') \: 
\varphi({\bf z},t") \right) \: 
\left( C^2({\bf y}, {\bf u}, t',t") + \varphi({\bf y},t') \: 
\varphi({\bf u},t") \right)  
} \\ 
& {\displaystyle 
+ \left( C^2({\bf x}, {\bf u}, t',t") + \varphi({\bf x},t') \: 
\varphi({\bf u},t") \right) \: 
\left( C^2({\bf y}, {\bf z}, t',t") + \varphi({\bf y},t') \: 
\varphi({\bf z},t") \right)  
} \ .  \ea \ee 

\subsection{Variational evaluation of the generating functional  }

We will use the time-dependent variational principle of  Balian and 
 V\'en\'eroni  \cite{1} to obtain an approximation for our quantity of
 interest 
 $Tr(D(t_0)A(J,t_0))$. This principle is an illustration of the general
 method developped in reference  \cite{1aa} to construct a variational
 principle suited to the evaluation of a desired quantity. The basic idea is
 to consider the equation  (\ref{2.1b}) which defines  $A(J,t_0)$ as a
 constraint and to introduce a lagrangian multiplier ${\cal D}(t_0)$ 
 associated to this constraint. 

We define the action functional : 
\be \label{2.4} 
{\cal Z} \left({\cal A}(t), {\cal D}(t) \right) = 
Tr \left({\cal A}(t_0) \: {\cal D}(t_0) \right) + {\cal Z}_{dyn} \ , \ee 
where   
\be \label{2.5} 
{\cal Z}_{dyn} = Tr \: \int_{t_0}^{\infty}  dt \: \: {\cal D}(t) \: 
\left\{ \frac{d {\cal A}(t)}{dt} 
- i \left[{\cal A}(t), \int d^3x {\cal H}({\bf x}) 
\right] + i {\cal A}(t) \: \left( \sum_j J_j(t) Q_j \right)  \right\} \ . \ee 
We have written in a compact form the term which involves the sources  $J$ : 
\be \ba{lll} \label{2.6} 
\sum_j J_j(t) Q_j =& {\displaystyle 
 \int d^3x \: \left(J^{\Phi}({\bf x},t) \Phi({\bf x}) + 
J^{\Pi}({\bf x},t) \Pi({\bf x}) \right) } \\ 
& {\displaystyle + \int d^3x \: d^3y  \: J^{\Phi \Pi}({\bf x}, {\bf y},t) 
\left( \Phi({\bf x})\Pi({\bf y}) + \Pi({\bf y}) \Phi({\bf x})  \right) } \\ 
& {\displaystyle  + \int d^3x \: d^3y \: 
\left(  J^{\Phi \Phi}({\bf x}, {\bf y},t) \Phi({\bf x}) 
\Phi({\bf y}) + J^{\Pi \Pi}({\bf x}, {\bf y} ,t) \Pi({\bf  x}) \Pi({\bf y}) 
\right) } 
\ . \ea \ee
The variational quantities of the action functional 
${\cal Z}$ are the observable-like operator  ${\cal A}(t)$ and  the
density-like operator ${\cal D}(t)$. We look for the stationarity of 
 ${\cal Z}$ with respect to variations of  ${\cal A}(t)$ and  ${\cal D}(t)$
 with the boundary conditions : 
 ${\cal A}(t=+ \infty) = A(t=+\infty)= 1$ and 
${\cal D} (t_0) = D(t_0)$. $D(t_0)$ is the initial statistical operator
which we assume to be given and equal to a Gaussian density matrix. By
construction of the variational principle, the generating functional for the
connected Green functions will be approximated by the value of the functional 
${\cal Z}$ at the stationary point :  
\be \label{2.7} 
W(J,t_0)=-i \: \ln {\cal Z}_{st} \ . \ee 

\subsection{ Choice of the variational spaces}

We restrict ourselves to trial operators  
${\cal A}(t)$ and  ${\cal D}(t)$ which are exponentials of quadratic and
linear forms of the field operators  $\Phi({\bf x})$ and $\Pi({\bf x})$ 
which we shall loosely call Gaussian operators.

A Gaussian state  ${\cal D}(t)$ is completely characterized by the vector 
$\alpha({\bf x},t)$  and the matrix  $\Xi({\bf x}, {\bf y},t)$ defined by : 
\be \label{2.7a} 
 \alpha({\bf x},t)=
 \pmatrix{\varphi({\bf x},t) \cr -i \pi({\bf x},t) \cr} \ , \ee
 \be \label{2.7b}
 \Xi({\bf x}, {\bf y},t)=\pmatrix{2G({\bf x},{\bf y},t) & -i 
   T({\bf x}, {\bf y},t)
 \cr -i T({\bf y}, {\bf x},t) & -2 S({\bf x},{\bf y},t) \cr } \ , \ee 
 where
\be \label{2.7c} 
\varphi({\bf x},t)=\frac{1}{n(t)} \: Tr({\cal D}(t) \Phi({\bf x})) \ , \ee 
\be \label{2.7d} 
\pi({\bf x},t) = \frac{1}{n(t)} \: Tr({\cal D}(t) \Pi({\bf x})) \ , \ee 
\be \label{2.7e} 
G({\bf x},{\bf y},t)= \frac{1}{n(t)} \: Tr({\cal D}(t) \bar  \Phi({\bf x}) 
\bar \Phi({\bf y})) \ , \ee 
\be \label{2.7f} 
S({\bf x}, {\bf y},t) = \frac{1}{n(t)} \: Tr({\cal D}(t) \bar \Pi({\bf x}) 
\bar  \Pi({\bf y}) ) \ , \ee 
\be \label{2.7g} 
T({\bf  x}, {\bf y},t) = \frac{1}{n(t)} \: Tr({\cal D}(t) 
( \bar \Phi({\bf  x}) \bar \Pi({\bf y}) + 
\bar  \Pi({\bf y}) \bar \Phi({\bf x}) )) 
\ , \ee 
\be \label{2.7h}
n(t)=Tr {\cal D}(t) \ , \ee
and, for any operator  $O$ :
$\bar O = O-Tr({\cal D}(t) O)$.
The hermiticity of ${\cal D}(t)$ implies that the matrices $G,S$ and $T$ are
real and that $G$ and $S$ are symmetric in the space indices.  

The Heisenberg invariant is equal to :  
\be \label{2.7i} \ba{ll}
I({\bf x}, {\bf y})=\int d^3z \: 
& {\displaystyle \left( 4 <\bar \Phi({\bf x}) \bar \Phi({\bf z})> 
<\bar \Pi({\bf z}) \bar \Pi({\bf y})> \right. } \\
& {\displaystyle \left. -<\bar \Phi({\bf x}) \bar \Pi({\bf z})+ 
  \bar \Pi({\bf z}) \bar \Phi({\bf x})>
\: <\bar \Phi({\bf z}) \bar \Pi({\bf y}) + \bar \Pi({\bf y}) 
\bar \Phi({\bf z})> \right) } \ea \, \ee 
\be \label{2.7j} 
I({\bf x}, {\bf y})=
\int d^3z \: \left(-\Xi_{11}({\bf x},{\bf z}) 
\: \Xi_{22}({\bf z},{\bf y}) + \Xi_{12}({\bf x},{\bf z}) \: 
\Xi_{12}({\bf z},{\bf y}) \right) \ . \ee 
For a pure state : 
\be 
\Xi_{11} \: \Xi_{22}- \Xi_{12}^2+1=0 \ee 
\be 
I({\bf x},{\bf y})=\delta^3({\bf x}-{\bf y}) \ee 
In Appendix A, we give the relations between our parametrization of a
Gaussian state and the parametrization used by the authors of reference 
\cite{1a}. 

We then introduce the operators 
${\cal T}_b = {\cal A}(t) \: {\cal D} (t)$ and
${\cal T}_c={\cal D}(t) \: {\cal A} (t)$ \cite{4}.
The operator  
${\cal T}_b$ is a Gaussian operator which is characterized by the following
expectation values :  
\be \label{2.7} 
n(t)=Tr {\cal T}_b=Tr{\cal T}_c \ , \ee 
\be \label{2.8} 
\varphi_b({\bf x},t)=\frac{1}{n(t)} \: Tr({\cal T}_b \Phi({\bf x})) \ , \ee 
\be \label{2.9} 
\pi_b({\bf x},t) = \frac{1}{n(t)} \: Tr({\cal T}_b \Pi({\bf x})) \ , \ee 
\be \label{2.10} 
G_b({\bf x},{\bf  y},t)= \frac{1}{n(t)} \: Tr({\cal T}_b \bar \Phi({\bf x}) 
\bar \Phi({\bf y})) \ , \ee 
\be \label{2.11} 
S_b({\bf x},{\bf y},t) = \frac{1}{n(t)} \: Tr({\cal T}_b \bar \Pi({\bf x}) 
\bar \Pi({\bf y}) ) \ , \ee 
\be \label{2.12} 
T_b({\bf x}, {\bf y},t) = \frac{1}{n(t)} \: Tr({\cal T}_b 
( \bar \Phi({\bf x}) \bar \Pi({\bf y}) + 
\bar \Pi({\bf y}) \bar \Phi({\bf x}) )) 
\ , \ee 
with 
$\bar \Phi({\bf x}) = \Phi({\bf x})-\varphi_b({\bf x},t)$ and   
$\bar \Pi({\bf x}) =\Pi({\bf x}) -\pi_b({\bf x},t)$. 

Similarly the operator  ${\cal T}_c$ is a Gaussian operator which is
characterized by the quantities  $n, \varphi_c, \pi_c,
G_c, S_c, T_b$ ( obtained by replacing  
 ${\cal T}_b$ by 
${\cal T}_c$ in (\ref{2.8})-(\ref{2.12})).  
For ${\cal D}(t)=1$, one has  
$\varphi_{b}({\bf x},t) =\varphi_{c}({\bf x},t)=\varphi_{a}({\bf x},t) $  
(and similar definitions for  $\pi_a, G_a, S_a, T_a$).  
For ${\cal A}(t) =1$, one has 
$\varphi_{b}({\bf x},t) =\varphi_{c}({\bf x},t)=\varphi_{d}({\bf  x},t) $ 
 (and similar definitions for    $\pi_d, G_d, S_d, T_d$). 
In Appendix C, we give the expressions of  
$\alpha_b, \alpha_c, \Xi_b$ and $ \Xi_c$ as functions of the independent
variational quantities 
$\alpha_a,  \Xi_a$ and $\alpha_d,  \Xi_d$ which characterize respectively 
${\cal A}$ and    ${\cal D}$.  
 
With this choice of the trial spaces for  ${\cal A}(t)$ and 
${\cal D}(t)$ , the Wick theorem allows us to express the functional 
${\cal Z}$ in the form :  
\be \ba{ll} \label{2.15}
 {\cal Z}=
& {\displaystyle n(t_0) + \int_{t_0}^{+ \infty} dt \: \left[ 
\frac{dn}{dt} \vert_{{\cal D}(t)=cte} - i \: n(t) \: \int d^3x \: 
\left( {\cal E}_c({\bf x},t) - {\cal E}_b({\bf x},t) \right) \right. } \\ 
& {\displaystyle \left. + i \: n(t) \: \int d^3x_{1} d^3x_2 \: 
{\cal K}_c ({\bf x}_1, {\bf x}_2,t) \right] } \ea \ , \ee 
where the energy density  ${\cal E}({\bf  x}, t)=Tr \left( {\cal D}(t) \: 
{\cal H}({\bf x}, t) \right)$ is given, for the Hamiltonian density 
(\ref{2.1c}), by : 
\be \ba{lll} \label{2.16} 
{\cal E}({\bf x},t)=
& {\displaystyle \frac{1}{2} \: \pi^2({\bf  x},t) + \frac{1}{2} \: 
(\vec \nabla \varphi({\bf x},t) )^2 + \frac{m_0^2}{2} \: \varphi^2({\bf x},t) + 
\frac{b}{24} \: \varphi^4({\bf x},t) } \\ 
& {\displaystyle + \frac{1}{2} \: S({\bf  x}, {\bf x},t) - \frac{1}{2} \: 
\Delta G({\bf x}, {\bf y},t) \vert_{{\bf x}={\bf y}} + \frac{m_0^2}{2} \: 
G({\bf x}, {\bf x},t) } \\ 
& {\displaystyle + \frac{b}{8} \: G^2({\bf x}, {\bf x},t) + 
\frac{b}{4} \: \varphi^2({\bf x},t) \: G({\bf  x}, {\bf x},t) }  \ea \ee 
and 
\be \label{2.16a} \ba{ll}
\ln n(t)= & {\displaystyle 
\ln n_a + \ln n_d -\frac{1}{2} \tilde \alpha_{b-c} \tau 
\alpha_{a-d} } \\ 
& {\displaystyle +\frac{1}{2} \ln \det \left( \sigma_1 (1+\tau)(\Xi_a+
\Xi_d)^{-1} (1+\tau) \right) } \ , \ea \ee
where $\alpha_{b-c}=\alpha_b-\alpha_c$ and $\tau$ and $\sigma_1$ are  
the following  $2 \times 2$ matrices : 
\be 
\sigma_1=\pmatrix{0 & 1 \cr 1 & 0 \cr} \: \: \ , \: \: 
\tau= \pmatrix{ 0 & 1 \cr -1 & 0 \cr } \ . \ee 
The last term of the functional (\ref{2.15}),   
$K_c = Tr \left( {\cal T}_c \: \sum_{j} J_j(t) \: Q_j \right) = 
\int d^3x_1 d^3x_2 \: {\cal K}_c({\bf x}_1,{\bf x}_2,t)$, involves the
sources $J$ and the expectation values indexed by $c$. 

With the parametrization we use, the boundary conditions 
 ${\cal D}(t_0)=D(t_0)$ and ${\cal A}(t=+\infty)=1$ correspond respectively : 
 \be \label{2.16a}
n_d(t_0)=n_0 \: \: , 
\: \: \alpha_d(t_0)=\alpha_0 \: \: , \: \: \Xi_d(t_0)=\Xi_0 \ , \ee
where 
$n_0, \alpha_0$ and $\Xi_0$ characterize the Gaussian initial state $D(t_0)$, 
and 
\be \label{2.16b}
n_a(t=+ \infty)=1 \: \: , 
\: \: \Xi_a^{-1}(t=+ \infty)=0 \: \: , \: \: 
\frac{1}{\Xi_a}\alpha_a(t=+ \infty)=0 
\ . \ee
We will write the conditions for the stationarity of the functional 
${\cal Z}$ with respect to the variations of  $n_d, \alpha_d, \Xi_d$ and 
$n_a, \alpha_a, \Xi_a$. 

\subsection{Dynamical equations with sources }

By varying the expression  (\ref{2.15}) with respect to $n_d, \alpha_d,
\Xi_d$, with the boundary conditions 
\be \label{2.17} 
n_d(t_0)=n_0 \: \: , 
\: \: \alpha_d(t_0)=\alpha_0 \: \: , \: \: \Xi_d(t_0)=\Xi_0 \ , \ee
we obtain the evolution equation for   $n_a, \alpha_a$ and   
$ \Xi_a$. Integrating  (\ref{2.15}) by parts and varying with respect to 
$n_a, \alpha_a, \Xi_a$ with the boundary conditions 
\be \label{2.18} 
n_a(t=+ \infty)=1 \: \: , \: \: \Xi_a^{-1}(t=+ \infty)=0 \: \: , \: \: 
\frac{1}{\Xi_a}\alpha_a(t=+ \infty)=0  
\ , \ee
we obtain the evolution equations for   $n_d, \alpha_d$ and $ \Xi_d$. 
In general the evolution equations for   $n_d, \alpha_d, \Xi_d$ and those for 
$n_a, \alpha_a,   \Xi_a$ are coupled. 
The solutions $n_d, \alpha_d, \Xi_d$ and $n_a, \alpha_a, \Xi_a$  depend on
the sources.  {\it Owing to this dependence, we will be abble to obtain non
  trivial approximations for the correlation functions in spite of the
  Gaussian nature of the trial operators. } 
 
By combining the evolution equation for  $n_d, \alpha_d, \Xi_d$ 
with those for   $n_a, \alpha_a, \Xi_a$, the dynamical equations for the
expectation values  $\alpha_b, \Xi_b$ and $ \alpha_c, \Xi_c$ can be written
in the following compact form : 
\be \label{2.19} 
i \: \dot \alpha_b = \tau \: w_b \ , \ee 
\be \label{2.20} 
i \: \dot \alpha_c = \tau \: \left( w_c - w^c_K \right) \ , \ee 
\be \label{2.21} 
i \: \dot \Xi_b = 2 \left( \Xi_b \: {\cal H}_b \: \tau - \tau \: 
{\cal H}_b \: \Xi_b \right) \ , \ee 
\be \label{2.22} 
i \: \dot \Xi_c = 2 \: \left( \Xi_c \: \left( {\cal H}_c -{\cal I}^c_K 
\right) \: \tau - \tau \: \left( {\cal H}_c - {\cal I}^c_K \right) \: \Xi_c 
\right) \ .  \ee
We have also  $\dot n(t)=0$. 

  The vector  $w_b$  and the matrix  ${\cal H}_b$ are defined from the
  variation of  $<H>_b=Tr({\cal T}_b H)$  : 
\be \label{2.24} 
\delta <H>_b = \int d^3x \: 
\tilde w_b({\bf x},t) \: \delta \alpha_b({\bf x},t) 
-\frac{1}{2} \: \int d^3x \: d^3y \: 
tr \left[{\cal H}_b({\bf x}, {\bf y},t) \: 
\delta \Xi_b({\bf y}, {\bf x},t) \right] \ .  \ee
Similarly,    $w_c, {\cal H}_c$ and    $w^c_K, {\cal I}^c_K$ are defined
from the variation of $<H>_c=Tr({\cal T}_c H)$ and from the variation of the
source term  $K_c$. 

Therefore, by using a convenient parametrization for the two variational
objects ${\cal D}(t)$ and  ${\cal A}(t)$, we have been abble to obtain
dynamical equations in a very compact form even for finite values of the
sources. The dynamical equations for $\alpha_d, \Xi_d$ and $\alpha_a, \Xi_a$ 
have a more complicated form. They are given in Appendix D where we give
also the explicit expressions of the vectors  $w$, $w^c_K$ and of the
matrices   ${\cal H}$ and  ${\cal I}^c_K$ in the case of  the  $ \Phi^4$
theory. In spite of their simple form, the  equations 
(\ref{2.19})-(\ref{2.22}) are coupled because the solutions  $\alpha_b, \Xi_b$ 
and $\alpha_c, \Xi_c$ do not satisfy simple boundary conditions. 

Let us note that the coupled equations for  $\alpha_d, \Xi_d$ ($\alpha_a,
\Xi_a$) do not conserve neither the energy $<H>_d$ ($<H>_a$) nor the entropy
$S_d$ ($S_a$) (see Appendix B for the  definition of the von-Neuman entropy
for bosons). On the other hand, the dynamical equations for 
$\alpha_b$ and $\Xi_b$ (eqs. (\ref{2.19}) and 
(\ref{2.21}) ) which have the same form as the mean-field TDHB equations 
conserve the pseudo-energy  $<H>_b$ and the pseudo-entropy  $S_b$. 

The expansion in powers of the sources of the stationarity conditions 
(\ref{2.19})- (\ref{2.22}) will provide approximate dynamical equations for
the expectation values and the correlation functions defined in Eqs. 
(\ref{2.3b})-(\ref{2.3f}). 

\subsection{Expansion in powers of the sources  } 

\subsubsection{ TDHB equations for the expectation values  }

 We will use the upper index$^{(0)}$ for the solutions of the dynamical
 equations when there are no sources. The limit with vanishing sources
 corresponds to 
 \be \label{2.25}
 \alpha^{(0)}_b=\alpha^{(0)}_c=\alpha^{(0)}_d  \quad \ , \quad 
\Xi^{(0)}_b = \Xi^{(0)}_c = \Xi^{(0)}_d  \ . \ee
From equations (\ref{C.4}) and  (\ref{C.5}) and from   
\be \label{2.26}
\alpha_b-\alpha_c=2 \: \tau \: \frac{1}{\Xi_a+\Xi_d} \: (\alpha_a -\alpha_d) 
\ , \ee 
we obtain that in the limit with vanishing sources 
\be \label{2.27}
\Xi_a^{-1}=0 \quad \ , \quad \frac{1}{\Xi_a} \: \alpha_a=0 \ . \ee 
We have also $w_b{}^{(0)}=w_c{}^{(0)}=w^{(0)}$ and  
${\cal H}_b{}^{(0)}={\cal H}_c{}^{(0)} ={\cal H}^{(0)}$. 
In this limit, the dynamical equations  (\ref{D.1}) and (\ref{D.2}) for 
$\alpha_d$ and $\Xi_d$ become (we suppress the index d) : 
\be \label{2.28} 
i \: \dot \alpha^{(0)} = \tau \: w^{(0)} \ , \ee 
\be \label{2.29} 
i \: \dot \Xi^{(0)} = - \left[ \left( \Xi^{(0)} + \tau \right) 
\: {\cal H}^{(0)} \: 
\left( \Xi^{(0)} - \tau \right) - \left( \Xi^{(0)} - 
\tau \right) \: {\cal H}^{(0)} \: 
\left( \Xi^{(0)} + \tau \right) \right] \ . \ee
These equations are the analog for the $ \Phi^4$ theory of the time-dependent 
Hartree-Bogoliubov  (TDHB) equations in nonrelativistic physics. They are
equivalent to the dynamical equations obtained in reference  \cite{1a} 
where the authors used an alternative form of the  Balian-V\'en\'eroni
variational principle suited to the evaluation of single-time expectation
values \cite{5}.  

In reference \cite{POISSON}, we have shown that these TDHB equations for the
$\Phi^4$ theory can be written as classical dynamical equations with a
nonsymplectic Poisson structure with the free-energy density playing the
role of a classical Hamiltonian. The Heisenberg invariant (\ref{2.7i})
appears as a structural invariant of the Poisson structure. 

The static HB solution satisfies  $ w_0=0$ that is, in the uniform case : 
\be \label{2.29aa}
(m_0^2+\frac{b}{6}\varphi_0+\frac{b}{2}G_0({\bf x},{\bf x}))\varphi_0=0 \ee
and the so-called gap  
\be \label{2.29a}
 \Xi_0=-\tau \: \coth(\beta  {\cal H}_0 \tau) \ee
which can also be written as : 
\be \label{2.29b}
\beta  {\cal H}_0=\frac{1}{2}\: \tau \ln \left(\frac{ \Xi_0+\tau}
{ \Xi_0-\tau}\right) \ . \ee
More explicitly, in the uniform case, by going to the momentum space : 
\be \label{2.29c}
 \Xi_0({\bf p})=\pmatrix{\frac{1}{\omega_{{\bf p}}} 
   \: \coth \frac{\beta \omega_{{\bf p}}}{2}
& 0 \cr 0 & -\omega_{{\bf p}} \: \coth \frac{\beta \omega_{{\bf p}}}{2} 
  \cr } \ , \ee
and for the $\Phi^4$ theory,
$\omega_{{\bf p}}=\sqrt{- g_0({\bf p})}$ with 
\be \label{2.29d}
g_0({\bf p})=-({\bf p}^2+m_0^2+\frac{b}{2}\varphi_0^2 +
\frac{b}{2}  G_0({\bf x}, {\bf x})) \ . \ee 

The solution $\varphi_0=0$ of equation (\ref{2.29aa}) corresponds to the
symmetric phase while the solution $\varphi_0 \neq 0$ corresponds to the
broken phase.  

We define the self-consistent mass $ m(\beta)$ depending on the temperature
as the solution of the gap equation : 
\be \label{2.29e} 
 m^2(\beta)=m_0^2+\frac{b}{2}\varphi_0^2 + \frac{b}{2} G_0({\bf x}, {\bf
  x}) \ee 
By using eq. (\ref{2.29aa}), we have also : 
\be \label{2.29f} 
 m^2(\beta)=c \left( m_0^2+\frac{b}{2} G_0({\bf x}, {\bf x}) \right) \ee
with $c=1$ in the symmetric phase and $c=-2$ in the broken phase. 

The gap equation (\ref{2.29e}) has ultraviolet divergences and requires  a
regularization. The divergent part $G_0^{{\rm div}}({\bf x}, {\bf x})$
of $G_0({\bf x}, {\bf x})$ comes from the zero temperature contribution. By
introducing a cutoff $\Lambda$ in momentum, 
\be \label{2.29g}
G_0^{{\rm div}}({\bf x}, {\bf x})=\frac{1}{8 \pi^2} \left[\Lambda^2-
\bar m^2 \log
\left( \frac{2 \Lambda}{\sqrt e \bar m} \right) \right] \ee
where $\bar m=m(T=0)$.   
The gap equation can be put in a finite form by choosing the following
renormalization conditions \cite{KERMAN} : 
\be \label{2.29h}
\epsilon \mu^2=\frac{m_0^2+\frac{b \Lambda^2}{16
    \pi^2}}{\frac{1}{c}+\frac{b}{16 \pi^2} \log (\frac{2 \Lambda }{\sqrt e
    \mu})} \: \: \ , \: \: \epsilon=0, \pm 1 \ee 
\be \label{2.29i} 
\frac{1}{g_R(\mu)}=\frac{2}{c b}+ \frac{1}{8 \pi^2} \log \left(\frac{2
  \Lambda}{\sqrt e \mu} \right) \ee
where $\mu$ is an arbitrary mass scale. The renormalized gap equation writes
at zero temperature : 
\be \label{2.29j}
\bar m^2=\epsilon \mu^2 + \frac{g_R(\mu)}{16 \pi^2} \: \bar m^2 \log
\left(\frac{\bar m^2}{\mu^2} \right) \ee 

In the next sections, the dynamical equations for the two-time correlation
functions will involve the bare coupling constant $b$. This one will be
understood as a function of the cutoff $\Lambda$. We will solve  the
dynamical equations  for a finite value of the cutoff. In the static case,
we will show that the final formula obtained for the two-time correlation
functions can be written with the renormalized parameters. 

At finite temperature, we can isolate the divergent part of $G_0({\bf
  x},{\bf x})$ according to : 
\be \label{2.29k}
G_0({\bf x}, {\bf x})=\int \frac{d^3k}{(2 \pi)^3} \: \frac{1}{2 \omega_k} +
\int \frac{d^3k}{(2 \pi)^3} \: \frac{1}{\omega_k} \frac{1}{e^{\beta
    \omega_k}-1} \ee 
Therefore, the gap equation at finite temperature writes : 
\be \label{2.29l} \ba{ll} 
m^2(\beta)=& {\displaystyle 
  \epsilon \mu^2 + \frac{g_R(\mu)}{16 \pi^2} m^2(\beta) 
  \log \left(\frac{m^2(\beta)}{\mu^2} \right) } \\ 
& {\displaystyle + \frac{g_R(\mu)}{16 \pi^2} m^2(\beta)  \int 
\frac{d^3k}{(2 \pi)^3} \: \frac{1}{\omega_k} \frac{1}{e^{\beta
    \omega_k}-1} } \ea \ee 

\subsubsection{Two-time correlation functions  :}

The first derivatives of  $W(J,t_0)=-i \: \ln {\cal Z}_{st}$ with respect to
the sources are equal to the expectations values with the index $c$. Indeed,
the functional  ${\cal Z}$ depend on the sources both explicitly and
implicitly since the approximate solutions for  ${\cal D}(t)$ and   
${\cal A}(t)$ depend on the sources. However, at the stationary point, only
the explicit dependence contributes to the first derivative :  
\be \label{2.30} 
\frac{d {\cal Z}}{d J_j(t)}=
\frac{\delta {\cal Z}}{\delta J_j(t)} = i \: Tr \left( {\cal D}(t) {\cal A}(t) 
Q_j \right) \ , \ee
which gives for instance  : 
\be \label{2.31}
\frac{\delta W }{\delta J_{\Phi}({\bf x},t)} = \varphi_c({\bf x},t) \ . \ee
The expressions for the second derivatives of $W$ are much more complicated. 
The introduction of sources coupled to the composite operators 
$\Phi({\bf x}) \Phi({\bf y}), \Phi({\bf x}) \Pi({\bf y})$ and  
$\Pi({\bf x}) \Pi({\bf y})$ together with eq.  (\ref{2.31})  allows us to
obtain dynamical equations for the two-time correlation functions with three
or four field operators merely from the expansion of the expectation values 
$\alpha_c$ and $\Xi_c$ at the first order in powers of the sources. 

From the first order corrections 
$\alpha_c - \alpha^{(0)}$ and $\Xi_c-\Xi^{(0)}$, we define the two-time
correlation functions    $\beta$ and $\Sigma$ 
($\beta$ is a vector and  $\Sigma$ is a matrix) : 
\be \ba{ll} \label{2.32} 
\alpha_c({\bf x},t)- 
& {\displaystyle \alpha^{(0)}({\bf x},t) \simeq    i \: \int_{t_0}^{\infty} 
dt'' \: \left \{ 
\int d^3y \: \beta^{\Phi}({\bf x}, {\bf y},t,t'') \: J^{\Phi}({\bf y} ,t'') 
+   \beta^{\Pi} \:  J^{\Pi} \right. 
} \\ 
&{\displaystyle  \left. + 
 \int d^3x_1 d^3x_2 \: \beta^{\Phi \Phi}({\bf x}, {\bf x}_1,{\bf x}_2,t,t'') 
\: J^{\Phi\Phi}({\bf x}_1,{\bf x}_2,t'')  + \beta^{\Phi \Pi} \: J^{\Phi \Pi} 
+ \beta^{\Pi \Pi} \: J^{\Pi \Pi} \right \} 
} \ea \ , \ee
\be \ba{ll} \label{2.33}
\Xi_c({\bf x}, {\bf y},t)- 
& {\displaystyle \Xi^{(0)}({\bf x}, {\bf y},t) \simeq  
 i \: \int_{t_0}^{+ \infty} dt''\: \left \{ \int d^3x_1 \: 
\Sigma^{\Phi}({\bf x}, {\bf y}, {\bf x}_1,t, t'') \: J^{\Phi}({\bf x}_1,t'') + 
\Sigma^{\Pi} \: J^{\Pi} \right. } \\
& {\displaystyle \left. +   \int d^3x_1 d^3x_2 \: 
\Sigma^{\Phi \Phi}({\bf x}, {\bf y},{\bf x}_1, {\bf x}_2,t, t'') 
\: J^{\Phi \Phi}({\bf x}_1,{\bf x}_2, t'') + 
\Sigma^{\Phi \Pi} \: J^{\Phi \Pi} +
\Sigma^{\Pi \Pi} \: J^{\Pi \Pi} \right \} 
}  \ea \ . \ee
We have :  
\be \label{2.34}
\beta_1^{\Phi \Phi}({\bf x},{\bf y},{\bf z},t,t") =\frac{1}{2} \:  
\Sigma^{\Phi}_{11}({\bf x}, {\bf y}, {\bf z}, t,t") \ . \ee 
The functions  $\beta$ and $\Sigma$ provide approximations for the exact
two-time correlation functions  $C^2$, $C^3$ and $C^4$ defined by Eqs. 
 (\ref{2.3d})-(\ref{2.3f}) : 
\be \label{2.35} 
 C^2_{\Phi \Phi}({\bf x}, {\bf y},t, t") \simeq 
\beta^{\Phi}_1({\bf x}, {\bf y}, t,t") \ , \ee  
\be \ba{lll} \label{2.36} 
C^3({\bf  x}, {\bf  y}, {\bf z},t,t") \simeq 
& {\displaystyle     
\beta^{\Phi \Phi}_1({\bf x}, {\bf y}, {\bf z}, t, t") } \\ 
& {\displaystyle -  
 \varphi^{(0)}({\bf y},t") \: 
\left(\beta_1^{\Phi}({\bf x}, {\bf z},t,t") + \varphi^{(0)}({\bf x},t) 
\varphi^{(0)}({\bf z},t") \right)  
} \\ 
& {\displaystyle -  \varphi^{(0)}({\bf z},t") \: 
\left(\beta_1^{\Phi}({\bf x}, {\bf y},t,t") + \varphi^{(0)}({\bf x},t) 
\varphi^{(0)}({\bf y},t") \right)  } \ea \ , \ee 
\be \ba{lll} \label{2.37a} 
C^4({\bf x}, {\bf y}, {\bf z},{\bf u}, t,t") \simeq 
& {\displaystyle \frac{1}{2} \: 
\Sigma_{11}^{\Phi \Phi}({\bf x}, {\bf y}, {\bf z}, {\bf u},t,t") } \\ 
& {\displaystyle - \left( \beta_1^{\Phi}({\bf x}, {\bf z},t,t'') + 
\varphi^{(0)}({\bf x},t) \: \varphi^{(0)}({\bf z},t") \right) \: 
 \left( \beta_1^{\Phi}({\bf y}, {\bf u},t,t'') + 
\varphi^{(0)}({\bf y},t) \: \varphi^{(0)}({\bf u},t") \right) } \\  
& {\displaystyle - \left( \beta_1^{\Phi}({\bf x}, {\bf u},t,t'') + 
\varphi^{(0)}(\vec x,t) \: \varphi^{(0)}({\bf u},t") \right) \: 
 \left( \beta_1^{\Phi}({\bf y}, {\bf z},t,t'') + 
\varphi^{(0)}({\bf y},t) \: \varphi^{(0)}({\bf z},t") \right) } \ea \ . \ee

The expansion in powers of the sources of the dynamical equations  
(\ref{2.20}) and  (\ref{2.22}) for $\alpha_c$ and $\Xi_c$ yields  : 
\be \label{2.37}
i \: \delta \dot \alpha_c = \tau \: \left(\delta w_c- \delta w^c_K \right)  
\ , \ee
\be \ba{ll} \label{2.38} 
i \: \delta \dot \Xi_c = 2 \: & {\displaystyle  \left[
\delta \Xi_c \: {\cal H}^{(0)} \: \tau - \tau \: {\cal H}^{(0)} \: 
\delta \Xi_c \right. } \\
& {\displaystyle \left.  + \Xi^{(0)} \: \left( \delta {\cal H}_c- 
\delta {\cal I}^c_K 
\right) \: \tau - 
\tau \: \left( \delta {\cal H}_c- \delta {\cal I}^c_K \right) \: 
\Xi^{(0)} \right] } \ea 
\ . \ee 
In these equations the matrix  ${\cal H}$ has to be evaluated for the TDHB
solutions  $\alpha^{(0)}, \Xi^{(0)}$ of eqs. (\ref{2.28}) and (\ref{2.29}). 
The variations of $w^c_K({\bf x},t)$ (eqs. (\ref{D.14})-(\ref{D.15})) 
and of  ${\cal I}^c_K({\bf x},{\bf y},t)$ (eqs. (\ref{D.16})-(\ref{D.18}))  
with respect to  $J(t")$ will give terms proportional to $\delta(t-t")$. 
From the variations    $\delta w^c$ and $\delta {\cal H}^c$, we define the
matrices  
 $t_{ij}, T_{i,jk}, r_{ij,k},R_{ij,kl}$ (which are the analogues of the
 random phase approximation (RPA) kernel of  Balian and V\'en\'eroni 
 \cite{1}) : 
\be \label{2.39} 
\delta w^c_i({\bf x},t)= \int d^3y \: t_{ij}({\bf x},{\bf y},t) 
\: \delta \alpha^c_j({\bf y},t) 
-\frac{1}{2} \: \int d^3y \: d^3z \:  T_{i,jk}({\bf x}, {\bf y},{\bf z}, t) 
\: \delta 
\Xi^c_{kj}({\bf z},{\bf y},t) \ , \ee 
\be \label{2.40} 
\delta {\cal H}^c_{ij}({\bf x},{\bf y},t)=\int d^3z \: 
r_{ij,k}({\bf x},{\bf y},{\bf z},t) \: 
\delta \alpha^c_k({\bf z},t) - \frac{1}{2} \: \int d^3z \: d^3u \:  
R_{ij,kl}({\bf x}, {\bf y},{\bf z}, {\bf u}, t) 
\: \delta \Xi^c_{lk}({\bf u}, {\bf z},t)  \ . \ee
These matrices depend on the  TDHB solutions $\alpha^{(0)}$ and
$\Xi^{(0)}$. Their expressions in the case of the $ \Phi^4$ theory are given
in Appendix D. With these notations, the dynamical equations for the
two-time and two-point correlation function  $\beta^{\Phi}$ and the two-time
and three-point correlation function  $\Sigma^{\Phi}$ write : 
\be \ba{ll} \label{2e41} 
i \: \frac{d}{dt} \beta^{\Phi} ({\bf x},{\bf y},t,t")= \tau & {\displaystyle 
\left[ t({\bf x},{\bf z},t) \: \beta^{\Phi}({\bf z},{\bf y},t,t") - 
\frac{1}{2} \: T_{,(jk}({\bf x}, {\bf z},{\bf u},t) \: \Sigma^{\Phi}_{kj)}
({\bf u}, {\bf z}, 
{\bf y},t,t") \right. } \\ 
& {\displaystyle \left. 
+ i \: \pmatrix{1 \cr 0 \cr } \delta({\bf x} - {\bf y}) \: 
\delta(t-t") \right] } \ea \ , \ee 
\be \ba{ll} \label{2e42} 
i \: \frac{d}{dt} \Sigma^{\Phi}_{ij} ({\bf x},{\bf y},{\bf x}_1,t,t")= 
& {\displaystyle 2 \: \left[ \Sigma^{\Phi} \: {\cal H}^{(0)} \: \tau 
 - \tau \: {\cal H}^{(0)} \: \Sigma^{\Phi}  \right. } \\
& {\displaystyle \left.   + \Xi^{(0)} \: r_{,(k} \: \beta^{\Phi}_{k)} \: \tau 
 - \tau \: r_{, (k} \: \beta^{\Phi}_{k)} \: \Xi^{(0)}  \right. } \\ 
& {\displaystyle \left. -\frac{1}{2} \: \Xi^{(0)}
 \: R_{,(kl} \: \Sigma^{\Phi}_{lk)} \: 
\tau + \frac{1}{2} \: \tau \: R_{,(kl} \: 
\Sigma^{\Phi}_{lk)} \Xi^{(0)} \right]_{ij} 
} \ea \ . \ee  
We obtain similar equations for the three-point functions  
$\beta^{\Phi \Phi}, \beta^{\Phi \Pi}$,  $\beta^{\Pi \Pi}$ and the four-point
functions  
$\Sigma^{\Phi \Phi}, \Sigma^{\Phi \Pi}$ and  $\Sigma^{\Pi \Pi}$. 

These dynamical equations are not sufficient since the boundary conditions
are
$ \Xi_a^{-1}(t=+\infty)=0, \frac{1}{\Xi_a} \alpha_a(t=+\infty)=0$ and
$\alpha_d({\bf x},t_0)=\alpha_0({\bf x}), \:
 \Xi_d({\bf x}, {\bf y},t_0)=\Xi_0({\bf x},{\bf y})$.
But, at this step, we have used the expansion of only one half of our
variational parameters. 
We will use also the dynamical equations for the two-time functions $m$ and $l$
defined from the expansion in powers of the sources of 
$\Xi_a^{-1}\alpha_a$ and of  $\alpha_a$ at the first order : 
  \be \label{2e43} 
  \frac{1}{\Xi_a}\alpha_a (t)\simeq i \: \int_{t_0}^{+ \infty} dt" \: 
  m^k(t",t) J_k(t") \ , \ee 
  \be \label{2e44} 
  \Xi_a^{-1}(t)\simeq i \: \int_{t_0}^{+ \infty} dt" \: l^k(t",t) J_k(t") \ . \ee 
  (We have ommitted the space indices) 
  
From the relations of Appendix C, we will be abble to obtain relations
between the functions  $\Sigma^k(t_0,t'')$ and  $l^k(t'',t_0)$ one the one
hand and the functions  $\beta^k(t_0,t'')$ and
$m^k(t'',t_0)-l^k(t'',t_0)\alpha_0$ on the other hand. 
Indeed,  that at the first order : 
\be \label{2e45} 
\delta \Xi_a^{-1}=-\frac{1}{\Xi_a}\: \delta \Xi_a \: \frac{1}{\Xi_a} 
\simeq - \frac{1}{\Xi_a+\Xi_d} \: \delta \Xi_a \: \frac{1}{\Xi_a + \Xi_d} \ ,
\ee
From eqs. (\ref{C.5}) and (\ref{C.2}), we obtain at the first order :  
\be \label{2e46} 
\delta \Xi_c=-(\Xi^{(0)}-\tau) \: \delta \Xi_a^{-1} \: (\Xi^{(0)}+ \tau) + 
\delta \Xi_d \ee 
and \be \label{2e47} 
\delta \alpha_c = (\Xi^{(0)}- \tau) \: \delta(\frac{\alpha_a}{\Xi_a+\Xi_d}) 
+ (\Xi^{(0)}-\tau) \: \frac{1}{\Xi_a+ \Xi_d} \: \delta \Xi_a \: 
\frac{1}{\Xi_a+ \Xi_d} \: \alpha^{(0)} + \delta \alpha_d \ . \ee
{\it At the initial time}, $\delta \Xi_d=0, \delta \alpha_d=0$ 
and we have the following relations between the two-time functions : 
\be \label{2e50} 
\Sigma^k(t_0,t")=-(\Xi_0-\tau) \: l^k(t",t_0) \: (\Xi_0+\tau) \ee 
\be \label{2e51} 
\beta^k(t_0,t")=(\Xi_0-\tau)\: (m^k(t",t_0)-l^k(t",t_0) \alpha_0
) \ee  

By using the dynamical equation for  $\Xi_a^{-1}$ at the first order  
\be  \label{2e52}
i \: \frac{d}{dt} \left( \Xi_a^{-1} \right) =
-{\cal H}_c+{\cal H}_b+{\cal I}_K^c +2 \: \Xi_a^{-1} \tau {\cal H}^{(0)} -
2 \: {\cal H}^{(0)} \tau \Xi_a^{-1}   \ ,  \ee
and the variations  $\delta w_c-\delta w_b$ and 
$\delta {\cal H}_c - \delta {\cal H}_b$, we obtain a backward dynamical
equation for  $l^k(t'',t)$ :   
  \be \ba{ll} \label{2e53}
i \: \frac{d}{dt} l^k(t",t)_{ij}=
& {\displaystyle 2 \: r_{ij,k} \: \tau_{kl} \: (m^k-l^k\alpha^{(0)})_l 
- R_{ij,kl} \: \left( \Xi^{(0)} \: l \: \tau - \tau \: l \:\Xi^{(0)}  
\right)_{lk} } \\
& {\displaystyle + \left[ 2 \: l^k \tau {\cal H}^{(0)} 
- 2 \: {\cal H}^{(0)} \: 
\tau \: l^k - i \: \frac{\delta {\cal I}^c_K}{\delta J_k}
\: \delta(t-t") \right]_{ij} } \ea \ ,
\ee
where the matrices  $r$ and $R$ have been defined in  (\ref{2.40}). 

Similarly, from the dynamical equation for
  $\Xi_a^{-1} \alpha_a$ at the first order, we obtain a backward dynamical
  equation for  $m^k(t'',t)$. It is actually more convenient 
  to consider the function
  $m^k(t'',t)-l^k(t'',t) \alpha^{(0)}(t)$ which is solution of :
  \be \ba{ll} \label{2e54}
  i \: \frac{d}{dt} \left( m^k(t",t)-l^k(t",t) \alpha^{(0)}(t) \right) =
  & {\displaystyle t \: \tau \: \left( m^k(t",t)-l^k(t",t) \alpha^{(0)}(t)
  \right) } \\
  & {\displaystyle -\frac{1}{2} \: T_{,jk}\: (\Xi^{(0)}\: l^k \: \tau
  -\tau \: l^k \: \Xi^{(0)})_{kj} + \frac{i}{2} \: \frac{\delta w_K^c}
  {\delta J_k} \: \delta(t-t") } \ea \ , \ee
where the matrices $t$ and $T$ have been defined in  (\ref{2.39}). 
The reason to use this function will appear with the introduction of
the function  $F(t',t'')$ bellow (see \ref{2e57}). 

The boundary conditions for these two equations are : 
   \be \label{2e55}
   l(t",t)=0, \: \:  m(t",t)=0 \quad{\rm pour}\quad t>t" \ee
   \be \label{2e55a}
   l^{\Phi}(t",t")=0 \: \: \ , \: \: l^{\Pi}(t",t")=0 \ee
   \be
   l^{\Phi \Phi}_{11}({\bf x},{\bf y},{\bf z},{\bf u},t",t")=
   -\frac{1}{2}(\delta^3({\bf x}-{\bf z}) 
   \delta^3({\bf y}-{\bf u})+\delta^3({\bf x}-{\bf u}) \delta^3({\bf y}-{\bf z}) ) \ee
   \be
   l^{\Phi \Pi}_{12}(t",t")=2i \: \:  \ , \: \: l^{\Phi \Pi}_{21}(t",t")=2i 
   \: \: \ , 
   \: \: l^{\Pi \Pi}_{22}(t",t")=1 \ee
The other elements of the matrix  $l$ are equal to zero.  
   \be \label{2e56} 
   m^{\Phi}_1({\bf x},{\bf y},t",t")=-\frac{1}{2}\delta^3({\bf x}-{\bf y}) 
   \: \: \ , \: \: 
   m^{\Pi}_2({\bf x},{\bf y},t",t")=-\frac{i}{2}\delta^3({\bf x}-{\bf y}) \ee
  \be \label{2e56a}
  m_1^{\Phi \Phi}({\bf x},{\bf x}_1,{\bf x}_2,t",t")-
  l^{\Phi \Phi}_{1i}(t",t") \alpha_i^{(0)}(t")=-\varphi^{(0)}({\bf x}_2,t") 
  \delta^3({\bf x}-{\bf x}_1) + {\rm sym.} \ee
\be \label{2e56b}
  m_1^{\Phi \Pi}({\bf x},{\bf x}_1,{\bf x}_2,t",t")-
  l^{\Phi \Pi}_{1i}(t",t") \alpha_i^{(0)}(t")=-\pi^{(0)}({\bf x}_2,t")
  \delta^3({\bf x}-{\bf x}_1) + {\rm sym.} \ee
  \be \label{2e56c}
  m_2^{\Phi \Pi}({\bf x},{\bf x}_1,{\bf x}_2,t",t")-
  l^{\Phi \Pi}_{2i}(t",t") \alpha_i^{(0)}(t")=-i\varphi^{(0)}({\bf x}_2,t")
  \delta^3({\bf x}-{\bf x_1}) + {\rm sym} \ee
\be \label{2e56d}
  m_2^{\Pi \Pi}({\bf x},{\bf x}_1,{\bf x}_2,t",t")-
  l^{\Pi \Pi}_{2i}(t",t") \alpha_i^{(0)}(t")=-i\pi^{(0)}({\bf x}_2,t")
  \delta^3({\bf x}-{\bf x_1}) + {\rm sym} \ee
where sym. means the same terms obtained by exchanging $x_2,x$ and $x_1$. 

The same matrices  $t,T,r,R$ coming from the expansion of  $<H>$ 
at the second order arround the TDHB solution appear in equations 
(\ref{2e41}),(\ref{2e42}) for  $\beta$ and $\Sigma$ and in equations 
(\ref{2e53}),(\ref{2e54}) for  $m$ and $l$. This leads to consider the
following quantity :   
\be \label{2e57} 
F^k(t',t")=2 \left( \tilde m^k(t',t)-\tilde \alpha^{(0)}(t) l^k(t',t) \right) 
\beta^k(t,t") 
-\frac{1}{2} \: tr\left( l^k(t',t)\Sigma^k(t,t")\right) 
\ee
where $\tilde m$ is the transpozed vector of  $m$. 
More explicitly, by reintroducing the space coordinates, 
$F^{\Phi}(t',t'')$ writes in the case of the functions  $m^{\Phi}, l^{\Phi},
\beta^{\Phi}$ and $\Sigma^{\Phi}$ : 
\be \label{2e57a} \ba{ll} 
F^{\Phi}(t',t'')= 
& {\displaystyle 2 \int d^3x \left( \tilde m^{\Phi}({\bf x}, {\bf 
x}_1,t',t)-\int d^3u \: \tilde \alpha^{(0)}({\bf u},t) l^{\Phi}({\bf  u}, 
{\bf  x},
{\bf  x}_1,t',t) \right) 
\beta^{\Phi}({\bf  x}_1, {\bf x}_2,t,t") } \\
& {\displaystyle -\frac{1}{2} \int d^3x \: d^3z \:
tr\left( l^{\Phi}({\bf x}, {\bf z}, {\bf x}_1,t',t)\Sigma^{\Phi}({\bf  z},
{\bf x}, {\bf x}_2,t,t")\right) } \ea \ee
The quantity $F^{k}$ is independent of the time $t$ : from the dynamical
equations  (\ref{2e41}),(\ref{2e42}) and (\ref{2e53}),(\ref{2e54}), one
shows that its derivative with respect to $t$ is equal to a sum of functions 
$\delta(t-t')$ and $\delta(t-t")$ :  
\be \label{2e58} \ba{ll}
i \frac{d}{dt} F^{k} =& {\displaystyle 
i \: \tilde \beta^k(t,t'') \frac{\delta w^c_K}{\delta J_k} 
\delta(t-t') + i 2 (\tilde m^k(t', t) - \tilde \alpha^{(0)}(t) l^k(t',t)) \tau 
\frac{\delta w^c}{\delta J_k} \delta(t-t'') } \\ 
& {\displaystyle 
-\frac{1}{2} tr \left\{i \frac{\delta {\cal I_K}}{\delta J_k} 
\Sigma(t,t'') \delta(t-t')+i l^k(t',t) 
\Xi^{(0)}(t) \frac{\delta {\cal I}_K}{\delta J_k} \tau \delta(t-t'') 
-i l^k(t',t) \tau \frac{\delta {\cal I}_K}{\delta J_k} \Xi^{(0)}(t) \delta(t-t'') 
\right \} } \ea \ . \ee
The fact that $F^k$ does not depend on the time t will save us to solve one
dynamical equation. 

By considering the local sources  $J^{\Phi}({\bf x},t)$ and by integrating
equation  (\ref{2e58}) from $t=+\infty$ to $t=t_0$, we obtain an
approximation for the two-time and two-point function. For  $t''>t'$ : 
\be \label{2e58a} \ba{ll}
\beta^{\Phi}_1(t',t'')=
& {\displaystyle 2\left[\tilde m^{\Phi}(t',t_0)-\tilde \alpha_0
l^{\Phi}(t',t_0)\right] \beta^{\Phi}(t_0,t'') } \\
& {\displaystyle -\frac{1}{2}tr \left[l^{\Phi}(t',t_0)
\Sigma^{\Phi}(t_0,t'') \right] } \ea \ . \ee 
In the uniform case, we have in momentum space :  
\be \label{2e58b} \ba{ll}
\beta_1^{\Phi}({\bf p},t',t'')= & {\displaystyle  2 \left(\tilde
m^{\Phi}(-{\bf p},t',t_0)- \tilde \alpha_0 l^{\Phi}(0,{\bf p},-{\bf p},t',t_0) 
\right)
\beta^{\Phi}({\bf p},t_0,t'') } \\
& {\displaystyle -\frac{1}{2} \int d^3k \: tr \left[l^{\Phi}({\bf p}+{\bf k}
,-{\bf k},-{\bf p},t',t_0)
\Sigma^{\Phi}({\bf k},-{\bf p}-{\bf k},{\bf p},t_0,t'')\right] } \ea
\ee
Our conventions for the Fourrier transforms are defined in Appendix D. 

These two last expressions will still be valid when we will optimize with
respect to the initial state in section 3. On the contrary, the following
expression will not be valid in section 3. By using   
(\ref{2e50}) and (\ref{2e51}), we obtain :  
 \be \label{2e59} \ba{ll}
 \beta_1^{\Phi}(t',t")=
 & {\displaystyle 2 (\tilde m^{\Phi}(t',t_0)-\tilde  \alpha_0 
 l^{\Phi}(t',t_0)) \: (\Xi_0-\tau) \: (m^{\Phi}(t",t_0)-l^{\Phi}(t",t_0) 
 \alpha_0 ) } \\ 
 & {\displaystyle +\frac{1}{2} tr \left[
 l^{\Phi}(t',t_0)(\Xi_0-\tau)l^{\Phi}(t",t_0)(\Xi_0+\tau) \right] } 
 \ea \ . \ee
 For $t'>t"$, one interchanges the operators  $\phi^H(t',t_0)$ and  
 $\phi^H(t",t_0)$. By using the expression of the retarded Green functions
 (see bellow), we show :   
 \be \label{2e60} \ba{ll} 
 \beta_1^{\Phi}(t',t")= 
 & {\displaystyle 2 (\tilde m^{\Phi}(t",t_0)-\tilde  \alpha_0 
 l^{\Phi}(t",t_0)) \: (\Xi_0-\tau) \: (m^{\Phi}(t',t_0)-l^{\Phi}(t',t_0) 
 \alpha_0 ) } \\ 
 & {\displaystyle +\frac{1}{2} tr \left[
 l^{\Phi}(t",t_0)(\Xi_0-\tau)l^{\Phi}(t',t_0)(\Xi_0+\tau) \right] } 
 \ea \ . \ee

For the correlation functions with the anti-T-product, it is sufficient to
replace in the functional  ${\cal Z}$ (\ref{2.15}) the source term 
${\cal K}_c$ by the analogous term ${\cal K}_b$. This is equivalent to
exchange the matrices  $\Xi_0-\tau$ and $\Xi_0+\tau$ in the previous formula
(or to exchange  $t'$ and $t''$). 

From the function  $F^k(t',t")$, we obtain also, by considering now the
bilocal sources  
$J^{\Phi \Phi}({\bf x},{\bf y},t)$, an expression which relates the
approximations for the two-time and three-point function  
$\beta_1^{\Phi \Phi}(t',t")$ and for the two-time and four-point function 
$\Sigma^{\Phi \Phi}_{11}(t',t")$. For $t">t'$ : 
\be \label{2e61} \ba{llll} 
& {\displaystyle   
\beta_1^{\Phi \Phi}({\bf x}_1, {\bf x}_1', {\bf x}_2',t',t") 
\varphi^{(0)}({\bf x}_2,t') + 
\beta_1^{\Phi \Phi}({\bf x}_2, {\bf x}_1', {\bf x}_2',t',t") 
\varphi^{(0)}({\bf x}_1,t') } \\
& {\displaystyle 
 - \frac{1}{4} \left( \Sigma_{11}^{\Phi \Phi}({\bf x}_1, {\bf x}_2, {\bf x}_1',
 {\bf x}_2', t',t'') + \Sigma_{11}^{\Phi \Phi}({\bf x}_2, 
 {\bf x}_1, {\bf x}_1', {\bf x}_2', t',t'') \right) = } \\ 
& {\displaystyle -
2 \int d^3x  \: \left( \tilde m^{\Phi \Phi}({\bf x}, {\bf x}_1, {\bf x}_2,
t',t_0)
- \int d^3u \: \tilde \alpha_0({\bf u}) l^{\Phi \Phi}({\bf u}, {\bf x}, {\bf 
x}_1, {\bf x}_2,t',t_0) \right) 
\beta^{\Phi \Phi}({\bf x}, {\bf x}_1', {\bf x}_2',t_0,t") } \\
& {\displaystyle + \frac{1}{2} \int d^3x \: d^3z \: 
tr \left[ l^{\Phi \Phi}({\bf x}, {\bf z}, {\bf x}_1, {\bf x}_2,t',t_0) 
\Sigma^{\Phi \Phi}({\bf z}, {\bf x}, {\bf x}_1', {\bf x}_2',t_0,t'') \right]
} \ea   \ . \ee 
In the uniform case, we have in momentum space :  
\be \label{2e61a} \ba{lllll}
& {\displaystyle \left[ \beta_1^{\Phi \Phi}({\bf p}_1,{\bf q},-({\bf p}_1+
  {\bf q}),t',t'')\delta^3({\bf p}_2)+ 
\beta_1^{\Phi \Phi}({\bf p}_2,{\bf q},-({\bf p}_2+{\bf q}),t',t'') 
\delta^3({\bf p}_1) \right] \varphi^{(0)}(t') } \\
& {\displaystyle -\frac{1}{4} \left[ \Sigma_{11}^{\Phi
\Phi}({\bf p}_1,{\bf p}_2,{\bf q},-({\bf p}_1+{\bf p}_2+{\bf q}),t',t'') +  
\Sigma_{11}^{\Phi
\Phi}({\bf p}_2,{\bf p}_1,{\bf q},-({\bf p}_1+{\bf p}_2+{\bf q}),t',t'')  
\right] = } \\
& {\displaystyle -2 \left[ \tilde m^{\Phi
\Phi}(-{\bf p}_1-{\bf p}_2,{\bf p}_1,{\bf p}_2,t',t_0)-\tilde \alpha_0 l^{\Phi
\Phi}(0,-{\bf p}_1-{\bf p}_2,{\bf p}_1,{\bf p}_2,t',t_0) \right] } \\ 
& {\displaystyle \times 
\beta^{\Phi
\Phi}({\bf p}_1+{\bf p}_2,{\bf q},-{\bf q}-{\bf p}_1-{\bf p}_2,t_0,t'') } \\
& {\displaystyle + \frac{1}{2} \int d^3k \: tr\left[l^{\Phi
\Phi}({\bf k}-{\bf p}_1-{\bf p}_2,-{\bf k},{\bf p}_1,{\bf p}_2,t',t_0) 
\Sigma^{\Phi
\Phi}({\bf k},-{\bf k}+{\bf p}_1+{\bf p}_2,{\bf q},-{\bf q}-{\bf p}_1-
{\bf p}_2,t_0,t'') \right] } \ea \ee

Let us recall that the function  $\Sigma_{11}^{\Phi }$  is related to 
$\beta_1^{\Phi \Phi}(t',t")$ by : $ \beta_1^{\Phi \Phi}=\frac{1}{2}
\Sigma_{11}^{\Phi}$. By comparing the dynamical equations for 
$\beta_i^{\Phi \Phi}(t',t'')$ and $\Sigma_{ij}^{\Phi }(t',t'')$ we obtain on
the other hand : $\beta_2^{\Phi \Phi}=\Sigma_{12}^{\Phi}+ \Sigma_{21}^{\Phi}$
and an expression for  $\Sigma_{22}^{\Phi }$ in terms of  $\beta_1^{\Phi
\Phi}$ and two integrales over the momenta which involve 
$\Sigma_{11}^{\Phi \Phi}$ and $\Sigma_{11}^{\Phi}=2 \beta_1^{\Phi \Phi}$. 
The functions  $\Sigma_{ij}^{\Phi}$ are therefore completly known if 
$\beta_1^{\Phi \Phi}$ and $\Sigma_{11}^{\Phi \Phi}$ are known. However, we
note a difficulty in the asymmetric phase since we have two equations 
(\ref{2e58a}) and (\ref{2e61}) for the three functions 
$\beta_1^{\Phi }(t',t"), \beta_1^{\Phi \Phi}(t',t")$ and 
$\Sigma^{\Phi \Phi}_{11}(t',t")$.
 
{\bf In the symmetric phase}, equations  (\ref{2e59}) and  (\ref{2e61}) 
simplify. Indeed, from the boundary condition 
$l^{\Phi}(t',t')=0$ and the dynamical equation (\ref{2e53}) for $l$, we
deduce  $l^{\Phi}(t',t)=0$ for $t\le t'$. Similarly, 
$m^{\Phi \Phi}(t',t)-l^{\Phi \Phi}(t',t) \alpha^{(0)}(t)=0$
for $t\le t'$. Therefore, we have in the symmetric phase and for  $t''>t'$ : 
\be \label{2e61a}
\beta_1^{\Phi}({\bf p},t',t'')=2 \: \tilde m^{\Phi}({\bf p},t',t_0)
\: (\Xi_0({\bf p})-\tau)\: m^{\Phi}({\bf p},t'',t_0) \ee 
and for the four-point function : \be \label{2e61b} \ba{ll}
\Sigma_{11}^{\Phi \Phi}({\bf p}_1,{\bf p}_2,{\bf p}_3,-\sum {\bf p}_i,t',t'')=
-\:  
\int \frac{d^3p}{(2 \pi)^3}  
& {\displaystyle tr \left[ l^{\Phi \Phi}({\bf p}-{\bf p}_1-{\bf p}_2,-{\bf p},
{\bf p}_1,{\bf p}_2,t',t_0) 
\right.} \\ 
& {\displaystyle \left. 
\Sigma_{11}^{\Phi \Phi}({\bf p},-{\bf p}+{\bf p}_1+{\bf p}_2,{\bf p}_3,
-\sum {\bf p}_i,t_0,t'') \right]
} \ea \ee
By using (\ref{2e50}) : 
\be \label{2e61c} \ba{ll} 
\Sigma_{11}^{\Phi \Phi}({\bf p}_1,{\bf p}_2,{\bf p}_3,-\sum {\bf p}_i,t',t'') 
= & {\displaystyle
\int \frac{d^3p}{(2 \pi)^3} tr\left[
 l^{\Phi \Phi}({\bf p}-{\bf p}_1-{\bf p}_2,-{\bf p},{\bf p}_1,{\bf p}_2,
 t',t_0) (\Xi_0({\bf p})-\tau) \right. } \\
& {\displaystyle \left.
l^{\Phi \Phi}({\bf p},-{\bf p}+{\bf p}_1+{\bf p}_2,{\bf p}_3,-\sum {\bf p}_i,
t'',t_0)
(\Xi_0(-{\bf p}+{\bf p}_1+{\bf p}_2)+\tau) \right] } \ea  \ee
Let us emphasize that it is  $\Xi_0$, the static HB solution and not
$\Xi^{(0)}$ which appears in expression  (\ref{2e61c}). 

For  $t'>t''$, we have, still in the symmetric phase : 
\be \label{2e61f}
\beta_1^{\Phi}({\bf p},t',t'')=2 \: \tilde m^{\Phi}({\bf p},t',t_0)
\: (\Xi_0({\bf p})+\tau)\: m^{\Phi}({\bf p},t'',t_0) \ee 
\be \label{2e61h} \ba{ll} 
\Sigma_{11}^{\Phi \Phi}({\bf p}_1,{\bf p}_2,{\bf p}_3,-\sum {\bf p}_i,t',t'')
= & {\displaystyle
\int \frac{d^3p}{(2 \pi)^3} tr \left[ 
 l^{\Phi \Phi}({\bf p}-{\bf p}_1-{\bf p}_2,-{\bf p},{\bf p}_1,{\bf
   p}_2,t'',t_0) (\Xi_0({\bf p})-\tau) \right. } \\ 
& {\displaystyle \left.  
l^{\Phi \Phi}({\bf p},-{\bf p}+{\bf p}_1+{\bf p}_2,{\bf p}_3,-\sum {\bf p}_i,
t',t_0) (\Xi_0(-{\bf p}+{\bf p}_1+{\bf p}_2)+\tau) \right] } \ea  \ee

To conclude, in order to obtain approximations for the causal 
correlation functions  $C^2,C^3$ and $C^4$, we have first to solve the TDHB
equations   (\ref{2.28}) and (\ref{2.29})
for  $\alpha^{(0)}(t)$ and $\Xi^{(0)}(t)$ and to evaluate the matrices 
$r,R,t$ and $T$ for the TDHB solutions. 
We then solve equations  (\ref{2e53}) and (\ref{2e54})  for 
$l(t",t)$ and $m(t'',t)-l(t'',t)\alpha^{(0)}(t) \: (t''>t\ge t_0)$
backward in time from 
$t''$ to $t_0$ with the boundary conditions  (\ref{2e55})-(\ref{2e56d}). 
Finally, we evaluate expression  (\ref{2e59}). 
Therefore, it is not necessary to solve the dynamical equations 
(\ref{2e41}) and (\ref{2e42}) for  $\beta(t,t")$ and  $\Sigma(t,t'')$ . 

The expressions we obtain in this way for three-point and four point
correlation functions are not completly satisfactory. Indeed, in the static
case, when  $\alpha^{(0)}(t)=
\alpha_0$ and  $\Xi^{(0)}(t)=\Xi_0$ are time-independent, the approximate
correlation functions do not depend only on the time difference 
$t'-t"$ as they should (the two-point function in the symmetric phase is a
particular case). We will cure this artefact by optimizing with respect to the
initial state in section 3. 

{\bf Retarded Green Functions  : }

  The function $\beta_1^{\phi}$ is an approximation for the Green function
  with the time-ordered Green function : 
  \be \label{2e63}  \ba{ll}
  C^2_{\Phi \Phi}({\bf  x},{\bf  y},t,t'')=
  &{\displaystyle Tr \left( T \Phi^H({\bf x},t',t_0) \:
    \Phi^H({\bf y},t'',t_0) D(t_0) \right) } \\
  &{\displaystyle - Tr\left( \Phi^H({\bf x},t',t'') D(t_0) \right) \:
  Tr\left( \Phi^H({\bf y},t'',t_0) D(t_0)\right)} \ea  \ . \ee
To obtain the approximations for the anticausal two-time functions, it is
sufficient to replace in the functional  (\ref{2.15}) the term  $K_c$ 
by the analogous term  $K_b$. Formula (\ref{2e59}) and  
(\ref{2e60}) are then interchanged.  

The retarded Green function is defined by :  
\be \label{2e64} 
\chi_{\Phi \Phi}({\bf x},{\bf  y},t',t'')= -i \: 
\Theta(t'-t'') Tr \left( [\Phi^H({\bf  x},t',t_0), 
\Phi^H({\bf y},t'',t_0)] D(t_0)\right) \ . \ee
It is interesting to obtain an approximation for the retarded Green
function. It is this Green function which appears in the study of the
response of the system to a small external perturbation. Moreover, the KMS
condition allows to relate the different components of the Green function in
the real time formalism for Quantum Field Theories at finite temperature 
\cite{AURENCHE}. These components can all be expressed from the retarded
Green function. The real parts of the time-ordered Green function  $C^2$ 
and of the retarded Green function  $\chi$  coincide. They describe
phenomena such as Debye screening in a plasma in the problem of charged
particules in an external electromagnetic field. But one has to consider the
imaginary part of the retarded Green function to describe damping phenomena 
\cite{DOLAN}. 
The optimization with respect to the initial state in section 3 will
allows to obtain approximated Green functions wich verify the KMS condition 
(the fluctuation-disspation theorem). 

From expression  (\ref{2e60}) and the analogous formula for the correlation
function with the anti-T product, we obtain an approximation for the
retarded Green function :  
\be  \ba{ll} \label{2e65} 
\chi_{\Phi \Phi}({\bf x},{\bf  y},t',t'')= -i \: \Theta(t'-t'')& 
{\displaystyle 
\left[- 4 \left(\tilde m^{\Phi}(t'',t_0)-\tilde \alpha_0 
l^{\Phi}(t'',t_0) \right) \: \tau \: \left( m^{\Phi}(t',t_0)-l^{\Phi}(t',t_0) 
\alpha_0\right)   \right.} \\
&  {\displaystyle  \left. 
 +tr\left( l^{\Phi}(t'',t_0) \Xi_0 l^{\Phi}(t',t_0) \tau
\right) - tr \left(l^{\Phi}(t'',t_0) \tau l^{\Phi}(t',t_0) \Xi_0
\right) \right] }  \ea \ee 
In the symetric phase : 
\be \label{2e65a}
\chi_{\Phi \Phi}(\vec x,\vec y,t',t'')=i \: \Theta(t'-t'') 
\:  4 \: \tilde m^{\Phi}(t'',t_0) \tau  m^{\Phi}(t',t_0) \ . \ee

The exact expression  (\ref{2e64}) remains identical if we translate the
refence time  $t_0$ to another time  $t$ such as $t''>t>t_0$. By using the
dynamical equation for  $\Xi^{(0)}(t)$ and equations
(\ref{2e53}) and (\ref{2e54}), we show that the approximation  (\ref{2e65})
has the same property. Indeed, if we call  $R(t',t'')$ the expression
obtained by replacing in the bracket of equation  (\ref{2e65}) $t_0$ by
$t$ with $t_0\le t<t''$, we have for $t'>t''$ :  
\be \label{2e65b} 
\frac{d R}{dt}=2 (\tilde m^{\Phi}(t',t)-\tilde \alpha^{(0)}(t))
l^{\Phi}(t',t) \tau \pmatrix{1 \cr 0 \cr} \: \delta(t-t'') \ee
We deduce : $\frac{dR}{dt}=0$ pour $t<t''$.
We thus have :  
\be \chi_{\Phi \Phi}({\bf x},{\bf y},t',t'')=-i \: \Theta(t'-t'')
R(t',t'') \ee

When the time $t'$ tends to $ t''$, the exact formula (\ref{2e64}) 
co\" \i ncides with the
commutator of two fields $\Phi^H(t',t_0)$ at equal time which is vanishing. 
The approximated expression gives also  $\chi_{\Phi \Phi}(t',t')=0$ since 
$m_1^{\Phi}(t'',t'')=-\frac{1}{2} $ and $ l^{\Phi}(t'',t'')=0$. 

Similarly, we can obtain approximations for the retarded Green function with
two field operators  $\Phi \Phi$ defined according to :  
\be \label{2e65c} 
\chi_{\Phi \Phi, \Phi \Phi}({\bf x}, {\bf y}, {\bf z}, {\bf u},t',t'')=-i \:  
\theta(t'-t'') \: Tr \left( [\Phi^H({\bf x},t',t_0) \Phi^H({\bf y}, t',t_0), 
\Phi^H({\bf z},t'',t_0) \Phi^H({\bf u},t'',t_0) ] D(t_0) \right) \ . \ee 
In the symmetric phase, from expression  (\ref{2e61h}) and from the analogous
formula for the correlation function with the anti-T product, we deduce an
approximation for the retarded Green function with two field operators : 
\be \label{2e65d} \ba{lll}
& {\displaystyle \chi_{\Phi \Phi, \Phi \Phi} ({\bf p}_1,{\bf p}_2,{\bf p}_3,
  -\sum {\bf p}_i,t',t'')= 
 -i \: \theta(t'-t'') \: \int \frac{d^3p}{(2 \pi)^3} \times  } \\  
& {\displaystyle   
tr[l^{\Phi \Phi}({\bf p}-{\bf p}_1-{\bf p}_2,-{\bf p},{\bf p}_1,{\bf p}_2,
t'',t_0)\: 
\Xi_0({\bf p})  \: l^{\Phi \Phi}({\bf p},-{\bf p}+{\bf p}_1+{\bf p}_2,
{\bf p}_3,-\sum {\bf p}_i,t',t_0)
\: \tau] } \\ 
& {\displaystyle -
tr[l^{\Phi \Phi}({\bf p}-{\bf p}_1-{\bf p}_2,-{\bf p},{\bf p}_1,{\bf p}_2,
t'',t_0)\: 
\tau  \: l^{\Phi \Phi}({\bf p},-{\bf p}+{\bf p}_1+{\bf p}_2,{\bf p}_3,
-\sum {\bf p}_i,t',t_0)
\: \Xi_0(-{\bf p}+{\bf p}_1+{\bf p}_2)]  }    \ea \ee
Again, the right member of this equation does not depend on the time $t_0$.
Indded, if we call  $S(t',t'')$ the expression obtained by replacing in the
right member  $t_0$ by $t$, with $t_0\le t<t''$, 
we have for $t'>t''$ : 
\be \label{2e65e} 
\frac{d S}{dt}=-2 \: tr[l^{\Phi \Phi}(t',t) \: \Xi_0 \: 
\frac{\delta {\cal I}_K}{\delta J^{\Phi \Phi}} \: \tau 
-l^{\Phi \Phi}(t',t) \: \tau \: 
\frac{\delta {\cal I}_K}{\delta J^{\Phi \Phi}} \: \Xi_0] \: \delta(t-t'') 
 \ , \ee
and we deduce : $\frac{d S}{dt}=0$ for $t<t''$. 
We have therefore : 
\be \label{2e65f} 
\chi_{\Phi \Phi, \Phi \Phi}({\bf p}_1,{\bf p}_2,{\bf p}_3,
-\sum {\bf p}_i,t',t'')= 
-i \: \theta(t'-t'') S(t',t'') \ee 

{\bf The limit when the two times t' and t'' co\" \i ncide : }

 When $t'\to t"$, $t'=t"=t+0$, expression  (\ref{2e59}) for the approximated
 two point function becomes :  
\be \ba{ll}  \label{2e66}
 \beta_1^{\Phi}(t,t) =& {\displaystyle 
   2 (\tilde m^{\Phi}(t,t_0)-\tilde  \alpha_0
 l^{\Phi}(t,t_0)) \: (\Xi_0-\tau) \: (m^{\Phi}(t,t_0)-l^{\Phi}(t,t_0)
 \alpha_0 ) }  \\
 & {\displaystyle  
 +\frac{1}{2} tr \left[
 l^{\Phi}(t,t_0)(\Xi_0-\tau)l^{\Phi}(t,t_0)(\Xi_0+\tau) \right] } \ea \ee 
At the initial time $t=t_0$, we get as we should the TDHB result : 
$\beta_1^{\Phi}(t_0,t_0)=G_0$.   
Indeed, $m^{\Phi}_1(t_0,t_0)=-1/2$ and  $l^{\Phi}(t_0,t_0)=0$.  
However for later times, this approximation is in general different from the
result of the TDHB evolution :  $\frac{1}{2} \Xi^{(0)}(t)=G^{(0)}(t)$. 
The  case at equilibrium 
in the symmetric phase is a special case since, as we
will show in section 3, we have : 
\be \label{2e66a}
\beta_1^{\Phi}(p,\Delta t=0)=G_0(p) \ee
This won't be true in the asymmetric phase. 
The result (\ref{2e66a}) is to relate to the fact that, at equilibrium in
the symmetric phase, the two approximations for the two-time and two-point 
function obtained with or without the optimization with respect to the
initial state co\"\i ncide (see below eq. (\ref{4.69})). 

Let us examine the four-point function when the two times co\" \i ncide. In
the symmetric phase, the limit  $t'=t''-0=t+0$ in expression 
 (\ref{2e61c}) for the four-point function $\Sigma_{11}^{\phi \phi}$ gives : 
\be \label{2e67} \ba{ll} 
\Sigma_{11}^{\Phi \Phi}({\bf p}_1,{\bf p}_2,{\bf p}_3,-\sum {\bf p}_i,t,t) 
= & {\displaystyle 
\int \frac{d^3p}{(2 \pi)^3} tr[
 l^{\Phi \Phi}({\bf p}-{\bf p}_1-{\bf p}_2,-{\bf p},{\bf p}_1,{\bf p}_2,t,t_0)
 (\Xi_0({\bf p})-\tau)  } \\
& {\displaystyle
l^{\Phi \Phi}({\bf p},-{\bf p}+{\bf p}_1+{\bf p}_2,{\bf p}_3,-\sum {\bf p}_i
,t_0,t)
(\Xi_0(-{\bf p}+{\bf p}_1+{\bf p}_2)+\tau) } \ea ] \ee
The boundary conditions for the function  $l^{\Phi \Phi}$ write
 \be \label{2e67a}
 l^{\Phi \Phi}_{11}({\bf p}_1,{\bf p}_2,{\bf p}_3,-\sum {\bf p}_i,t,t)=
 - \frac{1}{2} \left[ \delta^3({\bf p}_1+{\bf p}_3)+
 \delta^3({\bf p}_2+{\bf p}_3) \right] \ , \ee
 the other matrix elements of  $l^{\Phi \Phi}$ being equal to zero.
 We thus have at the initial time :
 \be \label{2e67b}
 \Sigma_{11}^{\Phi \Phi}({\bf p}_1,{\bf p}_2,{\bf p}_3,
 -\sum {\bf p}_i,t_0,t_0)=
 \frac{1}{2} \: \Xi^0_{11}(-{\bf p}_1)
 \Xi^0_{11}({\bf p}_2) \left[\delta^3({\bf  p}_1+ {\bf  p}_3)+
 \delta^3({\bf p}_2+ {\bf p}_3)
 \right] \ . \ee
The mean-field approximation for which $C_4=0$ corresponds to (see eq. 
(\ref{2.37a})) : 
 \be \label{2e68} \ba{ll}
 \Sigma_{11}^{\phi \phi}({\bf x},{\bf y},{\bf z},{\bf u},t,t)
 &{\displaystyle = 2\beta_1^{\phi}( {\bf  x},{\bf  z},t,t) 
 \beta_1^{\phi}({\bf y},{\bf  u},t,t) + 2 \beta_1^{\phi}({\bf  x},{\bf u},t,t) 
		 \beta_1^{\phi}({\bf y},{\bf  z},t,t) }
		 \\
&{\displaystyle =2 G^{(0)}({\bf x},{\bf z},t) G^{(0)}({\bf y},{\bf u},t)+
2 G^{(0)}({\bf x},{\bf u},t)
G^{(0)}({\bf y},{\bf z},t) }  \ea \ . \ee 
In momentum space, this corresponds to : 
\be \label{2e68a}
\Sigma^{\Phi \Phi}_{11}({\bf p}_1,{\bf p}_2,{\bf p}_3,-\sum {\bf p}_i,t,t)=2 
\beta_1^{\Phi}(-{\bf p}_1,t,t) \beta_1^{\Phi}({\bf p}_2,t) [ \delta({\bf p}_3+
{\bf p}_1)+\delta({\bf p}_3+{\bf p}_2)] \ee
At the initial time, we recover the result of the Wick theorem given by
expression  \ref{2e67b} in momentum space. However, for later times,
owing to the correlations we have taken into account in the dynamics, 
approximation
(\ref{2e67}) for the four-point correlation function is different from the
result of the Wick theorem. 

In nonrelativistic many-body problems, the mean-field TDHF calculations
correctly predict the expectation values but underestimate the fluctuations. 
Balian and V\'en\'eroni have showed the advantage to use a variational
principle adapted to calculate the fluctuations. For a pur state in the case
of heavy-ions collisions, one obtains in this way a better agreement with
experiments for the width of the mass distributions
\cite{BONCHE}. The example of the quartic oscillator in one dimension shows
also the usefullness of the method developped by  Balian and V\'en\'eroni 
to calculate the fluctuations of the operator  $x^2$, $<x^4>-<x^2>^2$, 
though staying in a mean-field approach  \cite{4}.

\newpage

\section{Optimization of the initial state}

\setcounter{equation}{0}

\subsection{Introduction}

   In physical situations, the initial state is not completly known. Only
   the expectation values of a few observables are known, for instance : 
\be \label{4.1}  
\left < \Phi({\bf  x}) \right >_{t_0} \: \: , \: \: 
\left < \Phi({\bf x}) \Phi({\bf  y}) \right >_{t_0} \: \: , \: \: 
\left < \Phi({\bf x}) \Phi({\bf  y}) \Phi({\bf  z}) \right >_{t_0} 
  \: \: , \: \: 
\left < \Phi({\bf  x}) \Phi({\bf  y}) \Phi({\bf  z})\Phi({\bf  u}) \right 
  >_{t_0}  
\ . \ee 

The exact state $D(t_0)$ can be written as   $\exp (-\beta \bar H )$.
 The operator $\bar H$ is
equal to the Hamiltonian of the system at thermal equilibrium 
but it can also contain other terms
corresponding to conserved quantities, like the charge operator 
for a complex scalar
field. For a nonequilibrium situation, the operator $\bar H$ is different
from the Hamiltonian $H$.   

We will approximate this initial state in the following way. We use the fact
that the operator  $e^{-\beta \bar H}$ is a solution of the Bloch equation : 
\be \label{4.2} 
\frac{d}{du} \: e^{-u \bar H} = -\bar H \: e^{-u \bar H}  \: \: ,     \: \:
0<u<\beta \ . \ee
Following the general method of  \cite{1}, \cite{1aa} to evaluate
variationally a given quantity, we introduce an ansatz operator  ${\cal D}(t)$
and we consider  equation (\ref{4.2}) as a constraint for 
 the complex times 
$t=t_0 + i (\beta-u)$ with the boundary condition  ${\cal D}(t_0+i \beta)=1$. 
We introduce also a Lagrange multiplier  ${\cal A}(t)$ associated to this
constraint. The functional to minimize writes : 
\be \label{4.3} 
{\cal Z}({\cal A}(t),{\cal D}(t))=
Tr \left( {\cal A}(t_0+ i \: 0) \: {\cal D}(t_0+i \:0 ) \right) 
- Tr \: \int^{t_0+i \beta}_{t_0} dt \: \: {\cal A}(t) \: 
\left( \frac{d{\cal D}(t)}{dt} + i \: \bar H \: {\cal D}(t) \right)
+ {\cal Z}_{dyn} \ , \ee
where ${\cal Z}_{dyn}$ is given by expression  (\ref{2.5}) of section 2. 
The ansatz operators  ${\cal D}(t)$ and ${\cal A}(t)$ are subject to the
boundary conditions  ${\cal A}(\infty)=1$ and ${\cal D}(t_0+i\: \beta)=1$. 

Let us note that for a complex value of $t$, ${\cal A}(t)$ is not close to
1 in the limit of vanishing sources. Indeed, the stationarity condition of 
${\cal Z}$ with respect to arbitrary variations of  ${\cal D}(t)$ writes : 
\be 
\frac{d {\cal A}(t)}{dt} -i {\cal A}(t) \bar H=0 \ee
and the solution of this equation for  $t=t_0+i(\beta-u)$ is equal to : 
${\cal A}(t)=A(t_0)\exp(-(\beta-u)\bar H)$. 

The approximate dynamical equation for  ${\cal D}(t)$ and ${\cal A}(t)$ are
coupled.  ${\cal D}(t_0)$ depends on  ${\cal A}(t)$
for $t>t_0$ and therefore on the sources  $J$ contrary to the exact state 
$ D(t_0)$.

\subsection{The variational spaces } 

For the variational spaces for  ${\cal D}(t)$ and ${\cal A}(t)$, we choose
again Gaussian operators characterized by the contractions  
 $\alpha_d({\bf x},t), \Xi_d({\bf x}, {\bf y},t)$ and  
$\alpha_a({\bf x},t), \Xi_a({\bf x},{\bf  y}, t)$ where the time  $t$ is on
the two segments 
  $[t_0+i \: \beta,t_0[$ and  $[t_0,+\infty [$ in the complex plane. 
  
The boundary condition  ${\cal D}(t_0+i\beta)=1$ writes : 
\be \label{4e4}
\Xi^{-1}_d(t_0+i \beta)=0 \quad \ , 
\quad \frac{1}{\Xi_d} \alpha_d(t_0+i \beta)=0 \ . \ee 

Let us write the functional  ${\cal Z}$ (\ref{4.3}) with the elementary
contractions :  
\be \label{4.5} 
{\cal Z}(n_a,\alpha_a,\Xi_a; n_d, \alpha_d, \Xi_d) = 
n(t_0) - 
\int_{t_0+i\: \beta}^{t_0} dt \: 
\left[ \frac{dn(t)}{dt} \vert_{{\cal A}(t)=cte} + i \: n(t) \: \int d^3x 
\: \bar {\cal E}_c ({\bf  x},t) \right] + {\cal Z}_{dyn} \ , \ee 
or, by integrating by part : 
\be \label{4.6} 
{\cal Z}(n_a,\alpha_a,\Xi_a; n_d, \alpha_d, \Xi_d) = 
n(t_0+ i \: \beta) + 
\int_{t_0+i\: \beta}^{t_0} dt \: 
\left[ \frac{dn(t)}{dt} \vert_{{\cal D}(t)=cte} - i \: n(t) \: \int d^3x 
\: \bar {\cal E}_c ({\bf  x},t) \right] + {\cal Z}_{dyn} \ . \ee 
We have used the notation : 
\be \label{4.6a} 
\int d^3x \: \bar {\cal E}_c({\bf x},t) = Tr \left( {\cal D}(t) \: {\cal A}(t) 
 \bar H \right) = <\bar H >_c \ . \ee

\subsection{The dynamical equations } 

For complex values of the time  $t_0+ i \: \beta <t<t_0$, the dynamical
equations for the contractions 
$\alpha_d, \Xi_d, \alpha_a$ and $\Xi_a$ write :  
\be  \label{4.7} 
2 i \: \dot \alpha_d = 
  -  \left( \Xi_d + \tau \right) \: \bar {\cal H}_c 
\:  \left( \Xi_d - \tau \right)       \: \tau \: \alpha_{b-c} 
+ \left( \Xi_d + \tau \right) \: \bar w_c \ , \ee 
\be \label{4.8} 
i \: \dot \Xi_d = -  \left( \Xi_d+ \tau \right) \: \bar  {\cal H}_c 
\:  \left( \Xi_d - \tau \right) \ , \ee 
\be  \label{4.9} 
2 i \: \dot \alpha_a = 
  \left( \Xi_a - \tau \right) \: \bar  {\cal H}_c 
 \left( \Xi_a + \tau \right) \: \tau \: \alpha_{c-b} 
- \left( \Xi_a-\tau \right) \: \bar w_c \ , \ee 
\be \label{4.10} 
i \: \dot \Xi_a =   \left( \Xi_a- \tau \right) \: \bar  {\cal H}_c 
\:  \left( \Xi_a + \tau \right) \ . \ee 
These dynamical equations are coupled through the boundary conditions
${\cal D}(t_0+ i \beta)=1, {\cal A}(t=+ \infty)=1$ which write : 
\be \label{4.10a} 
\Xi_d^{-1}(t_0+ i \beta)=0 \: \: , \: \: 
\frac{1}{\Xi_d}\alpha_d(t_0+i \beta)=0 \ , \ee 
\be \label{4.10b} 
\Xi_a^{-1}(t=+\infty)=0 \: \: , \: \: 
\frac{1}{\Xi_a}\alpha_a(t=+\infty)=0 \ . \ee 
 
\vspace*{0.5cm} 

For the contractions $ \alpha_b, \Xi_b$ et $\alpha_c, \Xi_c$, the dynamical
equation write :  
\be \label{4.11} 
\dot \alpha_b=0 \ , \ee 
\be \label{4.12} 
\dot \Xi_b=0 \ , \ee 
\be \label{4.13} 
i \: \dot \alpha_c = \tau \: \bar w_c  \ , \ee 
\be \label{4.14} 
i \: \dot \Xi_c= 2 \: \left( \Xi_c \bar {\cal H}_c \tau - \tau \bar 
{\cal H}_c \Xi_c \right) \ . \ee 

We have  $ \frac{d < \bar H>_c}{dt}=0$. 
We have introduce the vector  $ \bar w_c$ and the matrix $ \bar {\cal H}_c$ 
(analogous to $w$ and ${\cal H}$ ) to express the variation  
$\delta < \bar H>$ in terms of the variations 
$\delta \alpha_c$ and $\delta \Xi_c$. 
All the contractions $\alpha$ et $\Xi$ are continous at  $t=t_0$. 
In the complex time interval 
$[t_0+i \beta, t_0]$, $ \alpha_b$ and $\Xi_b$ are constant :  
$\alpha_b=\alpha_b(t_0), \Xi_b=\Xi_b(t_0)$.  

\subsection{Expansion in powers of the sources} 

We use the index  $^{(0)}$ for the solutions of the dynamical equations in
the limit with no source :  
$ \alpha_d(t)= \alpha_d^{(0)}(t), \Xi_d(t)=\Xi_d^{(0)}(t) , 
\alpha_a(t)=\alpha_a^{(0)}(t), \Xi_a(t)=\Xi_a^{(0)}(t)$.
 Since ${\cal A}(t) \neq 1 $ even in the limit with no source, this limit
 does not correspond to 
$\frac{1}{\Xi_a}\alpha_a=0, \Xi_a^{-1}=0$,
contrary to what happens when we do not optimize the initial state (see
section 2). Therefore,  $\Xi_b^{(0)} \neq \Xi_d^{(0)}$ and 
$\alpha_b^{(0)} \neq \alpha_d^{(0)} $. However, we can use a continuity
property at  $t=t_0$. in the limit with no source, 
${\cal A}(t_0+i0)={\cal A}(t_0+0) \to 1$. 
At the time  $t_0$, we thus have  : 
\be \label{4.17}
\alpha_b^{(0)}(t_0)=\alpha_c^{(0)}(t_0)=\alpha_d^{(0)}(t_0)=
\bar \alpha_0 \ , \ee 
\be \label{4.18} 
\Xi_b^{(0)}(t_0)=\Xi^{(0)}_c(t_0)=\Xi_d^{(0)}(t_0)=\bar \Xi_0 \ , \ee
where $\bar \alpha_0$ and $\bar \Xi_0$ are the HB static solutions
associated to the operator  $\bar H$ and verify : 
\be \label{4.19} 
\bar w_0=0 \ , \ee 
\be \label{4.20} 
\bar \Xi_0 \: \bar {\cal H}_0 \: \tau = \tau \: \bar {\cal H}_0 
\: \bar \Xi_0 \ . \ee
$\bar \Xi_0$ et
$\bar {\cal H}_0$ are related by equations similar to
 $(\ref{2.29a})$ and $ (\ref{2.29b})$. The expressions for the vector 
 $\bar w_0=\bar w (\bar \alpha_0, \bar \Xi_0)$ and the matrix 
 $\bar {\cal H}_0=\bar {\cal H}(\bar \alpha_0, \bar \Xi_0)$ 
are obtained from those in Appendix D by replacing  $f_0$ and 
$g_0$ by the {\it time-independent} functions $\bar f_0$ and $\bar g_0$.
(We recall that  $\bar {\cal H}$ comes from the variation of  $<\bar H>$
with respect to  $\Xi$ and has to be evaluated at  $\bar \alpha_0, \bar \Xi_0$
to obtain $\bar {\cal H}_0$. Since $\bar H$ is a time-independent operator,
$\bar {\cal H}_0$ is time-independent.) We stress the difference between the
quantities  $\Xi^{(0)}, \Xi_0$ and $\bar \Xi_0$. 

The expansion at zero order of equations  (\ref{4.7})-(\ref{4.10}) 
writes : 
\be \label{4.25} 
2 \: i \: \dot \alpha_d^{(0)} = \left ( \Xi_d^{(0)} + \tau \right) \: 
\bar w_0 \ , \ee 
\be \label{4.26} 
i \: \dot \Xi_d^{(0)} = - \left( \Xi_d^{(0)}+\tau \right) \: \bar {\cal H}_0 \: 
\left( \Xi_d^{(0)}- \tau \right) \ , \ee 
\be \label{4.27} 
2 \: i \: \dot \alpha_a^{(0)} = \left ( \Xi_a^{(0)} - \tau \right) \: 
\bar w_0 \ , \ee 
\be \label{4.28} 
i \: \dot \Xi_a^{(0)} = - \left( \Xi_a^{(0)}-\tau \right) \: \bar {\cal H}_0 \: 
\left( \Xi_a^{(0)}+ \tau \right) \ . \ee 

Since $\bar w_0=0$, we have for  $t_0+i \beta<t<t_0$ :
$\dot \alpha_d^{(0)}=0$ and  $\dot \alpha_a^{(0)}=0$. 
 By using the continuity at $t=t_0$, we thus obtain for  $0<u<\beta$ : 
\be \label{4.29} 
\alpha_d^{(0)}(t_0+i \beta-i \: u)= \bar \alpha_0 \ . \ee 
\be \label{4.30} 
(\frac{1}{\Xi_a}\alpha_a)^{(0)}(t_0+i \beta -i \:  u) =0 \  , \ee 

The solution of (\ref{4.26}) with the boundary condition 
$\Xi^{(0)^{-1}}_d(t_0+i \beta)=0$ is the following  
\be \label{4.31} 
\Xi^{(0)}_d(t_0+i \beta-i \: u  ) = - \tau \: 
\coth ( u \: \bar {\cal H}_0 \: \tau) 
\ , \ee
which can also be written as  : 
\be \label{4.32} 
\Xi^{(0)}_d(t_0+i \beta-iu)= \tau \left( 1- \frac{2}{1-\exp(-u \bar {\cal H}_0 2
\tau)} \right)  \ee 
or in the uniform case : 
\be \label{4.31a}
\Xi^{(0)}_d(t_0+i \beta-iu)=\pmatrix{\frac{1}{\omega_{{\bf p}}} \coth(\frac{u
\omega_{{\bf p}}}{2}) & 0 \cr 0 & -\omega_{{\bf p}} 
\coth(\frac{u \omega_{{\bf p}}}{2}) \cr} \ee
with, for the  $ \Phi^4$ theorie, 
$\omega_{{\bf p}}=\sqrt{-\bar g_0({\bf p})}$ and  
\be 
\bar g_0({\bf p})=-\left( {\bf p}^2+ \bar m_0^2+
\frac{b}{2} \bar \varphi_0^2+ \frac{b}{2}
\bar G_0({\bf  x}, {\bf  x}) \right) \ee
At $u=\beta$, we have to obtain the static HB solution associated to 
the Hamiltonian $\bar H$ , $\bar \Xi_0$ 
\be \label{4.33} 
\Xi_d^{(0)}(t_0+i \: 0)=\bar \Xi_0= 
- \tau \: \coth(\beta \: \bar {\cal H}_0 \: \tau ) \ , \ee 

The solution of (\ref{4.28}) writes : 
\be \label{4.34} 
\Xi_a^{(0)}(t_0+i \beta- i \: u)= - \tau \: \coth((\beta-u) \: 
\bar {\cal H}_0 \: 
\tau ) \ . \ee 
One checks that for  $u= \beta$, $\Xi_a^{(0)^{-1}}(t_0)=0$. 

At the first order, $\bar w^{(0)}_c(t_0)=\bar w_0(t_0)=0$ and 
$\bar \Xi^{(0)}_c(t_0)=\bar \Xi_0$. Therefore the right members of 
equations (\ref{4.13}) and (\ref{4.14}) are vanishing at $t_0$. We deduce
that  at the first order in the expansion in powers of the sources, 
$\alpha_c$ and $\Xi_c$ are constant in the interval  $[t_0+i \beta, t_0]$
and equal to  $\bar \alpha_0$ and  $\bar \Xi_0$ respectively. 
We can also obtain this
result from the relations of Appendix C and from the expressions obtained
above for 
$\alpha_d^{(0)}(t), \alpha_a^{(0)}(t), \Xi_d^{(0)}(t), \Xi_a^{(0)}(t)$. 

\vspace*{1cm} 

At the next order, the approximated initial state characterized by 
$\alpha_d(t_0), \Xi_d(t_0)$ is no more equal to the static HB solution 
$\alpha_0, \Xi_0$. 

Instead of using the expansion of equations 
(\ref{4.7}) and  (\ref{4.8}) for $\alpha_d$ and $\Xi_d$, it is more convenient 
to expand at the first order equations  (\ref{4.13}) and (\ref{4.14}) 
for $\alpha_c$ and $\Xi_c$. 
For $t_0+i \beta \leq t \leq t_0$ : 
\be  \label{4.42} 
i \: \frac{d}{dt} \beta^{\Phi} (\vec x,\vec y,t,t")= \tau  
\left[ \bar t \: \beta^{\Phi}(t,t") - 
\frac{1}{2} \: \bar T_{,(jk} \: \Sigma^{\Phi}_{kj)}
(t,t") \right]  \ , \ee 
\be \ba{ll} \label{4.43} 
i \: \frac{d}{dt} \Sigma^{\Phi}_{ij} (\vec x,\vec y,\vec x_1,t,t")= 
& {\displaystyle 2 \: \left[ \Sigma^{\Phi} \: \bar{\cal H}_0 \: \tau 
 - \tau \: \bar {\cal H}_0 \: \Sigma^{\Phi}  \right. } \\
& {\displaystyle \left.   + \bar \Xi_0 \: 
\bar r_{,(k} \: \beta^{\Phi}_{k)} \: \tau 
 - \tau \: \bar r_{, (k} \: \beta^{\Phi}_{k)} \: \bar \Xi_0  \right. } \\ 
& {\displaystyle \left. -\frac{1}{2} \: \bar \Xi_0
 \: \bar  R_{,(kl} \: \Sigma^{\Phi}_{lk)} \: 
\tau + \frac{1}{2} \: \tau \: \bar R_{,(kl} \: 
\Sigma^{\Phi}_{lk)} \bar \Xi_0 \right]_{ij}
} \ea \ . \ee
The matrices $\bar t, \bar T$ and $\bar r, \bar R$ associated to the operator 
$\bar H$ and depending on  $\bar \Xi_0$ and on $\bar \alpha_0$ are analogous
to the 
matrices $t,T,r$ and $R$. The quantities 
$\bar {\cal H}_0$, $\bar
\alpha_0, \bar \Xi_0$ are time-independent for  $t_0+i \beta \le t \le t_0$.
The matrices $\bar t , \bar T$ and $\bar r, \bar R$ are also time-independent 
for $t_0+i \beta \le t \le t_0$. Indeed, these are defined  from the
expansion of  $\bar w_c$ and $\bar {\cal H}_c$ 
in terms of the variations  $\delta \alpha_c$ and $\delta \Xi_c$ (see 
equations (\ref{2.39}) and (\ref{2.40})) and they have to be evaluated for
the values $\alpha_c^{(0)}$ and  $\Xi_c^{(0)}$. Now, as we have noticed
before, these quantities do not depend on the complex time and are equal
respectively to  $\bar \alpha_0$ and $\bar \Xi_0$. For the same reason, 
$\bar \Xi_0$ appears in the last two lines of 
(\ref{4.43}) and not  $\Xi_d^{(0)}$ as in equation (\ref{2e42}) which
describes the evolution of  $\Sigma^{\Phi}(t,t'')$ for real time. 

\subsection{Approximation for the two-point function} 

In the following we consider an uniform problem. In momentum space, 
equation (\ref{4.42}) for  $\beta^{\Phi}$ writes : 
\be \label{4.47} 
\frac{d^2}{dt^2} \: \beta_1^{\Phi}({\bf  p},t,t")=\bar g_0(-{\bf p}) \: 
\beta_1^{\Phi}({\bf p},t,t") + \frac{b}{4} \: \bar 
\varphi_0 \: \int_{{\bf  p}_1+ {\bf p}_2 
=- {\bf  p}} \frac{d^3p_1}{(2 \pi)^3} \: \frac{d^3p_2}{(2 \pi)^3}  \: 
\Sigma_{11}^{\Phi}({\bf  p}_1, {\bf  p}_2, {\bf  p},t,t") \ . \ee 

\be \label{4.48}
i \frac{d}{dt} \beta_1^{\Phi}({\bf p},t,t'')=-\beta_2^{\Phi}({\bf p},t,t'') \ee

{\bf Symmetric phase} 

 In the symmetric phase,  $\bar \varphi_0=0$ and the equation for the function 
 $\beta_1^{\Phi}({\bf p},t,t'')$ for $t_0+i\beta<t<t_0$ writes : 
\be \label{4.53} 
\frac{d^2}{dt^2} \beta_1^{\Phi}({\bf p},t,t'')=\bar g_0({\bf p}) 
\beta_1^{\Phi}({\bf p},t,t'')
\ , \ee
where 
\be
-\bar g_0({\bf p})={\bf p}^2+\bar m_0^2+\frac{b}{2}
\bar G_0({\bf x},{\bf x}) \ee
We can therefore relate $\beta_{1,2}^{\Phi}({\bf p},t_0,t'')$ 
to $\beta_{1,2}^{\Phi}
({\bf p},t_0+i\beta,t'')$ : 
\be \label{4.54} 
\pmatrix{\beta_1^{\Phi} \cr \beta_2^{\Phi}\cr} ({\bf p},t_0,t'')=A({\bf p}) \: 
\pmatrix{\beta_1^{\Phi} \cr \beta_2^{\Phi}\cr} ({\bf p},t_0+i \beta,t'') \ee
where $A({\bf p})$ is the matrix : 
\be \label{4.55}
A({\bf p})=\pmatrix{\cosh \omega_{{\bf p}} \beta & \frac{\sinh \omega_{{\bf
	p}} \beta}{\omega_{{\bf p}}} 
\cr \omega_{{\bf p}} \sinh \omega_{{\bf p}} \beta
& \cosh \omega_{{\bf p}} \beta \cr } \ee
with  $\omega_{{\bf p}}=\sqrt{-\bar g_0({\bf p})}$. 

We then use the fact that  ${\cal D}(t_0+ i \beta)=1$. We deduce :  
\be \label{4.56}
\alpha_c(t_0+i \beta)=\alpha_b(t_0+i \beta) \quad \ , \quad 
\Xi_c(t_0+i \beta)=\Xi_b(t_0+i \beta) \ . \ee
By introducing the functions $\delta$ and  $\Delta$ which come from the
expansion at the first order in powers of the sources of  $\alpha_b$ and 
$\Xi_b$ (and which are analogous to  $\beta$ and  $\Sigma$), we thus have : 
\be \label{4.57}
\beta(t_0+i\beta,t'')=\delta(t_0+i \beta,t'') \ee
\be \label{4.58}
\Sigma(t_0+i\beta,t'')=\Delta(t_0+i\beta,t'') \ee
Now $\alpha_b$ and  $\Xi_b$ are constants for  $t_0+i \beta<t<t_0$ (eq.
(\ref{4.11}) and (\ref{4.12})).  Consequently : 
\be \label{4.59}
\delta(t_0,t'')=\delta(t_0+i \beta,t'') \ee
\be \label{4.60} 
\Delta(t_0,t'')=\Delta(t_0+i \beta,t'') \ee
Let us compare the expansion of  $\Xi_b,\Xi_c$ and  $\alpha_b,
\alpha_c$ at  $t_0$. By using equation (\ref{C.6}) at 
$t=t_0$,  
\be \label{4.61}
\Delta_k(t_0,t'')=-(\bar \Xi_0+\tau) \: l_k(t'',t_0) \: 
(\bar \Xi_0-\tau) \ , \ee 
and from equation (\ref{C.1}) : 
\be \label{4.61a}
\delta_k(t_0,t'')=(\bar \Xi_0+\tau) (m_k(t'',t_0)-l_k(t'',t_0)\bar 
\alpha_0) \ . \ee
The comparison of these last two equations to  (\ref{2e50}) and  
(\ref{2e51}) leads to : 
\be \label{4.62}
\delta_k(t_0,t'')=\beta_k(t_0,t'')+2 \tau (m_k(t'',t_0)-l_k(t'',t_0)\alpha_0) 
\ee 
\be \label{4.63}
\Delta_k(t_0,t'')=\Sigma_k(t_0,t'')+2 \: 
\bar \Xi_0 \l_k(t'',t_0)\tau -2 \: \tau
l_k(t'',t_0) \bar \Xi_0 \ . \ee
Finally, we use : 
\be \label{4.64}
\beta(t_0+i \beta,t'')=\delta(t_0,t'') \ee
\be \label{4.65}
\Sigma(t_0+i \beta,t'')=\Delta(t_0,t'') \ee
In the symmetric phase, $l^{\Phi}=0$, therefore :  
\be \label{4.65a}
\beta^{\Phi}(t_0+i \beta, t'')=\beta^{\Phi}(t_0,t'')+2 \tau
m^{\Phi}(t'',t_0) \ee
From equation (\ref{4.54}), we obtain :  
\be \label{4.66} 
\beta^{\Phi}({\bf p},t_0,t'')=\left( A^{-1}({\bf p})-1 \right)^{-1} 2 \tau
m^{\Phi}({\bf p},t'',t_0) \ee

For $t''>t'$, the relation  (\ref{2e61a}) of section 2 is therefore
modified according to : 
\be \label{4.67}
\beta_1^{\Phi}({\bf p},t',t'')=
4 \: \tilde m^{\Phi}({\bf p},t',t_0)(A^{-1}({\bf p})-1)^{-1}
\tau m^{\Phi}({\bf p},t'',t_0) \ee
with 
\be \label{4.68}
(A^{-1}({\bf p})-1)^{-1}=\pmatrix{-\frac{1}{2} & -\frac{1}{2\omega_{{\bf p}}}
\coth ( \frac{\beta}{2}\omega_{{\bf p}}) \cr 
-\frac{\omega_{{\bf p}}}{2}\coth(\frac{\beta}{2}\omega_{{\bf p}}) & 
-\frac{1}{2} \cr}
\ee
For $t'>t''$, we have  : 
\be \label{4.69}
\beta_1^{\Phi}({\bf p},t',t'')=4 \: \tilde m^{\Phi}({\bf p},t',t_0)
\left((A^{-1}({\bf p})-1)^{-1}+1 \right)
\tau m^{\Phi}({\bf p},t'',t_0) \ee
We check again that in the limit $t'=t''=t_0$, we obtain the TDHB solution at 
$t_0$ : $\beta_1^{\Phi}({\bf p},t_0,t_0)=\frac{1}{2 \omega_{{\bf p}}}
\coth (\beta \omega_{{\bf p}})=G_0({\bf p},t_0)$. 

For the two-point function in the symmetric phase, one can check from
expression  (\ref{2.29c}) for the matrix  $\bar \Xi_0$
(and where the quantities which appear correspond now to  $\bar
H$ and no more $H$) that :  
\be \label{4.69aa}
\bar \Xi_0(p) -\tau=2 (A^{-1}(p)-1)^{-1} \tau \ee
Therefore expressions (\ref{2e61a}) and (\ref{4.67}) for $\beta_1^{\Phi}$
in the symmetric phase obtained without or with optimization with respect to
the initial state  co\" \i ncide when   $\bar \Xi_0=\Xi_0$ which is the case
at equilibrium. This is not true in the asymmetric phase. It will be also
more interesting to study the four-point function in section 3.6. 

\vspace*{0.5cm}

It is also interesting to calculate the approximation obtained for the
retarded Green function (\ref{2e64}) when we optimize both the dynamics and
the initial state. In the symmetric phase, we have for the time-ordered
correlation function for $t'>t''$ : 
\be 
\beta_1^{\Phi}(t',t'')=2\tilde m^{\Phi}(t'',t_0) \beta^{\Phi}(t_0,t') \ee
For the anti-T product and  $t'>t''$ : 
\be 
\beta_1^{\Phi}(t',t'')=2\tilde m^{\Phi}(t',t_0) \beta^{\Phi}(t_0,t'') \ee
By using (\ref{4.66}) and the fact that 
\be 
(A^{-1}(p)-1)^{-1} \tau + \left[\tau (A^{-1}(p)-1)^{-1}\right]^{T} 
=-\tau \ , \ee
we obtain : 
\be \label{4.69a}
\chi_{\Phi \Phi}(p,t',t'')=i \theta (t'-t'') \: 4 \tilde m^{\Phi}(p,t'',t_0)
\tau m^{\Phi}(p,t',t_0) \ee
which is identical to the formula  (\ref{2e65a}) obtained without the
optimization of the initial state. For the retarded Green function the
approach of section 2 is therefore sufficient. 

{\bf The case at equilibrium } 

In this case, the two-time functions should depend only on the difference
$t'-t''$. Unfortunatly, this is not true for the expressions we gave for the
three-point and four-point functions in section 2. At equilibrium,
$\alpha^{(0)}(t)=\alpha_0, \Xi^{(0)}(t)=\Xi_0$ and the matrices
$t,T,r,R$ are constant. In equations (\ref{2e53}) and  (\ref{2e54}) for
$l^k$ and  $m^k$, the shift  $t' \to t'+\delta t$ is therefore equivalent to
a shift of the initial time
$t_0\to t_0-\delta t$.
By writing the corresponding shifts of the functions $l^k$ and $m^k-l^k
\alpha_0$ and by inserting them in eq. (\ref{2e59}) or in eq. (\ref{2e61}),
one sees that, in general, our approximations are not invariant 
under the shift 
$t' \to t'+\delta t$, $t'' \to t''+\delta t$, $t_0$ being constant. They
contain a spurious dependence on the initial time  $t_0$. 

The case of the two-point function in the symmetric phase is special.
From equation  (\ref{2e54}) for $m^{\Phi}$, we have 
$\delta m^{\Phi} =i t \tau m^{\Phi} \delta t$ and the variation of
expression  (\ref{2e61a}) is equal to zero because $\delta \beta^{\Phi}_1$ 
is proportional to 
$-\Xi^0_{11}g_0+\Xi^0_{22}$ which is vanishing for the static solution. 
However, this is no more true for the four-point function  $\Sigma^{\Phi
\Phi}$ given by equation (\ref{2e61h}). The variation 
$\delta \Sigma$ involves two types of terms : one with the matrix  $R$ 
proportional to the coupling constant and another with the matrix 
${\cal H}_0$. Generally,   $\delta \Sigma$ is non vanishing. This spurious
dependence on the initial time will disappear when we take into account the
optimization with respect to the initial state : the two-time 
correlation functions depend only on the time difference  $t'-t''$ as they
should. 

  At equilibrium, in the symmetric phase, we can solve equation
(\ref{2e54}) for $m^{\Phi}$ : 
\be \label{4.70} 
m_1^{\Phi}({\bf p},t',t_0)=-\frac{1}{2}\: \cos(\omega_{{\bf p}}(t_0-t')) \ee 
\be \label{4.71} 
m_2^{\Phi}({\bf p},t',t_0)=-\frac{1}{2i} \: \frac{1}{\omega_{{\bf p}}}\: 
\sin(\omega_{{\bf p}} (t_0-t')) \ee
We define $\Delta t=t'-t''$. For  $\Delta t <0$, equation (\ref{4.67}) gives  
\be \label{4.72} 
\beta_1^{\Phi}({\bf p},\Delta t)=\frac{1}{2 \omega_{{\bf p}}} 
(\coth \frac{\beta
\omega_{{\bf p}}}{2} \: \cos \omega_{{\bf p}} \Delta t -\frac{1}{i} 
\: \sin \omega_{{\bf p}}
\Delta t ) \ee
For $\Delta t>0$, we use equation (\ref{4.69}) : 
\be \label{4.73} 
\beta_1^{\Phi}({\bf p},\Delta t)=\frac{1}{2 \omega_{{\bf p}}} 
(\coth \frac{\beta
\omega_{{\bf p}}}{2} \: \cos \omega_{{\bf p}} 
\Delta t +\frac{1}{i} \: \sin \omega_{{\bf p}}
\Delta t ) \ee
We check that $\beta_1^{\Phi}$ depends only on  $\Delta t=t'-t''$.
Expressions (\ref{4.72}) and  (\ref{4.73}) are interchanged by changing 
$\Delta t$ to $-\Delta t$, which was not explicit on formula 
(\ref{4.67}) and (\ref{4.69}). 

For $\Delta t=0$, we obtain $\beta_1^{\Phi}({\bf p},\Delta t=0)=G_0({\bf
  p})$. In this case, at equal time, the two-time and two-point function
co\"\i ncides with the mean-field HB solution for the two-point function.

By taking the Fourrier transform in time of equations 
(\ref{4.72}) and  (\ref{4.73}) : 
\be \label{4.73a}
\beta_1^{\Phi}({\bf p},\omega)=\int_{-\infty}^0 d(\Delta t) e^{i(\omega -i
\epsilon)\Delta t} \beta_1^{\Phi}({\bf p}, \Delta t) 
+ \int^{+\infty}_0 d(\Delta t) e^{i(\omega +i
\epsilon)\Delta t} \beta_1^{\Phi}({\bf p}, \Delta t) \ , \ee
we obtain : 
\be \label{4.73b}
\beta_1^{\Phi}({\bf p}, \omega)=i \left[ \frac{n_{{\bf p}}+1}
{\omega^2-(\omega_{{\bf p}}-i
\epsilon)^2}-\frac{n_{{\bf p}}}{\omega^2-(\omega_{{\bf p}}+i 
\epsilon)^2} \right] \ee
where $n_{{\bf p}}$ is the occupation number  : $2n_{{\bf p}}+1=
\coth \frac{\beta
  \omega_{{\bf p}}}{2}$. 

 We will now show that our approximation for the two-time correlation
 function with  two field operators satisfies to the fluctuation-dissipation
 theorem. This theorem can be written as a relation between the Fourrier
 transform in time of the correlation function 
 $C^2_{\Phi \Phi}({\bf x}, {\bf  x'},t,t')$ and the response function 
 \be \label{4.74}
 \chi_2({\bf  x}, {\bf x'},t,t')=\: Tr \left( [\Phi^{H}({\bf x},t,t_0), 
 \Phi^{H}({\bf x'},t',t_0)] D(t_0) \right) \ee
 in the following way  : 
 \be \label{4.75}
  \tilde \chi_2({\bf x}, {\bf x'},\omega)=(1-e^{-\beta \omega}) \: 
 \tilde C^2_{\Phi \Phi}({\bf x}, {\bf x'}, \omega) \ , \ee
where the Fourrier transform in time is defined according to :  
\be \label{4.76} 
\tilde \chi_2({\bf x}, {\bf x'},\omega)=\int_{-\infty}^{+\infty}
d(t-t') \: e^{i \omega (t-t')} \chi_2({\bf x}, {\bf x'},t-t') \ee
The correlation function  $C^2$ gives us informations about the fluctuations
in the equilibrium state while the response function  $\chi^2$ contains
informations on the dynamics of the system which has been driven  from the
equilibrium by a small external perturbation. The function  
$\chi^2$ appears also in the dissipation energy of the system. Therefore, the 
fluctuation-disspation theorem tells us that, in systems near equilibrium, the
transport properties which are linear in the external forces can be
calculated from the fluctuations at equilibrium  \cite{PARISI}. 

By using the definition 
\be \label{4.77} 
Tr (\Phi^H({\bf x},t',t_0) \Phi^H({\bf y},t'',t_0) D(t_0))-\varphi({\bf x},t') 
\varphi({\bf y},t'') \equiv <\Phi({\bf x}, \tau) \Phi({\bf  y}, \tau)> \ee
with $\tau=t'-t''$, the  fluctuation-dissipation theorem can be written in the
following way :  
\be \label{4.77a} \ba{ll}
\int_{-\infty}^{+\infty} d \tau \: e^{i \omega \tau} <\Phi({\bf x}, \tau) 
\Phi({\bf y})>=& {\displaystyle e^{\beta \omega} \int_{-\infty}^{+\infty} d
\tau \: e^{i \omega \tau} < \Phi({\bf  y}) \Phi({\bf x}, \tau)> } \\ 
& {\displaystyle = \int_{-\infty}^{+ \infty} d \tau \: e^{i \omega (\tau-i
\beta)} < \Phi({\bf x},\tau-i \beta) \Phi({\bf y})> } \ea \ee
In order to check the  fluctuation-dissipation theorem on our approximations
for the two-point correlation function, it is therefore sufficient to check
that expression  (\ref{4.72})  for $\Delta t=\tau$ is equivalent to
expression (\ref{4.73}) for  $\Delta t=-\tau-i \beta$, which can be done
easely.    

At equilibrium, expression  (\ref{2e65a}) for the retarded Green function
with two field operators in the symmetric phase gives : 
\be \label{4.77b}
\chi_{\Phi \Phi}({\bf p},t',t'')=-\theta(t'-t'')\frac{1}{\omega_{{\bf p}}} \sin
\omega_{{\bf p}}(t'-t'') \ee 

\subsection{Four-point function : a perturbative expansion }

 Still in the symmetric phase and for the uniform case, we consider now the
 four-point function  $\Sigma^{\Phi \Phi}$.

\subsubsection{ The response function  $\Pi_R({\bf  q}^2,t',t'')$  }

 When one couples a source  $J^{\Phi \Phi}({\bf x}, {\bf  y}, t')$ to the 
 the operator $\Phi^H({\bf  x},t',t_0) \Phi^H({\bf y},t'',t_0)$, the linear
 response theory gives the variation of  $G({\bf x}, {\bf 
 y},t')=Tr(\Phi^H({\bf  x},t',t_0) \Phi^H({\bf  y},t',t_0) D(t_0) )$ in
presence of the source : 
 \be \label{4.78}
 \delta G({\bf  x}, {\bf y},t')=\int_{t_0}^{+\infty} dt'' \int d^3z \: d^3u
 \: \: 
 \chi_{\Phi \Phi, \Phi \Phi}({\bf  x}, {\bf y}, {\bf z}, {\bf u}, t',t'') 
 J^{\Phi \Phi}({\bf z}, {\bf u},t'') \ee
 where $\chi_{\Phi \Phi, \Phi \Phi}$ is the retarded Green function with
 four field operators defined in eq. (\ref{2e65c}). In presence of a source of
 the following type : 
 \be \label{4.79} 
J^{\Phi \Phi}({\bf  x}_1, {\bf  x}_2,t'')=J^{\Phi \Phi}(t'') e^{-i {\bf q} \ .
{\bf x}_1} \delta^3({\bf x}_1-{\bf x}_2) \ , \ee
 \be \label{4.80} 
 \delta G({\bf x}, {\bf  y}, t') =\int_{t_0}^{+\infty} dt'' J^{\Phi \Phi}(t'')
 \int d^3z \: e^{-i {\bf  q} \ .  {\bf  z}} \chi_{\Phi \Phi, \Phi \Phi}({\bf
   x},{\bf  y}, {\bf  z},{\bf  z}, t',t'') \ee 
 
  By introducing the Fourrier transform , we obtain for  ${\bf x}={\bf  y}$ : 
 \be \label{4.81} 
 \delta G({\bf  x},{\bf  x},t') =
  \: e^{-i {\bf q} \ . {\bf x}} \int dt'' \: J^{\Phi
 \Phi} (t'') \int \frac{d^3 k}{(2 \pi)^3}  \frac{d^3p_1}{(2 \pi)^3} \: 
 \chi_{\Phi \Phi, \Phi \Phi}({\bf k}, -{\bf q}-{\bf k},{\bf p}_1,{\bf q}-
 {\bf p}_1,t',t'') \ee

The polarization function  $\Pi_R$ is defined according to :  
\be \label{4.82} 
\Pi_R({\bf  q}^2,t',t'')\equiv \frac{\delta G({\bf x}, {\bf x}, t')}
{e^{-i {\bf q} \ . {\bf x}} \delta J^{\Phi \Phi}(t'')} \bigg  \vert_{J^{\Phi
    \Phi}=0} \ee
i. e. :  
\be \label{4.83}
 \Pi_R({\bf  q}^2,t',t'')=\int \frac{d^3 k}{(2 \pi)^3}
 \frac{d^3p_1}{(2 \pi)^3} \:
 \chi_{\Phi \Phi, \Phi \Phi}({\bf k}, -{\bf q}-{\bf k},{\bf p}_1,{\bf q}-
{\bf  p}_1,t',t'') \ee
A variational approximation for the polarization function  $\Pi_R$
is obtained by using expression (\ref{2e65d}) for $\chi_{\Phi \Phi, \Phi
\Phi}$ as a function of  $l^{\Phi \Phi}$ and  $\Xi_0$.

In the static case,  $\Pi_R({\bf q}^2,t',t'')$ depends only on  $\Delta t=
t'-t''$ and is proportional to $\theta(t'-t'')$. We thus define its Fourrier
transform in time by : 
\be \label{4.84} 
\tilde \Pi_R({\bf q}^2,\omega)=\int_{0}^{+\infty} d(\Delta t) \: e^{i
\omega \Delta t+\epsilon \Delta t} \Pi_R({\bf  q}^2,\Delta t) \ee 

The polarization function  $\Pi_R({\bf q}^2,t',t'')$ can be recognized as
the Fourrier transform with respect to  ${\bf x}-{\bf y}$ of the retarded
Green function for the operators  $\Phi^2({\bf x})$ and  $\Phi^2({\bf y})$ : 
\be \label{4.85}
\chi_{\Phi^2, \Phi^2}({\bf x}, {\bf  x}, {\bf  y}, {\bf y},t',t'')=-i \theta
(t'-t'') \: Tr \left([\Phi^{H}({\bf  x},t',t_0)\Phi^{H}({\bf  x},t',t_0), 
\Phi^{H}({\bf y},t'',t_0)\Phi^{H}({\bf y},t'',t_0)] D(t_0) \right) \ee 
We have : 
\be \label{4.86} \ba{ll} 
\chi_{\Phi^2, \Phi^2}({\bf q}^2,t',t'')= & {\displaystyle 
\int d^3x \: e^{-i {\bf q} \ . ({\bf 
x}- {\bf y}) } \: \chi_{\Phi^2, \Phi^2}({\bf  x}, {\bf  x}, {\bf  y}, {\bf  y}
,t',t'')} \\
& {\displaystyle =\Pi_R({\bf q}^2,t',t'') } \ea \ee
This function is directly related to the viscuosity in the 
$\Phi^4$ theory. From the spectral representation : 
\be \label{4.87}
\chi_{\Phi^2, \Phi^2}({\bf p}, p_0)=\lim_{\epsilon \to 0^{+}}
\int_{-\infty}^{+\infty} d \omega \: \frac{\rho_{\Phi^2, \Phi^2}({\bf p},
\omega)}{p_0-\omega+i \epsilon} \ , \ee 
the viscuosity is defined by : 
\be \label{4.87a} 
\eta_{\Phi^2, \Phi^2} \equiv \lim_{{\bf  p}, p_0 \to 0} \left[
\frac{\rho_{\Phi^2, \Phi^2}({\bf  p}, p_0)}{p_0} \right] \ee
Calculations of  $\eta_{\Phi^2, \Phi^2}$ have been done in the limit of high
temperature and the result have been compared to the classical approximation 
\cite{HEINZ}.

Actually we will begin by calculating the following  non-retarded 
Green function : 
 \be \label{4.87b}
 \Pi({\bf q}^2,t',t'')
  \equiv \frac{1}{2} \frac{\delta \Xi^c_{11}({\bf  x},{\bf  x},t')}
 {e^{- i {\bf  q} \ . {\bf  x}} \delta J^{\Phi
 \Phi}( t'')} \bigg \vert_{J^{\Phi \Phi}=0} \ee
 which can be written with the four-point function 
 $\Sigma_{11}^{\Phi \Phi}$ as : 
 \be \label{4.87c}
 \Pi({\bf  q}^2,t',t'')=\frac{i}{2} \: \int \frac{d^3 k}{(2 \pi)^3} 
 \int \frac{d^3p_1}{(2 \pi)^3} \:
 \Sigma^{\Phi \Phi}_{11}({\bf k}, -{\bf q}-{\bf k},{\bf p}_1,{\bf q}-
 {\bf p}_1,t',t'') \ee
In the following, we will therefore consider the function  $\Sigma^{\Phi
\Phi}({\bf k},-{\bf q}-{\bf k},{\bf p}_1,{\bf q}-{\bf p}_1,t',t'')$. 
We will show the differences between the approximations for the function 
$\Pi({\bf  q}^2,t',t'')$ when we do or do not optimize with  respect to the
initial state. 

\vspace*{0.5cm}

\subsubsection{ Resolution of the dynamical equations for complex values of
  the time  }

In the symmetric phase, we have for  $t_0+i \beta \le t \le t_0$ : 
\be \label{4.88} 
i \frac{d}{dt}\Sigma_{11}^{\Phi \Phi}({\bf p}_1,{\bf p}_2,{\bf p}_3,
-\sum {\bf p}_i,t,t'')=
- \Sigma_{12}^{\Phi\Phi} -\Sigma_{21}^{\Phi\Phi}
\ee
\be \label{4.89}
i \frac{d}{dt}\Sigma_{12}^{\Phi \Phi}=\bar g_0(-p_2) \Sigma_{11}^{\Phi \Phi} 
-\Sigma_{22}^{\Phi \Phi}-\frac{b}{4} \bar \Xi^{0}_{11}(-p_1) \: I \ee
\be \label{4.90}
i \frac{d}{dt}\Sigma_{21}^{\Phi \Phi}=\bar g_0(p_1) \Sigma_{11}^{\Phi \Phi} 
-\Sigma_{22}^{\Phi \Phi}-\frac{b}{4}\bar \Xi^{0}_{11}(p_2) \: I \ee
\be \label{4.91}
i \frac{d}{dt}\Sigma_{22}^{\Phi \Phi}= \bar
g_0(-p_2) \Sigma_{21}^{\Phi \Phi}+ \bar g_0(p_1) \Sigma_{12}^{\Phi \Phi} \ee
where $I$ is the following integral :  
\be \label{4.92} 
I=\int_{{\bf p}_4+ {\bf  p}_5={\bf  p}_1+ {\bf p}_2} \frac{d^3 p_4}{(2 \pi)^3} 
\: \frac{d^3 p_5}{(2 \pi)^3}  
\:  \Sigma_{11}^{\Phi \Phi}({\bf p}_4,{\bf p}_5,{\bf p}_3,
-\sum {\bf p}_i,t,t'') \ee
These equations are analogous to  (\ref{2e42}), the  matrices ${\cal H}_0, r$
and $R$ being replaced by the time-independent matrices 
$\bar {\cal H}_0, \bar r$
and  $\bar R$ associated to  $\bar H$ and  $\Xi^{(0)}$ by $\bar \Xi_0$. We
see that we are not abble to obtain directly an expression analogous to 
(\ref{4.54}) which would relate the matrices  $\Sigma(t_0,t'')$ and 
$\Sigma(t_0+i \beta,t'')$ since the loop term  $\frac{b}{4} 
\bar \Xi^0_{11} \: I$ involve all the momenta. We will use a perturbative
expansion where we neglect this loop term to lowest order. From the expression
for  $\bar \Xi^0_{11}$, we see that this expansion will be valid when 
\be \label{4.92a}
\frac{b}{4} \coth \frac{\beta  m(\beta)}{2} \ll 1 \ee
where  $ m(\beta)$ is the  self-consistent mass defined by : 
\be \label{4.92b}
 m^2(\beta)=m_0^2+\frac{b}{2} \int \frac{d^3p}{(2 \pi)^3} \: \frac{1}{2
    \sqrt{{\bf p}^2+  m^2(\beta)}} \coth \left(\frac{\beta}{2}
  \sqrt{{\bf  p}^2+     m^2(\beta)} \right) \ee
At zero temperature, the condition  (\ref{4.92a})
for the validity of the perturbative expansion writes simply
$\frac{b}{4}\ll 1$. At high temperature it becomes :
$\frac{b}{2} \ll \beta  m(\beta)$ with $m^2(\beta)\simeq \bar m^2
+\frac{b}{24 \beta^2}$.  

To lowest order, by neglecting the loop term I, the solutions of equations 
(\ref{4.88})-(\ref{4.91}) write :  
\be \label{4.93}
\Sigma_{11}^{\Phi \Phi}({\bf k}, -{\bf q}-{\bf k}, {\bf p}_1, {\bf q}-{\bf
  p_1}, u,t'')=
A_1 \cosh (\Omega_3 u) + B_1 \sinh(\Omega_3 u) +C_1
\cosh(\Omega_4 u) + D_1 \sinh(\Omega_4 u) \ee
\be \label{4.94}
\Sigma_{12}^{\Phi \Phi}(u,t'')
=\omega_{{\bf q}+{\bf k}} \left[ A_1 \sinh(\Omega_3 u) + B_1
\cosh(\Omega_3 u) + C_1 \sinh (\Omega_4 u) + D_1 \cosh(\Omega_4 u) \right]\ee 
\be \label{4.95}
\Sigma_{21}^{\Phi \Phi}(u,t'')=\omega_{{\bf k}} 
\left[ A_1 \sinh(\Omega_3 u) + B_1
\cosh(\Omega_3 u) - C_1 \sinh (\Omega_4 u) - D_1 \cosh(\Omega_4 u) \right]\ee
\be \label{4.96} 
\Sigma_{22}^{\Phi \Phi}(u,t'')=\omega_{{\bf k}} \omega_{{\bf q}+{\bf k}} 
\left[A_1 \cosh (\Omega_3 u) + B_1 \sinh(\Omega_3 u) -C_1
\cosh(\Omega_4 u) - D_1 \sinh(\Omega_4 u) \right] \ee
where we have used the variable $u$ defined by :  $t=t_0+i(\beta-u)$ 
and we have introduced the frequencies 
\be \label{4.97} 
\Omega_3=\omega_{{\bf q}+{\bf k}}+ \omega_{\bf k} \ee 
\be \label{4.98} 
\Omega_4=\omega_{{\bf q}+{\bf k}} -\omega_{\bf k} \ee
$\omega_{\bf k}$ being defined by  $\omega_{\bf k}=\sqrt{-\bar g_0({\bf k})}$. 

We can therefore express $\Sigma^{\Phi \Phi}({\bf k}, -{\bf q}-{\bf k},
{\bf p}_1,
{\bf q}-{\bf p}_1,t_0+i \beta,t'')$ as a function of 
 $\Sigma^{\Phi \Phi}({\bf k}, -{\bf q}-{\bf k},{\bf p}_1,{\bf q}-{\bf p}_1,
 t_0,t'')$. 
The result is the following :  
\be \label{4.99}
\Sigma^{\Phi \Phi}({\bf k},-{\bf q}-{\bf k},{\bf p}_1,{\bf q}-{\bf p}_1,t_0+i
\beta,t'')=A^{-1}({\bf k}) \: \Sigma^{\Phi
\Phi}({\bf k}, -{\bf q}-{\bf k},{\bf p}_1,{\bf q}-{\bf p}_1,t_0,t'') 
\: (A^T)^{-1} ({\bf q}+{\bf k}) \ee
where the matrix $A({\bf k})$ is defined by  (\ref{4.55}) : 
\be \label{4.100}
A({\bf k})=\pmatrix{\cosh (\omega_{{\bf k}} \beta) & \frac{\sinh
    (\omega_{{\bf k}}
\beta)}{\omega_{{\bf k}}} \cr 
\omega_{{\bf k}} \sinh(\omega_{{\bf k}} \beta ) & \cosh (\omega_{{\bf k}} 
\beta) \cr } \ee 
\be \label{4.101} 
A^{-1}({\bf k})=\pmatrix{\cosh (\omega_{{\bf k}} \beta) & 
  -\frac{\sinh (\omega_{{\bf k}}
\beta)}{\omega_{{\bf k}}} \cr 
-\omega_{{\bf k}} \sinh(\omega_{{\bf k}} \beta ) & \cosh (\omega_{{\bf k}} 
\beta) \cr } \ee 

By introducing a column formed with the elements  $\Sigma_{ij}$,
we can write this expression in the following way, which will be usefull in
section 3.6.4 :  
\be \label{4.101a}
\pmatrix{\Sigma^{\Phi \Phi}_{11} \cr \Sigma^{\Phi \Phi}_{12} \cr 
\Sigma^{\Phi \Phi}_{21} \cr \Sigma^{\Phi \Phi}_{22} \cr }
({\bf k},-{\bf q}-{\bf k},{\bf p}_1,{\bf q}-{\bf p}_1,t_0,t'') = 
N({\bf k},{\bf q}+{\bf k}) 
\pmatrix{\Sigma^{\Phi \Phi}_{11} \cr \Sigma^{\Phi \Phi}_{12} \cr 
\Sigma^{\Phi \Phi}_{21} \cr \Sigma^{\Phi \Phi}_{22} \cr }
({\bf k},-{\bf q}-{\bf k},{\bf p}_1,{\bf q}-{\bf p}_1,t_0+i \beta,t'') \ee
where $N({\bf k},{\bf q}+{\bf k})$ is the following 4 by 4 matrice : 
\be \label{4.101b}
N({\bf k},{\bf q}+{\bf k})=\pmatrix{A & \frac{C}{\omega_{{\bf k}+{\bf q}}} 
  & \frac{D}{\omega_{{\bf k}}} & 
\frac{B}{\omega_{{\bf k}} \omega_{{\bf k}+{\bf q}}} \cr 
\omega_{{\bf k}+{\bf q}} C & A & \frac{\omega_{{\bf k}+{\bf q}}}{\omega_{\bf
    k}} B 
& \frac{D}{\omega_{\bf k}} \cr 
\omega_k D & \frac{\omega_{k}}{\omega_{{\bf k}+{\bf q}}} B & A 
& \frac{C}{\omega_{{\bf k}+{\bf q}}} \cr 
\omega_{{\bf k}} \omega_{{\bf k}+{\bf q}} B & \omega_{{\bf k}} D 
& \omega_{{\bf k}+{\bf q}} C & A \cr } \ee 
with 
\be 
A=\cosh(\omega_{{\bf k}+{\bf q}} \beta) \cosh(\omega_{{\bf k}} \beta) \ee 
\be
B=\sinh(\omega_{{\bf k}+{\bf q}} \beta) \sinh(\omega_{{\bf k}} \beta) \ee
\be
C=\sinh(\omega_{{\bf k}+{\bf q}} \beta) \cosh(\omega_{{\bf k}} \beta) \ee
\be 
D=\cosh(\omega_{{\bf k}+{\bf q}} \beta) \sinh(\omega_{{\bf k}} \beta) \ee

If  we do not optimize 
with respect to the initial state, an approximation for the 
four-point function 
$\Sigma_{11}^{\Phi \Phi}(t',t'')$ is given by equation 
(\ref{2e61c}) for $t''>t'$. 

On the other hand, if we optimize with respect to the initial state, 
$\Sigma_{11}^{\Phi \Phi}(t',t'')$ is given by 
(\ref{2e61b}) for  $t''>t'$ : 
\be \label{4.102} \ba{ll}
& {\displaystyle \Sigma_{11}^{\Phi \Phi}({\bf p}_1,{\bf p}_2,{\bf p}_3, 
  -\sum {\bf p}_i,t',t'') = 
  - \: \int \frac{d^3p}{(2\pi)^3} \times } \\ 
& {\displaystyle 
tr \left[ l^{\Phi \Phi}({\bf p}-{\bf p}_1-{\bf p}_2,-{\bf p},{\bf p}_1,
{\bf p}_2,t',t_0) \Sigma^{\Phi
\Phi}({\bf p},-{\bf p}+{\bf p}_1+{\bf p}_2,{\bf p}_3,-\sum {\bf p}_i,t_0,t'') 
\right] } \ea  \ee
and we use  (\ref{4.63}) and  (\ref{4.65}) to write  
\be \label{4.103} \ba{ll}
&{\displaystyle \Sigma^{\Phi \Phi}({\bf p}_1,{\bf p}_2,{\bf p}_3,
  -\sum {\bf p}_i,t_0+i \beta,t'')= 
\Sigma^{\Phi \Phi} ({\bf p}_1,{\bf p}_2,{\bf p}_3,-\sum {\bf p}_i,t_0,t'') } 
\\ 
& {\displaystyle + 2  \left[ \bar \Xi_0(-{\bf p}_1) l^{\Phi \Phi}
  ({\bf p}_1,{\bf p}_2,{\bf p}_3,
-\sum {\bf p}_i,t'',t_0) \tau -\tau l^{\Phi \Phi}({\bf p}_1,{\bf p}_2,
{\bf p}_3,-\sum {\bf p}_i,t'',t_0)
\bar \Xi_0({\bf p}_2) \right] } \ea \ee
or  : 
\be \label{4.103a} \ba{ll} 
& {\displaystyle 
A^{-1}({\bf k}) \Sigma^{\Phi \Phi}({\bf k},-{\bf q}-{\bf k},{\bf p}_1,
{\bf q}-{\bf p}_1,t_0,t'') (A^T)^{-1}({\bf q}+{\bf k})-
\Sigma^{\Phi \Phi}({\bf k},-{\bf q}-{\bf k},{\bf p}_1,{\bf q}-{\bf p}_1,
t_0,t'') = } \\
& {\displaystyle 2 \left[ \bar \Xi_0(-{\bf k}) l^{\Phi
\Phi}({\bf k},-{\bf q}-{\bf k},{\bf p}_1,{\bf q}-{\bf p}_1,t'',t_0) \tau 
-\tau l^{\Phi
\Phi}({\bf k},-{\bf q}-{\bf k},{\bf p}_1,{\bf q}-{\bf p}_1,t'',t_0) 
\bar \Xi_0(-{\bf q}-{\bf k}) \right] } \ea \ee
where $\bar \Xi_0$ is given by (\ref{2.29c}), $g_0$ being replaced by 
$\bar g_0$.

The matrix elements  $\Sigma_{ij}^{\Phi \Phi}(t_0,t'')$ can therefore be
written in terms of  $l^{\Phi \Phi}_{ij}(t'',t_0)$. By using 
 (\ref{4.102}), we finally obtain an approximation for the two-time and 
 four-point function  $\Sigma^{\Phi
\Phi}_{11}(t',t'')$ which involves only the function  $l^{\Phi \Phi}$. We
recall that this approximation has been obtained by neglecting the loop term 
I (\ref{4.92}) in the evolution equations of  
$\Sigma^{\Phi \Phi}$ for complex values of the time.  

\vspace*{0.5cm}

\subsubsection{ Resolution of the dynamical equations for  $l^{\Phi \Phi}$ 
in the static case } 

The dynamical equations for the matrix elements of  $l^{\Phi \Phi}$
are given by  (\ref{2e53}). In the static case,  $\Xi^{(0)}(t)$ is equal to
$\Xi_0$ given by (\ref{2.29c}),
$\alpha^{(0)}(t)=\alpha_0$  and  $g_0$ is time-independent. 
{\it In the symmetric phase }, equations (\ref{2e53}) 
write more explicitly in momentum space  : 
\be \label{4.104} 
i \frac{d}{dt}l^{\Phi \Phi}_{11}({\bf k},-{\bf q}-{\bf k},{\bf p}_i,
-{\bf p}_i+{\bf q},t'',t)= 
-g_0({\bf k}) l^{\Phi \Phi}_{21}-g_0({\bf q}+{\bf k}) l^{\Phi \Phi}_{12} + J 
\ee
\be \label{4.105} 
i \frac{d}{dt} l^{\Phi \Phi}_{12}=l^{\Phi \Phi}_{11}-g_0({\bf k}) l^{\Phi
\Phi}_{22} \ee 
\be \label{4.106} 
i \frac{d}{dt} l^{\Phi \Phi}_{21}=l^{\Phi \Phi}_{11}-g_0({\bf q}+{\bf k}) 
l^{\Phi \Phi}_{22} \ee 
\be \label{4.107} 
i \frac{d}{dt} l^{\Phi \Phi}_{22}=l^{\Phi \Phi}_{12}+l^{\Phi
\Phi}_{21} \ee
with the boundary conditions  (\ref{2e67a}). 
J is the following integral : 
\be \label{4.107a} \ba{ll} 
J=-\frac{b }{4} \int \frac{d^3p}{(2 \pi)^3} & {\displaystyle
\left[ \Xi_0(p) 
l^{\Phi \Phi}(-{\bf p},-{\bf q}+{\bf p},{\bf p}_i,-{\bf p}_i+{\bf q},t'',t) 
\tau \right. } \\
& {\displaystyle \left. -\tau l^{\Phi \Phi}(-{\bf q}-{\bf p},{\bf p},
{\bf   p}_i,-{\bf p}_i+{\bf q},t'',t) \Xi_0(p)  
\right]_{11} } \ea \ee 
\be \label{4.107b}
J=\frac{b }{4} \int \frac{d^3p}{(2 \pi)^3} \: \frac{1}{\omega_p}
\coth(\frac{\beta \omega_p}{2}) \left[ l^{\Phi
\Phi}_{12}(-{\bf p},-{\bf q}+{\bf p},{\bf p}_i,-{\bf p}_i+{\bf q},t'',t)+ 
l^{\Phi \Phi}_{21}(-{\bf q}-{\bf p},{\bf p},
{\bf p}_i,-{\bf p}_i+{\bf q},t'',t) \right] \ee
In the previous equations, in order to make the notations more compact, we
have ommitted when it was possible the arguments of $l^{\Phi \Phi}$. 
 Let us remark that, on the contrary to equations
(\ref{4.88})-(\ref{4.91}) for $\Sigma^{\Phi \Phi}$, the solution  $\Xi_0(p)$ 
appears inside the loop integral.  

Similarly to the resolution of the equations for  $\Sigma^{\Phi \Phi}$,
we will solve the equations for  $l^{\Phi \Phi}$, perturbatively, neglecting
first the loop term J. In the following section, we will take into account
of the loop term J to first order. To lowest order, the solutions of 
(\ref{4.104})-(\ref{4.107}) write : 
\be \label{4.109} 
l_{11}^{\Phi \Phi}({\bf k},-{\bf q}-{\bf k},{\bf p}_1,{\bf q}-{\bf p}_1,t',t)
=-\frac{1}{2} \left[ \delta^3(
{\bf k}+{\bf p}_1)+\delta^3(-{\bf  q}-{\bf  k}+{\bf  p}_1) \right] \:
\cos(\omega_{{\bf k}+{\bf q}}(t-t')) \cos(\omega_{{\bf k}}(t-t')) \ee
\be \label{4.110}
l_{12}^{\Phi \Phi}({\bf k},-{\bf q}-{\bf k},{\bf p}_1,{\bf q}-{\bf p}_1,t',t)=
-\frac{1}{2i \omega_{{\bf k}+{\bf q}}} 
\left[ \delta^3(
{\bf k}+{\bf  p}_1)+\delta^3(-{\bf  q}-{\bf  k}+{\bf  p}_1) \right] \:
\sin(\omega_{{\bf k}+{\bf q}}(t-t')) \cos(\omega_{{\bf k}}(t-t')) \ee
\be \label{4.111}
l_{21}^{\Phi \Phi}({\bf k},-{\bf q}-{\bf k},{\bf p}_1,{\bf q}-{\bf p}_1,t',t)
=-\frac{1}{2i \omega_{{\bf k}}} 
\left[ \delta^3(
{\bf k}+{\bf  p}_1)+\delta^3(-{\bf  q}-{\bf  k}+{\bf  p}_1) \right] \:
\cos(\omega_{{\bf k}+{\bf q}}(t-t')) \sin(\omega_{{\bf k}}(t-t')) \ee
\be \label{4.112}
l_{22}^{\Phi \Phi}({\bf k},-{\bf q}-{\bf k},{\bf p}_1,{\bf q}-{\bf p}_1,t',t)
=\frac{1}{2 \omega_{{\bf k}}\omega_{{\bf k}+{\bf q}}} 
\left[ \delta^3(
{\bf k}+{\bf  p}_1)+\delta^3(-{\bf  q}-{\bf  k}+{\bf  p}_1)
\right] \:
\sin(\omega_{{\bf k}+{\bf q}}(t-t')) \sin(\omega_{\bf k}(t-t')) \ee

Formula (\ref{4.102}) and  (\ref{4.103a}) then give, for  $\Delta
t=t'-t''<0$ : 
\be \label{4.113}
\Sigma_{11}^{\Phi \Phi}({\bf k},-{\bf q}-{\bf k},{\bf p}_1,{\bf q}-{\bf p}_1,
t',t'')=2 \left[ \delta^3(
{\bf k}+{\bf  p}_1)+\delta^3(-{\bf  q}-{\bf  k}+{\bf  p}_1)
\right] \:
\beta_1^{\Phi}({\bf k},\Delta t) \beta_1^{\Phi}({\bf q}+{\bf k},\Delta t) \ee
where $\beta_1^{\Phi}({\bf k}, \Delta t)$ is given by (\ref{4.72}). 
We thus have at the lowest order : 
\be \label{4.113a} 
\Pi^{(0)}({\bf q}^2,\Delta t)=2 i  \int \frac{d^3k}{(2 \pi)^3} 
\beta_1^{\Phi}({\bf k},\Delta t) \beta_1^{\Phi}({\bf q}+{\bf k},\Delta t) \ee
When we neglect the loop terms $I$ and $J$ in the dynamical equations for 
$\Sigma^{\Phi \Phi}$ for the complex time and in the dynamical equations for 
$l^{\Phi \Phi}$ for the time 
$t>t_0$, we thus recover the Wick theorem as we should. Let us stress that
we have neglected the interactions which appear in the terms  I and J but
the interactions remain in the frequencies  $\omega_k$ which involve the 
renormalized self-consistent mass $ m(\beta)$ (eq. (\ref{2.29l})).  

At this order, $\Sigma_{11}^{\Phi \Phi}$ depends only on the time difference 
$\Delta t=t'-t''$ and our approximation satisfies the
fluctuation-dissipation theorem for the two-time and four-point correlation
function. 

If we apply formula  (\ref{2e61c}), obtained without optimization with
respect to the initial state, with the solutions  $l^{\Phi \Phi}$ given by 
(\ref{4.109})-(\ref{4.112}), we find again the result of the Wick theorem.
When we neglect the terms with the interaction in the dynamical equations for 
$\Sigma^{\Phi \Phi}$ and $l^{\Phi \Phi}$, the two approximations for 
$\Sigma_{11}^{\Phi \Phi}(t',t'')$ obtained with or without optimization with
respect to the initial state  therefore co\" \i ncide. Accordingly, 
at zero order, we can use instead of expression  (\ref{4.103a}) the simpler
expression  : 
 \be \label{4.114} 
 \Sigma^{\Phi \Phi}({\bf k},-{\bf q}-{\bf k},{\bf p}_1,-{\bf p}_1+{\bf q}
 ,t_0,t'')=- \left(\bar \Xi_0({\bf k})-\tau
 \right) l^{\Phi \Phi}({\bf k},-{\bf q}-{\bf k},{\bf p}_1,{\bf q}-{\bf p}_1
 ,t'',t_0) \left(\bar
 \Xi_0(-{\bf q}-{\bf k})+\tau\right) \ee
The existence of such an expression with matrices (to distinguish from a
vectorial expression like   (\ref{4.101a})  was not obvious at the
beginning. 

For the polarization function  $\Pi_R({\bf q}^2,t',t'')$, we obtain to 
lowest order : 
\be \label{4.114a} \ba{ll}
\Pi_R^{(0)}({\bf  q}^2,t',t'')=&{\displaystyle 
-\theta(t'-t'') \int \frac{d^3p}{(2 \pi)^3} \:
\frac{1}{\omega_{{\bf p}} \omega_{{\bf q}+{\bf p}}} } \\ 
&{\displaystyle \left[ \coth \frac{\beta \omega_{{\bf p}}}{2} 
    \cos \omega_{{\bf p}} \Delta t
\sin \omega_{{\bf p}+{\bf q}} \Delta t + 
\coth \frac{\beta \omega_{{\bf p}+{\bf q}}}{2} 
\cos \omega_{{\bf p}+{\bf q}} \Delta t
\sin \omega_{{\bf p}} \Delta t \right] } \ea \ee 
Its Fourrier transform in time writes : 
\be \label{4.114b} \ba{ll} 
\Pi_R^{(0)}& {\displaystyle 
({\bf q}^2, \omega)= \int \frac{d^3p}{(2 \pi)^3} \:
\frac{1}{\omega_{{\bf p}} \omega_{{\bf q}+{\bf p}}} } \\
& {\displaystyle \left[ (n_{{\bf p}+{\bf q}}+n_{{\bf p}}+1)
\frac{\omega_{{\bf p}+{\bf q}}+\omega_{{\bf p}}}
{(\omega+i\eta)^2-(\omega_{{\bf p}+{\bf q}}+\omega_{{\bf p}})^2}-
(n_{{\bf p}+{\bf q}}-n_{{\bf p}}) 
\frac{\omega_{{\bf p}+{\bf q}}-\omega_{{\bf p}}}
{(\omega+i\eta)^2-(\omega_{{\bf p}+{\bf q}}-\omega_{{\bf p}})^2}
\right] } \ea \ee 
A similar expression appears for instance in \cite{DEVEGA}. Here, we  have 
derived it variationally. We stress again that this formula is
nonperturbative in the sense that it contains the renormalized 
self-consistent mass.

\subsubsection{Resolution of the dynamical equations with the
  interaction terms I and J at  first order } 

At first order, we will solve the dynamical equations 
(\ref{4.88})-(\ref{4.91}) for 
\hfill\break $\Sigma^{\Phi \Phi}({\bf k},-{\bf q}-{\bf k},{\bf p}_1, 
-{\bf p}_1+{\bf q}
,u,t'')$ by replacing in the loop integral I the solution 
$\Sigma^{\Phi \Phi}$ obtained at the lowest order. We thus have to solve a
first order differential sytem with a nonhomogeneous term which involves : 
\be \label{4.115}
I(u)=\int \frac{d^3l}{(2 \pi)^3} \left[a \cosh(\Omega_5 u) +b \sinh(\Omega_5
u) +c \cosh(\Omega_6 u) + d \sinh(\Omega_6 u) \right] \ee 
with (see the analogous definitions  (\ref{4.97}) and (\ref{4.98}) ) :  
\be \label{4.116} \Omega_5=\omega_{{\bf q}+{\bf l}}+\omega_{{\bf l}} \ee
\be \label{4.117} \Omega_6=\omega_{{\bf q}+{\bf l}}-\omega_{{\bf l}} \ee
The constants  a,b,c and d depend on the momenta  ${\bf  l}$ and  ${\bf  q}$.
They can be expressed in terms of the matrix elements 
$\Sigma_{ij}({\bf l},-{\bf q}-{\bf l},{\bf p}_i,{\bf q}-{\bf p}_i,t_0,t'')$
and  $\cosh(\Omega_5 \beta),
\sinh(\Omega_6 \beta), \cosh(\Omega_6 \beta)$, $ \sinh(\Omega_6 \beta)$. 
At the order we consider, we can use  (\ref{4.114}) and express 
$a,b,c,d$ with the matrix elements $l^{\Phi
\Phi}_{ij}({\bf l},-{\bf q}-{\bf l},{\bf p}_i,{\bf q}-{\bf p}_i,t'',t_0)$, 
with $\coth(\frac{\beta \Omega_5}{2}), \coth(\frac{\beta \Omega_6}{2})$ and
with the occupation numbers  $n_{\bf l}$ et $n_{{\bf l}+{\bf q}}$ defined by
: 
\be \label{ 4.118}
2 n_{{\bf l}}+1=\coth(\frac{\beta \omega_{{\bf l}}}{2}) \ . \ee

The relation between  $\Sigma_{ij}^{\Phi \Phi}(t_0,t'')$ and  
$\Sigma_{ij}^{\Phi \Phi}(t_0 + i \beta,t'')$ writes now : 
\be \label{4.119} 
\pmatrix{\Sigma^{\Phi \Phi}_{11} \cr \Sigma^{\Phi \Phi}_{12} \cr 
\Sigma^{\Phi \Phi}_{21} \cr \Sigma^{\Phi \Phi}_{22} \cr }
({\bf k},-{\bf q}-{\bf k},{\bf p}_1,{\bf q}-{\bf p}_1,t_0,t'') = 
N({\bf k},{\bf q}+{\bf k}) 
\pmatrix{\Sigma^{\Phi \Phi}_{11} \cr \Sigma^{\Phi \Phi}_{12} \cr 
\Sigma^{\Phi \Phi}_{21} \cr \Sigma^{\Phi \Phi}_{22} \cr }
({\bf k},-{\bf q}-{\bf k},{\bf p}_1,{\bf q}-{\bf p}_1,t_0+i \beta,t'')  + 
W(t_0,t'') \ee
where the matrix  $N({\bf k},{\bf q}+{\bf k})$ given by  (\ref{4.101b}) does
not contain explicitly the coupling constant and  W is a vector whose
components are proportional to the coupling constant $b$. 

By using the fact that the first term can be written as 
(\ref{4.114}), we write in matrix notation : 
\be \label{4.120} \ba{ll} 
\Sigma^{\Phi \Phi}({\bf k},-{\bf q}-{\bf k},{\bf p}_1,{\bf q}-{\bf
  p}_1,t_0,t'')= & {\displaystyle
- \left(\bar \Xi_0({\bf k})-\tau
 \right) l^{\Phi \Phi}({\bf k},-{\bf q}-{\bf k},{\bf p}_1,{\bf q}-{\bf
  p}_1,t'',t_0) } \\ 
& {\displaystyle \times \left(\bar
 \Xi_0(-{\bf q}-{\bf k})+\tau\right) + X(t_0,t'') } \ea  \ee
 where
 \be \label{4.121}
 X(t_0,t'')=\pmatrix{X_1 & X_2 \cr X_3 & X_4 \cr} \ee
 and 
 \be \label{4.122}
 X_i=(1-N)^{-1}_{ij} W_j \ee
 $X(t_0,t'')$ depends on the momenta  ${\bf k},{\bf k}+{\bf q}$. The
 elements   $X_i$ can be expressed relatively simply with the integrals 
 $I_n, J_n, K_n, L_n, M_n$ of Appendix E  :
 \be \label{4.123} \ba{llll} 
 X_1=\frac{b}{8 \omega_{{\bf k}} \omega_{{\bf k}+{\bf q}}} 
 \{  & {\displaystyle
 (n_{{\bf k}+{\bf q}}+n_{{\bf k}}+1) \left[ -\Omega_3
 (I_1+I_2+K_1+K_2+K_3+K_4) \right. } \\  
 & {\displaystyle \left. -\coth(\frac{\beta \Omega_3}{2})
 (J_1+J_2-M_1-M_2-M_3-M_4)\right] } \\
 & {\displaystyle (n_{{\bf k}}-n_{{\bf k}+{\bf q}}) \left[-\Omega_4
 (I_3+I_4+L_1+L_2+L_3+L_4) \right. } \\ 
 & {\displaystyle \left. -\coth(\frac{\beta \Omega_4}{2})
 (J_4+J_3-N_1-N_2-N_3-N_4)\right] } \ea  \} \ee
 \be \label{4.124} \ba{llll}
 X_2=\frac{b}{8 \omega_{{\bf k}}} \{  & {\displaystyle
(n_{{\bf k}+{\bf q}}+n_{{\bf k}}+1) 
\left[ -\Omega_3 \coth(\frac{\beta \Omega_3}{2})
 (I_1+I_2-K_1-K_2-K_3-K_4) \right. } \\
 & {\displaystyle \left. -
 (J_1+J_2+M_1+M_2+M_3+M_4)\right] } \\
 & {\displaystyle (n_{{\bf k}}-n_{{\bf k}+{\bf q}})
 \left[-\Omega_4 \coth(\frac{\beta \Omega_4}{2})
 (I_3+I_4-L_1-L_2-L_3-L_4) \right. } \\ 
 & {\displaystyle \left. -
 (J_4+J_3+N_1+N_2+N_3+N_4)\right] } \ea \}  \ee
 \be \label{4.125} \ba{llll}
 X_3=\frac{b}{8 \omega_{{\bf k}+{\bf q}}} \{  & {\displaystyle
(n_{{\bf k}+{\bf q}}+n_{{\bf k}}+1) 
\left[ -\Omega_3 \coth(\frac{\beta \Omega_3}{2})
 (I_1+I_2-K_1-K_2-K_3-K_4) \right. } \\  
 & {\displaystyle \left. -
 (J_1+J_2+M_1+M_2+M_3+M_4)\right] } \\
 & {\displaystyle - (n_{{\bf k}}-n_{{\bf k}+{\bf q}})
 \left[-\Omega_4 \coth(\frac{\beta \Omega_4}{2})
 (I_3+I_4-L_1-L_2-L_3-L_4) \right. }\\ 
 & {\displaystyle \left. -
 (J_4+J_3+N_1+N_2+N_3+N_4)\right] } \ea \}  \ee
 \be \label{4.126} \ba{llll}
 X_4=\frac{b}{8 }  \{  & {\displaystyle
 (n_{{\bf k}+{\bf q}}+n_{\bf k}+1) \left[ -\Omega_3
 (I_1+I_2+K_1+K_2+K_3+K_4) \right. } \\ 
 & {\displaystyle \left. -\coth(\frac{\beta \Omega_3}{2})
 (J_1+J_2-M_1-M_2-M_3-M_4)\right] } \\
 & {\displaystyle -(n_{{\bf k}}-n_{{\bf k}+{\bf q}}) \left[-\Omega_4
 (I_3+I_4+L_1+L_2+L_3+L_4) \right. } \\ 
 & {\displaystyle \left. -\coth(\frac{\beta \Omega_4}{2})
 (J_4+J_3-N_1-N_2-N_3-N_4)\right] } \ea  \} \ee
By using the expressions of Appendix E, we obtain the elements  $X_i$ 
in terms of $l^{\Phi \Phi}_{ij}$. 

 The approximation for the function  $\Pi({\bf q}^2,t',t'')$ is given by :
 \be \label{4.127a} 
\Pi({\bf q}^2,t',t'')=\frac{i}{2} \int \frac{d^3 k}{(2 \pi)^3} 
\frac{d^3 p_1}{(2 \pi)^3} \Sigma_{11}^{\Phi
 \Phi}({\bf k},{\bf q}-{\bf k},{\bf p}_1,{\bf q}-{\bf p}_1,t',t'') \ee
or 
\be \label{4.127} \ba{ll}
\Pi({\bf  q}^2,t',t'') = -\frac{i}{2}
  \int & {\displaystyle \frac{d^3 k}{(2 \pi)^3}\frac{d^3 p_1}{(2 \pi)^3}
 \frac{d^3 p}{(2 \pi)^3} \times  } \\ 
 & {\displaystyle 
tr \left[ l^{\Phi \Phi}({\bf p}+{\bf q},-{\bf p},{\bf k},-{\bf q}-{\bf k},
t',t_0)
 \Sigma^{\Phi\Phi}({\bf p},-{\bf p}-{\bf q},{\bf p}_1,{\bf q}-{\bf p}_1,
 t_0,t'') \right] } \ea \ee

It remains now to solve the dynamical equations for  $l^{\Phi
 \Phi}$ taking into account of the loop term J (\ref{4.107a}). We will do it
in the static case. Equations
 (\ref{4.104})-(\ref{4.107}) form a nonhomogeneous first order differential 
 system. The  solutions $l^{\Phi
 \Phi}_{ij}({\bf k},-{\bf q}-{\bf k},{\bf p}_i,-{\bf p}_i+{\bf q},t',t'')$ 
are given in Appendix F. 
The approximation to first order for  $\Pi({\bf  q}^2,t',t'')$ is
composed of two terms which are for  $t'<t''$ :
 \be \label{4.128}\ba{lll}
  & {\displaystyle \Pi({\bf q}^2,t',t'')=
 \frac{i}{2} \int \frac{d^3 k}{(2 \pi)^3} \frac{d^3 p_1}{(2 \pi)^3}
 \frac{d^3 p}{(2 \pi)^3} \times } \\
 & {\displaystyle  \:
 tr \left[l^{\Phi \Phi} ({\bf p}+{\bf q},
 -{\bf p},{\bf k},-{\bf q}-{\bf k},t',t_0) (\bar
 \Xi_0({\bf p})-\tau)l^{\Phi \Phi}({\bf p},-{\bf p}-{\bf q},{\bf p}_1,{\bf q}-
{\bf  p}_1,t'',t_0) (\bar
 \Xi_0({\bf p}+{\bf q})+\tau) \right] } \\
 & {\displaystyle - \:
 tr[l^{\Phi \Phi} ({\bf p}+{\bf q},-{\bf p},{\bf k},-{\bf q}-{\bf k},t',t_0) 
 X(t_0,t'') ] }
 \ea \ee
The first term corresponds to the approximation one obtains without
optimization with respect to the initial state (by replacing  $\bar \Xi_0$ 
by $\Xi_0$) . When one optimizes with respect to initial state, one has to
take into account of the second term. We insert in the previous formula the
expressions obtained to  first order for the solutions  $l^{\Phi \Phi}$ 
in the static case. We obtain the following result at the first order, for 
$\Delta t=t'-t''<0$  :
 \be \label{4.129} \ba{lllll}
 \Pi^{(1)}({\bf  q}^2,t',t'')= & {\displaystyle -\frac{i}{2} \frac{b }{2}
  \int \frac{d^3 k}{(2 \pi)^3} \frac{d^3 p_1}{(2 \pi)^3}
 \frac{1}{\omega_{{\bf k}} \omega_{{\bf k}+{\bf q}} \omega_{{\bf p}_1} 
   \omega_{-{\bf p}_1+{\bf q}} }  \times } \\
 & {\displaystyle (n_{{\bf k}+{\bf q}}+n_{\bf k}+1) \left((n_{-{\bf
       p}_1+{\bf q}}+n_{{\bf p}_1}+1)
 \frac{\Omega_7}{\Omega_3^2-\Omega_7^2}
 -(n_{-{\bf p}_1+{\bf q}}-n_{{\bf p}_1}) 
 \frac{\Omega_8}{\Omega_3^2-\Omega_8^2}\right) } \\
 & {\displaystyle \times \left(\coth(\frac{\beta \Omega_3}{2}) \cos
 (\Omega_3 \Delta t) +i \sin(\Omega_3 \Delta t) \right) } \\
 & {\displaystyle -(n_{{\bf k}+{\bf q}}-n_{\bf k}) \left((n_{-{\bf p}_1+{\bf
       q}}+n_{{\bf p}_1}+1)
 \frac{\Omega_7}{\Omega_4^2-\Omega_7^2}
 -(n_{-{\bf p}_1+{\bf q}}-n_{{\bf p}_1}) 
 \frac{\Omega_8}{\Omega_4^2-\Omega_8^2}\right) } \\
& {\displaystyle \times \left(\coth(\frac{\beta \Omega_4}{2}) \cos
 (\Omega_4 \Delta t) +i \sin(\Omega_4 \Delta t) \right) } \ea \ee
 with 
 \be \label{4.130}
 \Omega_3=\omega_{{\bf k}+{\bf q}}+ \omega_{\bf k} \: \: \ , \: \:
 \Omega_4=\omega_{{\bf k}+{\bf q}}-\omega_{{\bf k}} \ee
 \be \label{4.131}
 \Omega_7=\omega_{-{\bf p}_1+{\bf q}}+ \omega_{{\bf p}_1} \: \: \ , \: \:
 \Omega_8=\omega_{-{\bf p}_1+{\bf q}}-\omega_{-{\bf p}_1} \ee
 This is the essential result of this section. One has to add to this
 formula the term (\ref{4.113a}) obtained to lowest order. In the static case,
 we obtain an approximation for the correlation function 
 $\Pi({\bf  q}^2,t',t'')$ which depends only on the time difference 
 $\Delta t=t'-t''$. If we do not optimize with respect to the initial state,
 which corresponds to consider only the first term in  (\ref{4.128}), the
 approximation obtained for  $\Pi({\bf  q}^2,t',t'')$ contains a spurious
 dependence on the initial time  $t_0$ which writes  :
 \be \label{4.132} \ba{lllllllllllllllll}
 P^{(1)}=& {\displaystyle
 -\frac{i}{2} \int \frac{d^3 k}{(2 \pi)^3} \frac{d^3 p_1}{(2 \pi)^3}
 \frac{b }{4 \omega_{{\bf k}} \omega_{{\bf k}+{\bf q}} \omega_{{\bf p}_1} 
   \omega_{-{\bf p}_1+{\bf q}} } \times } \\
 & {\displaystyle + \frac{1}{2}(n_{{\bf k}+{\bf q}}+n_{{\bf k}}+1) 
   (n_{-{\bf p}_1+{\bf q}}+n_{{\bf p}_1}+1)
 \frac{1}{\Omega_3^2-\Omega_7^2} \left(\Omega_7 \coth(\frac{\beta
 \Omega_3}{2}) -\Omega_3 \coth(\frac{\beta
 \Omega_7}{2}) \right) } \\
 & {\displaystyle \times \left[ \cos \Omega_7(t_0-t'') \cos \Omega_3(t_0-t')+
 \cos \Omega_7(t_0-t') \cos \Omega_3(t_0-t'') \right] } \\
& {\displaystyle + \frac{1}{2}(n_{{\bf k}+{\bf q}}+n_{{\bf k}}+1) 
  (n_{-{\bf p}_1+{\bf q}}+n_{{\bf p}_1}+1)
 \frac{1}{\Omega_3^2-\Omega_7^2}\left(\Omega_3 \coth(\frac{\beta
 \Omega_3}{2}) -\Omega_7 \coth(\frac{\beta
 \Omega_7}{2}) \right) } \\
 & {\displaystyle \times \left[ \sin \Omega_7(t_0-t'') \sin \Omega_3(t_0-t')+
 \sin \Omega_7(t_0-t') \sin \Omega_3(t_0-t'') \right] } \\
 & {\displaystyle - \frac{1}{2}(n_{{\bf k}+{\bf q}}-n_{{\bf k}}) 
   (n_{-{\bf p}_1+{\bf q}}+n_{{\bf p}_1}+1)
 \frac{1}{\Omega_4^2-\Omega_7^2}\left(\Omega_7 \coth(\frac{\beta
 \Omega_4}{2}) -\Omega_4 \coth(\frac{\beta
 \Omega_7}{2}) \right) } \\
 & {\displaystyle \times \left[ \cos \Omega_7(t_0-t'') \cos \Omega_4(t_0-t')+
 \cos \Omega_7(t_0-t') \cos \Omega_4(t_0-t'') \right] } \\
 & {\displaystyle - \frac{1}{2}(n_{{\bf k}+{\bf q}}-n_{{\bf k}}) 
   (n_{-{\bf p}_1+{\bf q}}+n_{{\bf p}_1}+1)
 \frac{1}{\Omega_4^2-\Omega_7^2}\left(-\Omega_7 \coth(\frac{\beta
 \Omega_7}{2}) +\Omega_4 \coth(\frac{\beta
 \Omega_4}{2}) \right) } \\
 & {\displaystyle \times \left[ \sin \Omega_7(t_0-t'') \sin \Omega_4(t_0-t')+
 \sin \Omega_7(t_0-t') \sin \Omega_4(t_0-t'') \right] } \\
 & {\displaystyle - \frac{1}{2}(n_{{\bf k}+{\bf q}}+n_{{\bf k}}+1) 
   (n_{-{\bf p}_1+{\bf q}}-n_{{\bf p}_1})
 \frac{1}{\Omega_3^2-\Omega_8^2}\left(\Omega_8 \coth(\frac{\beta
 \Omega_3}{2}) -\Omega_3 \coth(\frac{\beta
 \Omega_8}{2}) \right) } \\
 & {\displaystyle \times \left[ \cos \Omega_8(t_0-t'') \cos \Omega_3(t_0-t')+
 \cos \Omega_8(t_0-t') \cos \Omega_3(t_0-t'') \right] } \\
& {\displaystyle - \frac{1}{2}(n_{{\bf k}+{\bf q}}+n_{{\bf k}}+1) 
  (n_{-{\bf p}_1+{\bf q}}-n_{{\bf p}_1})
 \frac{1}{\Omega_3^2-\Omega_8^2}\left(\Omega_3 \coth(\frac{\beta
 \Omega_3}{2}) -\Omega_8 \coth(\frac{\beta
 \Omega_8}{2}) \right) } \\
 & {\displaystyle \times \left[ \sin \Omega_8(t_0-t'') \sin \Omega_3(t_0-t')+
 \sin \Omega_8(t_0-t') \sin \Omega_3(t_0-t'') \right] } \\
 & {\displaystyle + \frac{1}{2}(n_{{\bf k}+{\bf q}}-n_{{\bf k}}) 
   (n_{-{\bf p}_1+{\bf q}}-n_{{\bf p}_1})
 \frac{1}{\Omega_4^2-\Omega_8^2}\left(\Omega_8 \coth(\frac{\beta
 \Omega_4}{2}) -\Omega_4 \coth(\frac{\beta
 \Omega_8}{2}) \right) } \\
 & {\displaystyle \times \left[ \cos \Omega_8(t_0-t'') \cos \Omega_4(t_0-t')+
 \cos \Omega_8(t_0-t') \cos \Omega_4(t_0-t'') \right] } \\
 & {\displaystyle + \frac{1}{2}(n_{{\bf k}+{\bf q}}-n_{\bf k}) 
   (n_{-{\bf p}_1+{\bf q}}-n_{{\bf p}_1})
 \frac{1}{\Omega_4^2-\Omega_8^2}\left(\Omega_4 \coth(\frac{\beta
 \Omega_4}{2}) -\Omega_8 \coth(\frac{\beta
 \Omega_8}{2}) \right) } \\
 & {\displaystyle \times \left[ \sin \Omega_8(t_0-t'') \sin \Omega_4(t_0-t')+
 \sin \Omega_8(t_0-t') \sin \Omega_4(t_0-t'') \right] } \ea \ee
 When we optimize with respect to initial state, the second term of 
 (\ref{4.128}) $tr[l^{\Phi \Phi}(t',t_0) X(t_0,t'')]$ cancels exactly the
 term  $P^{(1)}$. One checks also that at  $t'=t''=t_0$, this second term is
 vanishing.  

 In order to obtain the expression of  $\Pi({\bf  q}^2,t',t'')$, we have
 used a symmetry in the exchange of  ${\bf  k}$ and $-{\bf  p}_1$ of the
 function to integrate in  $\Pi({\bf  q}^2,t',t'')$.  It is important to
 note that the function  $\int \frac{d^3 p_1}{(2 \pi)^3}
 \Sigma_{11}^{\Phi\Phi}({\bf k},-{\bf q}-{\bf k},{\bf p}_1,{\bf q}-{\bf p}_1,
t',t'')$ remains dependent on $t_0$.  

For  $t'>t''$, expression (\ref{4.128}) becomes :
 \be \label{4.133}\ba{lll}
  & {\displaystyle \Pi({\bf  q}^2,t',t'')=
 \frac{i}{2} \int \frac{d^3 k}{(2 \pi)^3} \frac{d^3 p_1}{(2 \pi)^3}
 \frac{d^3 p}{(2 \pi)^3} \times  } \\
 & {\displaystyle  \: 
 tr \left[l^{\Phi \Phi} ({\bf p}+{\bf q},-{\bf p},{\bf k},-{\bf q}-{\bf k},
 t'',t_0) (\bar
 \Xi_0({\bf p})-\tau)l^{\Phi \Phi}({\bf p},-{\bf p}-{\bf q},{\bf p}_1,{\bf q}
 -{\bf p}_1,t',t_0) (\bar
 \Xi_0({\bf p}+{\bf q})+\tau) \right] } \\
 & {\displaystyle - \: 
 tr[l^{\Phi \Phi} ({\bf p}+{\bf q},-{\bf p},{\bf k},-{\bf q}-{\bf k},t'',t_0) 
 X(t_0,t') ] } \ea \ee
and  one  shows that one obtains  $\Pi^{(1)}({\bf  q}^2, \Delta t)$ from 
(\ref{4.129}) by replacing  $\Delta t$ by $-\Delta t$. 

Similarly to the case of the Green function with two field operators 
$\Phi({\bf  x}) \Phi({\bf  y})$, we can check the fluctuation-dissipation
theorem for the Green function with four field operators 
$\Phi^2({\bf  x}) \Phi^2({\bf  y})$. The expression of $\Pi({\bf  q}^2,
\tau)$ for  $\Delta t>0$ is indeed equivalent to expression  (\ref{4.129}) 
of $\Pi({\bf q}^2, -\tau-i \beta)$ for  $\Delta t<0$.   

\vspace*{0.5cm}

Let us study the retarded Green function with two field operators 
$\chi_{\Phi \Phi, \Phi \Phi}(t',t'')$ defined by (\ref{2e65c}) when we
optimize with respect to the initial state. For  $t'>t''$ and with the
time-ordered product  : 
\be \ba{ll}
\Sigma_{11}^{\Phi \Phi}({\bf k},-{\bf q}-{\bf k},{\bf p}_1,{\bf q}-{\bf p}_1,
t',t'') = 
& {\displaystyle - \int \frac{d^3p}{(2 \pi)^3} tr \left[
l^{\Phi \Phi}({\bf p}-{\bf q},-{\bf p},{\bf k},-{\bf q}-{\bf k},t'',t_0) 
\right. } \\ 
& {\displaystyle \left. \Sigma^{\Phi
\Phi}({\bf p},-{\bf p}-{\bf q},{\bf p}_1,{\bf q}-{\bf p}_1,t_0,t') \right] } 
\ea \ee
For $t'>t''$ and the anti-T product :  
\be \ba{ll}
\Sigma_{11}^{\Phi \Phi}({\bf k},-{\bf q}-{\bf k},{\bf p}_1,{\bf q}-{\bf p}_1,
t',t'') = 
& {\displaystyle - \int \frac{d^3p}{(2 \pi)^3} tr \left[
l^{\Phi \Phi}({\bf p}-{\bf q},-{\bf p},{\bf k},-{\bf q}-{\bf k},t',t_0) 
\right. } \\ 
& {\displaystyle \left. \Sigma^{\Phi
\Phi}({\bf p},-{\bf p}-{\bf q},{\bf p}_1,{\bf q}-{\bf p}_1,t_0,t'') 
\right] } \ea \ee
We use formula  (\ref{4.120}) obtained at the first order in perturbation.
The retarded Green function  
$\chi_{\Phi \Phi, \Phi \Phi}(t',t'')$ is given by expression 
(\ref{2e65d}) obtained by optimizing only the dynamics and a term equal to : 
\be \ba{ll}
& {\displaystyle
-\int \frac{d^3p}{(2 \pi)^3} tr[l^{\Phi \Phi}({\bf p}-{\bf q},-{\bf p},{\bf
  k},-{\bf q}-{\bf k},t'',t_0) X(t_0,t')  ] } \\
& {\displaystyle + \int \frac{d^3p}{(2 \pi)^3} tr[l^{\Phi \Phi}({\bf p}-{\bf
    q},-{\bf p},{\bf k},-{\bf q}-{\bf k},t'',t_0)
X(t_0,t')  ] } \ea \ee
where $tr[l^{\Phi \Phi}(t',t_0) X(t_0,t'')]$ is equal to the term  $P^{(1)}$
(eq. (\ref{4.132})). We see from the expression of  $P^{(1)}$ by
interchanging $t'$ and  $t''$ that the difference  
$tr[l^{\Phi \Phi}(t'',t_0) X(t_0,t')]-tr[l^{\Phi \Phi}(t',t_0) X(t_0,t'')]$
is vanishing. 

To conclude, for the the linear response formula which involves the retarded
Green functions, we obtain the same result with or without the optimization
with respect to initial state.  It is sufficient to solve the backward
dynamical equations  (\ref{4.104}-\ref{4.107}) 
in real time for $l^{\Phi \Phi}$. 

The first order contribution to the polarization function  $\Pi_R$ writes : 
\be \label{4.133a} \ba{lll}
\Pi_R^{(1)}&{\displaystyle 
({\bf q}^2,t',t'')=-\theta(t'-t'') \: \frac{b}{4} \int 
\frac{d^3p}{(2 \pi)^3} \: \frac{1}{\omega_{{\bf p}} \omega_{{\bf p}+{\bf q}}} 
\times } \\
& {\displaystyle \left[ (n_{{\bf p}+{\bf q}}+n_{{\bf p}}+1) 
  \sin \Omega_1 \Delta t \: 
\int \frac{d^3k}{(2 \pi)^3} \: \frac{1}{\omega_{{\bf k}} \omega_{{\bf
      q}-{\bf k}}}
[(n_{{\bf q}-{\bf k}}+n_{{\bf k}}+1) 
\frac{\Omega_3}{\Omega_1^2-\Omega_3^2}-(n_{{\bf q}-{\bf k}}-n_{{\bf k}})
\frac{\Omega_4}{\Omega_1^2-\Omega_4^2} ] \right. } \\
&{\displaystyle \left. -(n_{{\bf p}+{\bf q}}-n_{{\bf p}}) 
  \sin \Omega_2 \Delta t \:
\int \frac{d^3k}{(2 \pi)^3} \: \frac{1}{\omega_{{\bf k}} \omega_{{\bf
      q}-{\bf k}}}
[(n_{{\bf q}-{\bf k}}+n_{{\bf k}}+1) 
\frac{\Omega_3}{\Omega_2^2-\Omega_3^2}-(n_{{\bf q}-{\bf k}}-n_{{\bf k}})
\frac{\Omega_4}{\Omega_2^2-\Omega_4^2} ] \right] } \ea \ee 
with 
\be \label{4.133b}
\Omega_1=\omega_{{\bf p}+{\bf q}}+\omega_{{\bf p}} 
\: \: \, \: \: \Omega_2=\omega_{{\bf p}+{\bf q}}-\omega_{{\bf p}}
\ee
\be \label{4.133c} 
\Omega_3=\omega_{{\bf q}-{\bf k}}+\omega_{{\bf k}} \: \: 
\, \: \: \Omega_4=\omega_{{\bf q}-{\bf k}}-\omega_{{\bf k}}
\ee
and  $\Delta t=t'-t''$. 

By taking its Fourrier transform in time and adding the zero order term and
the first order term, we obtain :  
\be \label{4.133d} \ba{lllll}
&{\displaystyle \Pi_R ({\bf q}^2, \omega) = \int \frac{d^3p}{(2 \pi)^3} 
\: \frac{1}{\omega_{{\bf p}} \omega_{{\bf p}+{\bf q}}} \times } \\
& {\displaystyle \left[ (n_{{\bf p}+{\bf q}}+n_{{\bf p}}+1) 
  \frac{\Omega_1}{(\omega+i\eta)^2-\Omega_1^2} \right. } \\
& {\displaystyle \left. 
\left(1+\frac{b}{4}
\int \frac{d^3k}{(2 \pi)^3} \: \frac{1}{\omega_{{\bf k}} \omega_{{\bf
      q}-{\bf k}}}
[(n_{{\bf q}-{\bf k}}+n_{{\bf k}}+1) 
\frac{\Omega_3}{\Omega_1^2-\Omega_3^2}-(n_{{\bf q}-{\bf k}}-n_{{\bf k}})
\frac{\Omega_4}{\Omega_1^2-\Omega_4^2} ] \right) \right. } \\
&{\displaystyle \left. -(n_{{\bf p}+{\bf q}}-n_{{\bf p}}) 
 \frac{\Omega_2}{(\omega+i
\eta)^2-\Omega_2^2} \right. } \\ 
&{\displaystyle \left. \left(1+\frac{b}{4}
\int \frac{d^3k}{(2 \pi)^3} \: \frac{1}{\omega_{{\bf k}} \omega_{{\bf
      q}-{\bf k}}}
[(n_{{\bf q}-{\bf k}}+n_{{\bf k}}+1) 
\frac{\Omega_3}{\Omega_2^2-\Omega_3^2}-(n_{{\bf q}-{\bf k}}-n_{{\bf k}})
\frac{\Omega_4}{\Omega_2^2-\Omega_4^2} ] \right) \right] } \ea \ee
Although the coupling constant appears linearly in the numerator, this
formula includes nonperturbative contributions since the frequencies
involves the self-consistent renormalized mass $m^2(\beta)$ eq.
(\ref{2.29l}). 

The divergent part of the integrals at high momentum is entirely determined
by the contribution at  $T=0$. This one writes :  
\be \label{4.133e} \ba{ll}
\Pi_R ({\bf  q}^2, \omega)&{\displaystyle = \int \frac{d^3p}{(2 \pi)^3} 
\: \frac{1}{\omega_{{\bf p}} \omega_{{\bf p}+{\bf q}}} 
\: \frac{\omega_{{\bf p}+{\bf q}}+\omega_{{\bf p}}}{(\omega+i
\eta)^2-(\omega_{{\bf p}+{\bf q}}+\omega_{{\bf p}})^2} \times  } \\
& {\displaystyle \left(1 +\frac{b}{4} 
\int \frac{d^3k}{(2 \pi)^3} \: \frac{1}{\omega_{{\bf k}} \omega_{{\bf
      q}+{\bf k}}} \:
\frac{\omega_{{\bf q}+{\bf k}}+\omega_{{\bf k}}}{(\omega_{{\bf p}+{\bf q}}+
  \omega_{{\bf p}})^2
-(\omega_{{\bf q}+{\bf k}}+\omega_{{\bf k}})^2} \right) } \ea \ee
(We remark that there is no term  $i\eta$ in the denominator of the second
integral.)

At zero temperature, $\Pi_R({\bf  q}^2,\omega)$ depends only on the
variable $s=\omega^2-{\bf  q}^2$. At the lowest order,   
\be \label{4.133f}
\Pi_R^{(0)}(s)=
\int \frac{d^3p}{(2 \pi)^3} 
\: \frac{\omega_{{\bf p}+{\bf q}}+\omega_{\bf p}}{\omega_{{\bf p}} 
  \omega_{{\bf p}+{\bf q}}} \:
\frac{1}{(\omega+i\eta)^2-(\omega_{{\bf p}+{\bf q}}+\omega_{{\bf p}})^2} \ee
which is related to the explicitly Lorentz invariant form : 
\be \label{4.133g} 
\Pi_F(s)=i \int \frac{d^4p}{(2 \pi)^4} \: \frac{1}{p^2-\bar m^2+i \epsilon} \:
\frac{1}{(p+q)^2-\bar m^2+i \epsilon} \ee
where $q=(\omega, {\bf  q})$. We have $Re \Pi^{(0)}_R(s)=Re \Pi_F(s)$ and
$ Im \Pi^{(0)}_R(s)= \mbox{sign}(\omega) Im \Pi_F(s)$. $\bar m$ is the
renormalized self-consistent mass at zero temperature given by eq.
(\ref{2.29j}).  
By using for instance the results of  Kerman and Lin \cite{KERMAN},
\be \label{4.133g}
\Pi_R^{(0)}(s)=-\frac{1}{4 \pi^2} \log\left(\frac{2 \Lambda}{\sqrt e
\mu}\right) +\frac{1}{4 \pi^2}\log \left(\frac{\bar m}{\mu} \right)-
\frac{1}{8 \pi^2} f(s) -\theta (s-4 \bar m^2) \frac{i}{8 \pi}
\sqrt{\frac{s-4 \bar m^2}{s}} \ee
where $\Lambda$ is a momemtum  cut-off, $\mu$ is an arbitrary mass scale 
and the function  $f(s)$ is defined
by : 
\be \label{4.133h}
f(s) =2+\sqrt{\frac{s-4\bar m^2}{s}} \log \frac{\sqrt s -
  \sqrt{s-4\bar m^2}}{\sqrt s+
\sqrt{s-4 \bar m^2}} \quad  \: \mbox{si} \quad \:  s>4 \bar m^2 \ee 
\be \label{4.133i}
f(s) =2-2 \sqrt{\frac{-4 \bar m^2}{s}} 
\tan^{-1} \sqrt{\frac{s}{-4 \bar m^2}} \quad  \: 
\mbox{si} \quad \: 
0<s<4 \bar m^2 \ee 
\be \label{4.133j}
f(s) =2+\sqrt{\frac{s-4 \bar m^2}{s}} \log \frac{\sqrt{s-4 \bar m^2}-\sqrt
s}{\sqrt{s+4 \bar m^2}+\sqrt s} \quad  \: \mbox{si} \quad \: s<0 \ee 
Adding the zero order term and the first order term, we have at zero
temperature : 
\be \label{4.133k} 
\Pi_R(s)=\Pi_R^{(0)}(s) \left (1-\frac{b}{16 \pi^2} \log 
\left(\frac{2 \Lambda}{\sqrt e
\mu} \right) \right) \ee
and  $\Pi_R^{(0)}(s)$ is given by expression (\ref{4.133g}). We see that at
the order we have considered, the logarithmic divergence appears squared. By
resumming the perturbative serie, we obtain : 
\be \label{4.133l}
\Pi_R(s)=\frac{\Pi_R^{(0)}(s)}{1+ \frac{b}{16 \pi^2} \log 
\left(\frac{2 \Lambda}{\sqrt e \mu} \right) } \ee
We can define a renormalized  coupling constant  $g_R(\mu)$ by  
\be \label{4.133m}
2 g_R(\mu) \Pi^{(0)}_R(s)=b \Pi_R(s) \ee 
i. e.  : 
\be 
\frac{1}{2 g_R(\mu)} =
\frac{1}{b}+ \frac{1}{16 \pi^2} \log \left(\frac{2 \Lambda}{e\sqrt \mu} 
\right) \ee
This renormalization is identical to the one obtained by Kerman and Lin in
the symmetric phase  \cite{KERMAN}. 

\vspace*{0.5cm} 

\section{Conclusion}

In this paper, we have shown how to derive variational approximations for
two-time correlation functions in $\Phi^4$ theory. We have studied
in particular the two-time correlation functions with two field operators
$\Phi({\bf x})$ and with four field operators $\Phi({\bf x})$ in the
symmetric phase. By using the variational principle introduced by Balian and
Veneroni \cite{1}, we have shown how to take into account correlations both
in the dynamics and in the initial state while keeping uncorrelated
ans\"atze for the variational objects. The calculation sheme we have
presented is valid for arbitrary time-dependent problems. As an
illustration, we have calculated explicitly the two-time correlation
functions with two field operators and with four field operators at
equilibrium in the symmetric phase. The approximation we have obtained for
the two field operators correlation function is identical in this case with
the usual mean-field result. This won't be true in the asymmetric phase or
for a time-dependent problem. The approximation for the two-time and four
field operators correlation function contains already in the simple case we
have considered corrections to the standard mean-field result. For the
non-retarded correlation functions, it is important to optimize with respect
to the initial state in order not to obtain a spurious dependence on the
initial time. However, for the retarded correlation functions, the same
result is obtained whether or not we optimize with respect to the initial
state. Our variational approximations for the two-time correlation functions
satisfy the fluctuation-dissipation theorem. For the four field operators
correlation function, we were obliged to solve the dynamical equations
perturbatively. The zero order approximation was already known in the
litterature but we have shown here how to derive it variationally. We have
calculated also the first order approximation. At zero temperature, it is
possible to resum the perturbative serie and to define in this way a
renormalized coupling constant. We stress that our approximations remain
nonperturbative since they involve a self-consistent renormalized mass. 

There are several important questions  for future works. First, how
can one  obtain variational approximations which include damping. It would
be very interesting to obtain a variational nonperturbative approximation
for the viscosity defined in eq. (\ref{4.87a}). Damping effects are
probably related to a non symmetrical form in ${\bf x}$ and ${\bf y}$ for the
  matrix $T$ defined in eq. (\ref{2.7g}) to characterize our gaussian
  variational ans\"atze and to a choice for the operator $\bar H$ different
  from the Hamiltonian $H$. A second interesting problem is the evolution of
  a scalar field in self-interaction in a time-dependent metric. The
  difficulty in this case is to solve the backward dynamical equations in
  order to obtain variational approximations for the two-time correlation
  functions. 

\vspace*{1cm} 

{\bf Aknowledgements}
  
I would like to thank Roger Balian and Marcel V\'en\'eroni for very
fruitful discussions. I thank also Olivier Martin for a critical reading of
a part of the manuscript.   
 
\vspace*{1cm}

{\bf Appendix A : Characterization of a Gaussian state}

\setcounter{equation}{0}

\renewcommand{\theequation}{A.\arabic{equation}}

 We call a Gaussian operator an operator which is an exponential of
 quadratic and linear forms of the field operators 
 $\Phi({\bf  x})$ and  $\Pi({\bf x})$. A Gaussian operator  ${\cal D}(t)$ 
is completely characterized by the vector  $\alpha({\bf  x},t)$ and the
matrix  $\Xi({\bf  x}, {\bf  y},t)$ defined by the equations 
 (\ref{2.7a})-(\ref{2.7h}) of the section  2.3.  

Instead of the quantities  $G,T$ and  $S$, the authors of reference 
\cite{1a} have introduced the quantities  $G_{\xi}, \Sigma_{\xi}$ and  $\xi$,
$\xi$ being the degree of mixing (for a pur state, $\xi=1$) : 
\be \label{A.9} 
Tr({\cal D}(t)\bar \Phi({\bf  x}) \bar \Phi({\bf  y}))=\left(G^{1/2}_{\xi}
(1-\xi)^{-1}G_{\xi}^{1/2}\right)({\bf  x}, {\bf  y},t) \ , \ee 
\be \label{A.10}
Tr({\cal D}(t)\bar \Pi({\bf  x}) \bar \Pi({\bf  y}))=\frac{1}{4}\: \left(
G^{-1/2}_{\xi}(1+\xi)G^{-1/2}_{\xi}\right)({\bf  x}, {\bf  y},t) + 4 \left( 
\Sigma_{\xi}G^{1/2}_{\xi}(1-\xi)^{-1}G^{1/2}_{\xi}\Sigma_{\xi} \right)
({\bf  x}, {\bf  y},t) \ , \ee 
\be \label{A.11}
Tr({\cal D}(t)\bar \Phi({\bf  x}) \bar \Pi({\bf  y}))=
\frac{i}{2}\: \delta^3({\bf x} 
-{\bf  y})+2\left(G_{\xi}^{1/2}(1-\xi)^{-1}G^{1/2}_{\xi}\Sigma_{\xi}\right) 
({\bf  x},{\bf  y},t) \ . \ee

It is usefull to relate our notations to those of the authors of  \cite{1a}.
For a pure state  ($\xi=0$), 
\be \label{A.12}
T({\bf  x}, {\bf  y},t)=2\: (G \Sigma+\Sigma G)({\bf  x}, {\bf  y},t) \ , \ee
\be \label{A.13} 
S({\bf x}, {\bf  y},t)=\frac{1}{4} \: G^{-1}({\bf x}, {\bf y},t)+4 \: (\Sigma G
\Sigma)({\bf  x}, {\bf  y},t) \ . \ee 
For an uniform configuration, we have the following relations : 
\be \label{A.14} 
G({\bf p})=G_{\xi}({\bf p})(1-\xi({\bf p}))^{-1} \ , \ee 
\be \label{A.15} 
S({\bf p})=\frac{1}{4}\: G_{\xi}^{-1}({\bf p})\: (1+\xi({\bf p}))+
4 \Sigma^2_{\xi}({\bf p})G_{\xi}({\bf p})
\frac{1}{1-\xi({\bf p})} \ , \ee 
\be \label{A.16}
T({\bf p})=4 G_{\xi}({\bf p})\: \frac{1}{1-\xi({\bf p})} \:
\Sigma_{\xi}({\bf p}) \ , \ee
and the  Heisenberg invariant is related to the degree of mixing according
to : 
\be \label{A.16a}
I({\bf p})=\frac{1+\xi({\bf p})}{1-\xi({\bf p})} \ee 

\vspace*{1cm}

{\bf Appendix B : Expression of the von Neuman entropy for bosons in the
  Hartree-Bogoliubov approximation }

\setcounter{equation}{0}

\renewcommand{\theequation}{B.\arabic{equation}}

The  von-Neuman entropy is given as a function of the density operator ${\cal D}$ 
by  :  
\be \label{B.1}
S=-\frac{Tr {\cal D} \log{\cal D}}{Tr {\cal D}} + \log Tr {\cal D}\ . \ee
In the HB approximation, ${\cal D}$ is the exponential of linear and
quadratic forms of the field operator  $\Phi$ and  $\Pi$. The entropy
density then writes in terms of the matrix  $\Xi$ : 
\be \label{B.2} \ba{ll}
{\cal S}=\int \frac{d^3p}{(2 \pi)^3}  
\: & {\displaystyle \frac{1}{4}\: tr\left\{ \left(\frac{1}{2}\: (1-\tau)\: \Xi
\: (1-\tau) -1 \right) \: \log \left(\frac{\Xi+\tau}{\Xi-\tau}\right)
\right\} } \\ 
& {\displaystyle + \frac{1}{2} \: \log \det \left\{ \frac{1}{4}\: 
\pmatrix{ 1& 1 \cr 1 & -1\cr } \: \Xi \: (1-\tau) -\frac{1}{2} \: 
\pmatrix{0&1\cr 1 & 0\cr } \right\} }  \ea \ . \ee
where $\tau$ is the $2\times 2$ matrix  
\be 
\tau =\pmatrix{0 & 1 \cr -1 & 0 \cr} \ee 
It can also be written in terms of the Heisenberg invariant  I (\ref{2.7i}) : 
\be \label{B.3} 
{\cal S}=\int \frac{d^3p}{(2 \pi)^3} 
\: \left[ \frac{1}{2}\:\sqrt I \log (\frac{\sqrt I+1}{\sqrt I-1})  
+ \frac{1}{2} \: \log (I-1)  \right]    \ . \ee 

For the static  HB solution characterized by  $\alpha^0, \Xi^0$, we have : 
$I=-\Xi^0_{11} \Xi^0_{22}$. It is easy to convince oneself that  expression 
(\ref{B.3}) is in this case identical to : 
\be \label{B.6}
S=\int \: \frac{d^3p}{(2 \pi)^3} \: [(n_{{\bf p}}+1)\: \log(n_{{\bf p}}+1)
-n_{{\bf p}} \: \log n_{{\bf p}}] \
, \ee
where the occupation number  $n_{{\bf p}}$ is related to the degree of mixing 
$\xi(p)$  according to  : 
\be \label{B.7}
2n_{{\bf p}}+1=\sqrt{\frac{1+\xi({\bf p})}{1-\xi({\bf p})}}=\sqrt{I({\bf p})} 
\ . \ee
By using the parametrization  $\xi({\bf p})=\frac{1}{\cosh \beta \omega_{{\bf
    p}}}$,
we have also :  $2 n_{{\bf p}}+1=\coth(\frac{\beta \omega_{{\bf p}}}{2})$.
In the  $\Phi^4$ theory, $\omega_{{\bf p}}=\sqrt{-g_0({\bf p})}$ and  
\be 
g_0({\bf p})=-\left({\bf p}^2+m_0^2+\frac{b}{2}\: \varphi_0^2 + 
\frac{b}{2} \: G_0({\bf  x}, {\bf  x}) \right)  \ . \ee 
The static solution satisfies : 
\be \label{B.8}
\Xi^0_{11}({\bf p})\: g_0({\bf p})=\Xi^0_{22}({\bf p}) \ , \ee 
and  \be \label{B.9}
\Xi^0_{11}({\bf p})\: \Xi^0_{22}({\bf p})=
-\frac{1+\xi({\bf p})}{1-\xi({\bf p})} \ . \ee
In terms of the frequency  $\omega_{{\bf p}}$ : 
\be \label{B.10} 
\Xi_{11}^0({\bf p})=\frac{1}{\omega_{{\bf p}}} \: 
\coth (\frac{\beta}{2}\omega_{{\bf p}}) \ee 
\be \label{B.11} 
\Xi_{22}^0({\bf p})=-\omega_{{\bf p}}\: \coth (\frac{\beta}{2}\omega_{{\bf
    p}}) \ee 
\be \label{B.12} 
I({\bf p})=\coth^2 (\frac{\beta}{2}\omega_{{\bf p}}) \ee 

The matrix of the second derivatives of the entropy evaluated for the static
solution  $\Xi_0$ is given by the following elements :  
\be \label{B.13}
\frac{\partial^2 {\cal S}}{\partial \Xi_{11}({\bf p}) \partial \Xi_{11}({\bf
    p})} \bigg\vert_{\Xi_0}= - \frac{\omega^2_{{\bf p}}}{4} \: \left[\frac{x}{\coth
x}+\frac{1}{\coth^2 x-1}  \right]  \ee
\be \label{B.14}
\frac{\partial^2 {\cal S}}{\partial \Xi_{11}({\bf p}) \partial \Xi_{22}({\bf
    p})} \bigg\vert_{\Xi_0}= - \frac{1}{4} \: \left[ \frac{x}{\coth
x}-\frac{1}{\coth^2 x-1}  \right] \ee
\be \label{B.15}
\frac{\partial^2 {\cal S}}{\partial \Xi_{22}({\bf p}) \partial \Xi_{22}({\bf
    p})} \bigg\vert_{\Xi_0}= - \frac{1}{4 \omega^2_{{\bf p}}} \: \left[ \frac{x}{\coth
x}+\frac{1}{\coth^2 x-1}  \right] \ee
\be \label{B.16}
\frac{\partial^2 {\cal S}}{\partial \Xi_{12}({\bf p}) \partial \Xi_{12}({\bf
    p})} \bigg\vert_{\Xi_0}=\frac{x}{\coth x}  \ee
where we have introduced the notation  $x=\beta \omega_{{\bf p}}/2$.

\vspace*{1cm}

{\bf Appendix C : Parametrization of the product of two Gaussians } 

\setcounter{equation}{0}

\renewcommand{\theequation}{C.\arabic{equation}}

In this appendix, we give the  expressions of the expectation values 
$\alpha_b, \alpha_c, \Xi_b$ and   $\Xi_c$ in terms of  
$\alpha_d, \alpha_a, \Xi_d$ and   $\Xi_a$. 

\be \label{C.1} 
\alpha_b=\left( \Xi_a-\tau \right) \: \frac{1}{\Xi_a+\Xi_d} \: 
\alpha_d + \left(\Xi_d + \tau \right) \: \frac{1}{\Xi_a+\Xi_d} \: 
\alpha_a \ , \ee 
\be \label{C.2} 
\alpha_c = \left( \Xi_d - \tau \right) \: \frac{1}{\Xi_a + \Xi_d} \: 
\alpha_a + \left( \Xi_a + \tau \right) \: \frac{1}{\Xi_a + \Xi_d} \: 
\alpha_d \ , \ee
where   $\tau$ is  $2 \times 2$ matrix : 
\be \label{C.3} 
\tau= \pmatrix{ 0 & 1 \cr -1 & 0 \cr } \ . \ee   

For the matrices  $\Xi$, we have the following relations :    
\be \label{C.4}
\Xi_b- \tau = \left( \Xi_a -\tau \right) \: \frac{1}{\Xi_a + 
\Xi_d} \: \left( \Xi_d - \tau \right) \ , \ee
\be \label{C.5} 
\Xi_c - \tau = \left( \Xi_d - \tau \right) \: 
\frac{1}{\Xi_a + \Xi_d} \: \left( \Xi_a - \tau \right) \ . \ee 

\be \label{C.6} 
\Xi_b+ \tau = \left( \Xi_d +\tau \right) \: \frac{1}{\Xi_a + 
\Xi_d} \: \left( \Xi_a + \tau \right) \ , \ee
\be \label{C.7} 
\Xi_c + \tau = \left( \Xi_a + \tau \right) \: 
\frac{1}{\Xi_a + \Xi_d} \: \left( \Xi_d+ \tau \right) \ . \ee

\vspace*{1cm} 

{\bf Appendix D} 
 
\setcounter{equation}{0} 

\renewcommand{\theequation}{D.\arabic{equation}}

In this  appendix, we give the evolution equations for 
 $n_d, \alpha_d, \Xi_d$  $n_a, \alpha_a, \Xi_a$.
 \be \label{D.0}
 i \: \frac{\dot n_d}{n_d}=F_c^{(d)}-F_b^{(d)}-F_{Kc}^{(d)}
 \ee
\be \ba{ll} \label{D.1} 
2 i \: \dot \alpha_d = 
& {\displaystyle  - \left[ \left( \Xi_d + \tau \right) \left( {\cal H}_c 
-{\cal I}^c_K \right) \left( \Xi_d - \tau \right) - \left( \Xi_d - 
\tau \right) \: {\cal H}_b \: \left( \Xi_d + \tau \right) \right] \: \tau \: 
\alpha_{b-c} } \\ 
& {\displaystyle + \Xi_d \left( w_c-w_b \right) + \tau \: \left( w_c+w_b 
\right) - \left( \Xi_d + \tau \right) \: w^c_K } \ea  \ee 
\be \label{D.2} 
i \: \dot \Xi_d = - \left[ \left( \Xi_d+ \tau \right) \left( {\cal H}_c 
-{\cal I}^c_K \right) \left( \Xi_d - \tau \right) - \left( \Xi_d - 
\tau \right) \: {\cal H}_b \: \left( \Xi_d + \tau \right) \right]
\ee
\vspace*{0.5cm}
\be \label{D.2a}
 i \: \frac{\dot n_a}{n_a}=F_b^{(a)}-F_c^{(a)}+F_{Kc}^{(a)}
 \ee
\be \ba{ll} \label{D.3} 
2 i \: \dot \alpha_a = 
& {\displaystyle   \left[ \left( \Xi_a - \tau \right) \left( {\cal H}_c 
-{\cal I}^c_K \right) \left( \Xi_a + \tau \right) - \left( \Xi_a+ 
\tau \right) \: {\cal H}_b \: \left( \Xi_a - \tau \right) \right] \: \tau \: 
\alpha_{c-b} } \\ 
& {\displaystyle + \Xi_a \left( w_b-w_c \right) + \tau \: \left( w_c+w_b 
\right) - \left( \Xi_a - \tau \right) \: w^c_K } \ea \ , \ee 
\be \label{D.4} 
i \: \dot \Xi_a =  \left[ \left( \Xi_a- \tau \right) \left( {\cal H}_c 
-{\cal I}^c_K \right) \left( \Xi_a + \tau \right) - \left( \Xi_a + 
\tau \right) \: {\cal H}_b \: \left( \Xi_a - \tau \right) \right]
\ee
 
The  expressions for the vector $w$ and the matrices  ${\cal H}$ and 
${\cal I}$ are the following :  
\be \label{D.5} 
\tilde w_b({\bf  x},t)_1 = \frac{\delta <H>_b}{\delta \varphi_b({\bf  x},t)} 
= - f_b({\bf x},t) \: \: \ , \: \: 
\tilde w_b({\bf  x},t)_2 = i \: \frac{\delta <H>_b}{\delta \pi_b({\bf  x},t)} 
= i \pi_b({\bf  x},t) \ , \ee
\be \label{D.6} 
{\cal H}_b({\bf  x}, {\bf  y},t)_{ij} = 
-2 \frac{\delta <H>_b}{\delta \Xi_b({\bf  x}, {\bf  y},t)_{ji}} \ , \ee 
\be \label{D.7} 
{\cal H}_b({\bf  x}, {\bf  y},t)_{11}= - \frac{\delta <H>_b}{\delta 
G_b({\bf  y}, {\bf  x},t)} = \frac{1}{2} \: g_b({\bf  x},{\bf  y},t) \ , \ee
\be \label{D.8} 
 {\cal H}_b({\bf  x}, {\bf  y},t)_{22}= + \frac{\delta <H>_b}{\delta 
S_b({\bf  y}, {\bf  x},t)} = \frac{1}{2} \: \delta({\bf  x}-{\bf  y}) \ , \ee
\be \label{D.9}
 {\cal H}_b({\bf x}, {\bf  y},t)_{12}= 2 i \: \frac{\delta <H>_b}{\delta
T_b({\bf  y}, {\bf  x},t)} = 0\ . \ee
The quantities $F$ which appear in equations  (D.1) and  (D.3) are defined
by : 
\be \ba{ll}
F_b^{(d)}=& {\displaystyle \int d^3x \: \left( {\cal E}_b({\bf  x},t)-\tilde
w_b({\bf  x},t) \alpha_{b-d}({\bf  x},t) \right) } \\
& {\displaystyle + \frac{1}{2} \int d^3x  d^3y \: tr \left[ {\cal H}_b(
{\bf x}, {\bf  y},t) (\Xi_b-\Xi_d +2 \alpha_{b-d}\tilde \alpha_{b-d})
({\bf  y}, {\bf x},t) \right] } \ea \ee 
\be \ba{ll}
F_{Kc}^{(d)}=& {\displaystyle \int d^3x \: \left( {\cal K}_c({\bf  x},t)-
  \tilde
w_{Kc}({\bf x},t) \alpha_{c-d}({\bf  x},t) \right) } \\
& {\displaystyle + \frac{1}{2} \int d^3x  d^3y \:
  tr \left[ {\cal I}_{Kc}({\bf
x}, {\bf  y},t) (\Xi_c-\Xi_d +2 \alpha_{c-d}\tilde \alpha_{c-d})({\bf  y},
{\bf x},t) \right] } \ea \ee
For the $\Phi^4$ theory, we have  : 
\be \label{D.10} 
f_b({\bf x},t) =- \left( -\Delta + m_0^2 +\frac{b}{6} \: 
\varphi^2_b({\bf  x},t) 
+ \frac{b}{2} \: G_b({\bf  x}, {\bf  x},t)\right) \: \varphi_b({\bf  x}, t) 
\ , \ee 
\be \label{D.11} 
g_b({\bf  x},{\bf  y},t) = - \left( - \Delta + m_0^2 + \frac{b}{2} \: 
\varphi^2_b({\bf  x},t) + \frac{b}{2} \: G_b({\bf  x},{\bf  x},t) \right) \: 
\delta({\bf  x}- {\bf  y}) \ . \ee
From the variations of the source term  $K_c$, we obtain :   
\be \label{D.12} 
\tilde w^c_K({\bf  x},t)_1= \frac{\delta K_c}{\delta \varphi_c({\bf  x},t)} 
\: \: \ , \: \: 
\tilde w^c_K({\bf  x},t)_2= i \: \frac{\delta K_c}{\delta \pi_c({\bf  x},t)} 
\ , \ee
\be \label{D.13} 
{\cal I}^c_K({\bf  x}, {\bf  y},t)_{ij} = -2 \frac{\delta K_c}{\delta 
\Xi_c ({\bf  y}, {\bf  x},t)_{ji}} \ . \ee 
\be \label{D.14} 
w^c_K({\bf  x},t)_1=J^{\Phi}({\bf  x},t) + 2 \: \int d^3x_2 \:  \left( 
J^{\Phi \Phi} ({\bf  x}, {\bf  x}_2,t) \: \varphi_c({\bf x}_2,t) + 
J^{\Phi \Pi} ({\bf x}, {\bf x}_2,t) \: \pi_c({\bf  x}_2,t) \right) \ , \ee
\be \label{D.15} 
w^c_K({\bf  x},t)_2=
i \: J^{\Pi}({\bf  x},t) + 2  \: i \: \int d^3x_2  \: \left( 
J^{\Phi \Pi} ({\bf  x}, {\bf  x}_2,t) \: \varphi_c({\bf  x}_2,t) + 
J^{\Pi \Pi} ({\bf  x}, {\bf  x}_2,t) \: \pi_c({\bf  x}_2,t) \right) \ , \ee
\be \label{D.16} 
{\cal I}^c_K({\bf x}, {\bf  y},t)_{11}= -J^{\Phi \Phi} ({\bf  x}, {\bf  y},t) 
\ , \ee 
\be \label{D.17}
{\cal I}^c_K({\bf  x}, {\bf  y},t)_{12}= -2 i \: 
J^{\Phi \Pi} ({\bf  x}, {\bf  y},t) \ , \ee
\be \label{D.18} 
{\cal I}^c_K({\bf  x}, {\bf  y},t)_{22}= J^{\Pi \Pi} ({\bf  x}, {\bf  y},t) 
  \ . \ee 

The expressions for the matrices  $t, T,r$ and $R$ which appear in the
dynamical equations for the two-time correlation functions are the following
:  
\be \label{D.19} 
t_{11}({\bf  x},{\bf  y}, t)=
-g^{(0)}({\bf  x},{\bf  y},t) \: \: , \: \: t_{22}({\bf  x}, {\bf 
y},t)=-\delta({\bf  x}-{\bf  y}) \ , \ee 
\be \label{D.20} 
T_{1,11}({\bf  x}, {\bf  y},{\bf  z},t) =- 
\frac{b}{2} \: \varphi^{(0)}({\bf  x},t) \delta({\bf  x}-{\bf  y}) 
\: \delta^3({\bf x} -{\bf  z}) 
\: \: , \: \: T_{2,jk}=0 \ , \ee 
\be \label{D.21} 
r_{11,1}({\bf  x},{\bf  y},{\bf  z},t)=T_{1,11}({\bf  x}, {\bf  y}, 
{\bf  z},t) \ee 
\be \label{D.22} 
R_{11,11}({\bf  x},{\bf  y},{\bf  z}, {\bf  u}, t) 
= \frac{b}{4} \: \delta^3({\bf  x}-{\bf  y}) \: 
\delta^3({\bf  x}-{\bf z}) \: \delta({\bf  x} -{\bf  u}) \ . \ee 
The other elements are equal to zero. 

We use the following definitions for the Fourrier transform :  
\be \label{D.23} 
W({\bf x}_1,...,{\bf  x}_n)=
\int \frac{d^3p_1}{(2 \pi)^3} ...\frac{d^3p_n}{(2 \pi)^3}  
\: \exp(i \sum_{j=1}^n {\bf  p}_j \: . \: {\bf  x}_j) 
\: \tilde W({\bf  p}_1,...,{\bf  p}_n) \ , \ee 
\be \label{D.24} 
\tilde W({\bf  p}_1,...,{\bf  p}_n)=\int d^3x_1 ...  d^3x_n  
\: \exp(-i \sum_{j=1}^n {\bf p}_j \: . \: {\bf x}_j) \: 
W({\bf  x}_1,...,{\bf  x}_n) \ , \ee 
\be \label{D.25} 
\tilde W({\bf  p}_1,{\bf  p}_2,...,{\bf  p}_n)=
(2 \pi)^3 \: \delta^3({\bf  p}_1+{\bf  p}_2+...+{\bf  p}_n) \:
W({\bf  p}_1,{\bf  p}_2,...,{\bf  p}_n) \ . \ee
So for the two-point function  $\beta^{\Phi}$ : 
\be \label{D.26} 
\beta^{\Phi}_1({\bf  x},{\bf  x}_1,t,t")= \int \frac{d^3p}{(2 \pi)^3} \: 
e^{i \: {\bf  p} .  
  ({\bf  x}_1 - {\bf  x})} \: \beta_1^{\Phi}({\bf p},t,t") \ . \ee 

\vspace*{1cm} 

{\bf Appendix E}
 
\setcounter{equation}{0} 

\renewcommand{\theequation}{E.\arabic{equation}}

We give in this appendix the expressions of the integrals 
 $I_n,J_n,K_n,L_n,M_n,N_n$, $n=1,2,3,4$ 
 in termes of the constants  $a,b,c,d$ introduced in  (\ref{4.115})
 and of  $l^{\Phi \Phi}_{ij}$. 
We introduce the frequencies  
\be 
\Omega_3=\omega_{{\bf k}+{\bf q}}+\omega_{{\bf k}} \: \: \ , 
\: \: \Omega_4=\omega_{{\bf k}+{\bf q}}-\omega_{{\bf k}}
\ee
\be 
\Omega_5=\omega_{{\bf l}+{\bf q}}+\omega_{{\bf l}} 
\: \: \ , \: \: \Omega_6=\omega_{{\bf l}+{\bf q}}-\omega_{\bf l} \ee
We define :  
\be \label{E.1}
 I_1=\int \frac{d^3l}{(2 \pi)^3} \frac{a}{\Omega_3^2-\Omega^2_5} \: \: \ , 
 \: \: 
 I_2=\int \frac{d^3l}{(2 \pi)^3} \frac{c}{\Omega_3^2-\Omega^2_6}
 \ee 
\be \label{E.2}
 I_3=\int \frac{d^3l}{(2 \pi)^3} \frac{a}{\Omega_4^2-\Omega^2_5} 
 \: \: \ , \: \: 
 I_4=\int \frac{d^3l}{(2 \pi)^3} \frac{c}{\Omega_4^2-\Omega^2_6}
 \ee 
\be \label{E.3}
 J_1=\int \frac{d^3l}{(2 \pi)^3} \frac{b \Omega_5}{\Omega_3^2-\Omega^2_5} 
 \: \: \ , \: \: 
 J_2=\int \frac{d^3l}{(2 \pi)^3} \frac{d \Omega_6}{\Omega_3^2-\Omega^2_6}
 \ee 
\be \label{E.4}
 J_3=\int \frac{d^3l}{(2 \pi)^3} \frac{b \Omega_5}{\Omega_4^2-\Omega^2_5} 
 \: \: \ , \: \: 
 J_4=\int \frac{d^3l}{(2 \pi)^3} \frac{d \Omega_6}{\Omega_4^2-\Omega^2_6}
 \ee 
\be \label{E.5}
 K_1=\int \frac{d^3l}{(2 \pi)^3} \frac{a}{\Omega_3^2-\Omega^2_5} \cosh(\beta
 \Omega_5) \: \: \ , \: \: 
 K_2=\int \frac{d^3l}{(2 \pi)^3} \frac{b}{\Omega_3^2-\Omega^2_5} \sinh(\beta
 \Omega_5) 
 \ee 
\be \label{E.6}
 K_3=\int \frac{d^3l}{(2 \pi)^3} \frac{c}{\Omega_3^2-\Omega^2_6} \cosh(\beta
 \Omega_6) \: \: \ , \: \:
 K_4=\int \frac{d^3l}{(2 \pi)^3} \frac{d}{\Omega_3^2-\Omega^2_6} \sinh(\beta
 \Omega_6)
 \ee
The integrals $L_n$ are obtained from the integrals  $K_n$ by replacing 
$\Omega_3$ by  $\Omega_4$. 
\be \label{E.7}
 M_1=\int \frac{d^3l}{(2 \pi)^3} \frac{b \Omega_5}{\Omega_3^2-\Omega^2_5}
 \cosh(\beta \Omega_5)  
 \: \: \ , \: \: 
 M_2=\int \frac{d^3l}{(2 \pi)^3} \frac{a \Omega_5}{\Omega_3^2-\Omega^2_5}
 \sinh(\beta \Omega_5) 
 \ee 
\be \label{E.8}
 M_3=\int \frac{d^3l}{(2 \pi)^3} \frac{d \Omega_6}{\Omega_3^2-\Omega^2_6}
 \cosh(\beta \Omega_6)  
 \: \: \ , \: \: 
 M_4=\int \frac{d^3l}{(2 \pi)^3} \frac{c \Omega_6}{\Omega_3^2-\Omega^2_6}
 \sinh(\beta \Omega_6) 
 \ee 
The integrals $N_n$ are obtained from the integrals  $M_n$ by replacing 
$\Omega_3$ by  $\Omega_4$. 

The  expressions of sums of integrals in terms of  $l^{\Phi
\Phi}_{ij}$ are usefull : 
\be \label{E.9} \ba{ll} 
I_1+I_2=V \int \frac{d^3l}{(2 \pi)^3} & {\displaystyle \left \{ 
 -\frac{n_{{\bf l}+{\bf q}}+n_{\bf l}+1}
{\Omega_3^2-\Omega_5^2} \left[ (\frac{1}{\omega_{{\bf l}}
\omega_{{\bf l}+{\bf q}}} l_{11}^{\Phi \Phi} +l_{22}^{\Phi \Phi}) 
\coth(\frac{\beta
\Omega_5}{2}) +\frac{1}{\omega_{{\bf l}}} l^{\Phi \Phi}_{12}
+\frac{1}{\omega_{{\bf l}+{\bf q}}}l^{\Phi \Phi}_{21} \right] \right.  } \\ 
& {\displaystyle  \left. + \frac{n_{{\bf l}+{\bf q}}-n_{{\bf l}}}
  {\Omega_3^2-\Omega_6^2} 
\left[ (\frac{1}{\omega_{{\bf l}}
\omega_{{\bf l}+{\bf q}}} l_{11}^{\Phi \Phi} -l_{22}^{\Phi \Phi}) 
\coth(\frac{\beta
\Omega_6}{2}) +\frac{1}{\omega_{\bf l}} l^{\Phi \Phi}_{12}
-\frac{1}{\omega_{{\bf l}+{\bf q}}}l^{\Phi \Phi}_{21} \right] \right \}
} \ea \ee
$I_3 +I_4$ is obtained from  $I_1 +I_2$ by replacing $\Omega_3$
by  $ \Omega_4$.
The arguments of the functions $l^{\Phi \Phi}_{ij}$ in these integrals are
$l^{\Phi \Phi}_{ij}({\bf l},-{\bf q}-{\bf l},{\bf p}_1,{\bf q}-{\bf p}_1,
t'',t_0)$. 
\be \label{E.10} \ba{lll}  
J_1+J_2= & {\displaystyle V \int \frac{d^3l}{(2 \pi)^3} \times } \\ 
& {\displaystyle 
 \frac{\Omega_5}{\Omega_3^2-\Omega_5^2} 
(n_{{\bf l}+{\bf q}}+n_{{\bf l}}+1) \left[ (\frac{1}{\omega_{{\bf l}}
\omega_{{\bf l}+{\bf q}}} l_{11}^{\Phi \Phi} +l_{22}^{\Phi \Phi}) 
+(\frac{1}{\omega_{{\bf l}}} l^{\Phi \Phi}_{12}
+\frac{1}{\omega_{{\bf l}+{\bf q}}}l^{\Phi \Phi}_{21})  \coth(\frac{\beta
\Omega_5}{2}) \right]  } \\ 
& {\displaystyle  - \frac{\Omega_6}{\Omega_3^2-\Omega_6^2} 
  (n_{{\bf l}+{\bf q}}-n_{{\bf l}})
\left[ (\frac{1}{\omega_{{\bf l}}
\omega_{{\bf l}+{\bf q}}} l_{11}^{\Phi \Phi} -l_{22}^{\Phi \Phi}) 
+(\frac{1}{\omega_{{\bf l}}} l^{\Phi \Phi}_{12}
-\frac{1}{\omega_{{\bf l}+{\bf q}}}l^{\Phi \Phi}_{21}) \coth(\frac{\beta
\Omega_6}{2}) \right]  } \ea \ee
$J_3 +J_4$ is obtained from  $J_1 +J_2$ by replacing  $\Omega_3$ by 
 $ \Omega_4$.

\be \label{E.11} \ba{lll} 
K_1+K_2+K_3+K_4=& {\displaystyle -V \int \frac{d^3l}{(2 \pi)^3} \times } \\
& {\displaystyle 
 \frac{n_{{\bf l}+{\bf q}}+n_{{\bf l}}+1}{\Omega_3^2-\Omega_5^2} 
 \left[ (\frac{1}{\omega_{{\bf l}}
\omega_{{\bf l}+{\bf q}}} l_{11}^{\Phi \Phi} 
+l_{22}^{\Phi \Phi}) \coth(\frac{\beta
\Omega_5}{2}) -(\frac{1}{\omega_{{\bf l}}} l^{\Phi \Phi}_{12}
+\frac{1}{\omega_{{\bf l}+{\bf q}}}l^{\Phi \Phi}_{21}) \right]  } \\ 
& {\displaystyle  + \frac{n_{{\bf l}+{\bf q}}-n_{{\bf l}}}
  {\Omega_3^2-\Omega_6^2} 
\left[ -(\frac{1}{\omega_{{\bf l}}
\omega_{{\bf l}+{\bf q}}} 
l_{11}^{\Phi \Phi} -l_{22}^{\Phi \Phi}) \coth(\frac{\beta
\Omega_6}{2}) +\frac{1}{\omega_{{\bf l}}} l^{\Phi \Phi}_{12}
-\frac{1}{\omega_{{\bf l}+{\bf q}}}l^{\Phi \Phi}_{21} \right]  } \ea \ee
$L_1+L_2+L_3 +L_4$ is obtained from $K_1+K_2+K_3+K_4$ by replacing 
$\Omega_3$ by $ \Omega_4$.

\be \label{E.12} \ba{lll}  
M_1+M_2+M_3+M_4 & {\displaystyle =-V \int \frac{d^3l}{(2 \pi)^3} \times } \\ 
& {\displaystyle 
 \frac{\Omega_5}{\Omega_3^2-\Omega_5^2} 
(n_{{\bf l}+{\bf q}}+n_{{\bf l}}+1) \left[ (\frac{1}{\omega_{{\bf l}}
\omega_{{\bf l}+{\bf q}}} l_{11}^{\Phi \Phi} +l_{22}^{\Phi \Phi}) 
-(\frac{1}{\omega_{{\bf l}}} l^{\Phi \Phi}_{12}
+\frac{1}{\omega_{{\bf l}+{\bf q}}}l^{\Phi \Phi}_{21})  \coth(\frac{\beta
\Omega_5}{2}) \right]  } \\ 
& {\displaystyle  + \frac{\Omega_6}{\Omega_3^2-\Omega_6^2} 
  (n_{{\bf l}+{\bf q}}-n_{{\bf l}})
\left[ -(\frac{1}{\omega_{{\bf l}}
\omega_{{\bf l}+{\bf q}}} l_{11}^{\Phi \Phi} -l_{22}^{\Phi \Phi}) 
+(\frac{1}{\omega_{{\bf l}}} l^{\Phi \Phi}_{12}
-\frac{1}{\omega_{{\bf l}+{\bf q}}}l^{\Phi \Phi}_{21}) \coth(\frac{\beta
\Omega_6}{2}) \right]  } \ea \ee
$N_1+N_2+N_3 +N_4$ is obtained from  $M_1+M_2+M_3+M_4$ by replacing 
$\Omega_3$ by $ \Omega_4$.

\vspace*{1cm}

{\bf Appendix F}
 
\setcounter{equation}{0} 

\renewcommand{\theequation}{F.\arabic{equation}}

In this appendix, we give the solutions of the backward dynamical equations 
for  $l^{\Phi \Phi}_{ij}$ at the first order . By noting : 
\be 
\Omega_3=\omega_{{\bf k}+{\bf q}}+ \omega_{{\bf k}} \: \: \ 
, \: \: \Omega_4=\omega_{{\bf k}+{\bf q}}-
\omega_{{\bf k}} \ee 
\be 
\Omega_5=\omega_{{\bf p}_i-{\bf q}}+ \omega_{{\bf p}_i} \: \: \ 
, \: \: \Omega_6=\omega_{{\bf p}_i-{\bf q}}-
\omega_{{\bf p}_i} \ee 
\be F=\frac{b}{4} \frac{1}{\omega_{{\bf p}_i} \omega_{{\bf q}-{\bf p}_i}} \ee

\be \label{F.1} \ba{llllll} 
& {\displaystyle l^{\Phi \Phi}_{11}({\bf k},-{\bf q}-{\bf k},{\bf p}_i,
  -{\bf p}_i+{\bf q},t',t))= } \\
& {\displaystyle -\frac{1}{2} [\delta^3({\bf  k}+{\bf  p}_i) + 
  \delta^3(-{\bf  q} -
{\bf  k} + {\bf  p}_i) ] \cos \omega_{{\bf q}+{\bf k}}(t-t') 
\cos \omega_{{\bf k}}(t-t') } \\ 
& {\displaystyle -\frac{F}{2} (n_{{\bf q}-{\bf p}_i}+n_{{\bf p}_i}+1) \left[
\frac{\Omega_5}{\Omega_3^2-\Omega_5^2} \cos \Omega_3(t-t') +
\frac{\Omega_5}{\Omega_4^2-\Omega_5^2} \cos \Omega_4(t-t') \right] } \\ 
& {\displaystyle +\frac{F}{2} (n_{{\bf q}-{\bf p}_i}-n_{{\bf p}_i}) \left[
\frac{\Omega_6}{\Omega_3^2-\Omega_6^2} \cos \Omega_3(t-t') +
\frac{\Omega_6}{\Omega_4^2-\Omega_6^2} \cos \Omega_4(t-t') \right] } \\ 
& {\displaystyle +\frac{F}{2} (n_{{\bf q}-{\bf p}_i}+n_{{\bf p}_i}+1)
\left(\frac{\Omega_5}{\Omega_3^2-\Omega_5^2}+
\frac{\Omega_5}{\Omega_4^2-\Omega_5^2}\right)\cos \Omega_5 (t-t') } \\
& {\displaystyle 
-\frac{F}{2} (n_{{\bf q}-{\bf p}_i}-n_{{\bf p}_i})
\left(\frac{\Omega_6}{\Omega_3^2-\Omega_6^2}+
\frac{\Omega_6}{\Omega_4^2-\Omega_6^2}\right)\cos \Omega_6 (t-t') } \ea \ee 
\be \label{G.2} \ba{llllll} 
& {\displaystyle l^{\Phi \Phi}_{12}({\bf k},-{\bf q}-{\bf k},{\bf p}_i,-{\bf
    p}_i+{\bf q},t',t))= } \\ 
& {\displaystyle -\frac{1}{2i \omega_{{\bf k}+{\bf q}}} 
[\delta^3({\bf  k}+{\bf  p}_i) + \delta^3(-{\bf  q} -
{\bf  k} + {\bf  p}_i) ] \sin \omega_{{\bf q}+{\bf k}}(t-t') 
\cos \omega_{{\bf k}}(t-t') } \\ 
& {\displaystyle -\frac{F}{2 i \omega_{{\bf k}+{\bf q}}} (n_{{\bf q}-{\bf
      p}_i}+n_{{\bf p}_i}+1) \left[
\frac{\Omega_5}{\Omega_3^2-\Omega_5^2} \sin \Omega_3(t-t') +
\frac{\Omega_5}{\Omega_4^2-\Omega_5^2} \sin \Omega_4(t-t') \right] } \\ 
& {\displaystyle +\frac{F}{2 i \omega_{{\bf k}+{\bf q}}} (n_{{\bf q}-{\bf
      p}_i}-n_{{\bf p}_i}) \left[
\frac{\Omega_6}{\Omega_3^2-\Omega_6^2} \sin \Omega_3(t-t') +
\frac{\Omega_6}{\Omega_4^2-\Omega_6^2} \sin \Omega_4(t-t') \right] } \\ 
& {\displaystyle +\frac{F}{2 i \omega_{{\bf k}+{\bf q}}} (n_{{\bf q}-{\bf
      p}_i}+n_{{\bf p}_i}+1)
\left(\frac{\Omega_3}{\Omega_3^2-\Omega_5^2}+
\frac{\Omega_4}{\Omega_4^2-\Omega_5^2}\right)\sin \Omega_5 (t-t') } \\
& {\displaystyle 
-\frac{F}{2 i \omega_{{\bf k}+{\bf q}}} (n_{{\bf q}-{\bf p}_i}-n_{{\bf p}_i})
\left(\frac{\Omega_3}{\Omega_3^2-\Omega_6^2}+
\frac{\Omega_4}{\Omega_4^2-\Omega_6^2}\right)\sin \Omega_6 (t-t') } \ea \ee 
\be \label{G.3} \ba{llllll} 
& {\displaystyle l^{\Phi \Phi}_{21}({\bf k},-{ \bf  q}-{\bf k},
  {\bf p}_i,-{\bf p}_i+{\bf q},t',t))= } \\
& {\displaystyle -\frac{1}{2i \omega_{{\bf k}}} 
[\delta^3({\bf  k}+{\bf  p}_i) + \delta^3(-{\bf  q} -
{\bf  k} + {\bf  p}_i) ] \cos \omega_{{\bf q}+{\bf k}}(t-t') \sin
\omega_{{\bf k}}(t-t') } \\ 
& {\displaystyle -\frac{F}{2 i \omega_{{\bf k}}} (n_{{\bf q}-{\bf
      p}_i}+n_{{\bf p}_i}+1) \left[
\frac{\Omega_5}{\Omega_3^2-\Omega_5^2} \sin \Omega_3(t-t') -
\frac{\Omega_5}{\Omega_4^2-\Omega_5^2} \sin \Omega_4(t-t') \right] } \\ 
& {\displaystyle +\frac{F}{2 i \omega_{{\bf k}}} (n_{{\bf q}-{\bf
      p}_i}-n_{{\bf p}_i}) \left[
\frac{\Omega_6}{\Omega_3^2-\Omega_6^2} \sin \Omega_3(t-t') -
\frac{\Omega_6}{\Omega_4^2-\Omega_6^2} \sin \Omega_4(t-t') \right] } \\ 
& {\displaystyle +\frac{F}{2 i \omega_{{\bf k}}} (n_{{\bf q}-{\bf p}_i}+
  n_{{\bf p}_i}+1)
\left(\frac{\Omega_3}{\Omega_3^2-\Omega_5^2}-
\frac{\Omega_4}{\Omega_4^2-\Omega_5^2}\right)\sin \Omega_5 (t-t') } \\
& {\displaystyle 
-\frac{F}{2 i \omega_{{\bf k}}} (n_{{\bf q}-{\bf p}_i}-n_{{\bf p}_i})
\left(\frac{\Omega_3}{\Omega_3^2-\Omega_6^2}-
\frac{\Omega_4}{\Omega_4^2-\Omega_6^2}\right)\sin \Omega_6 (t-t') } \ea \ee 
\be \label{G.4} \ba{llllll} 
& {\displaystyle l^{\Phi \Phi}_{22}({\bf k},-{\bf q}-{\bf k},{\bf p}_i,-{\bf
    p}_i+{\bf q},t',t))= } \\ 
& {\displaystyle +\frac{1}{2 \omega_{{\bf k}} \omega_{{\bf k}+{\bf q}}} 
[\delta^3({\bf  k}+{\bf  p}_i) + \delta^3(- {\bf q} -
{\bf  k} + {\bf  p}_i) ] \sin \omega_{{\bf q}+{\bf k}}(t-t') 
\sin \omega_{\bf k}(t-t') } \\ 
& {\displaystyle -\frac{F}{2 \omega_{{\bf k}} \omega_{{\bf k}+{\bf q}}} 
(n_{{\bf q}-{\bf p}_i}+n_{{\bf p}_i}+1) \left[
\frac{\Omega_5}{\Omega_3^2-\Omega_5^2} \cos \Omega_3(t-t') -
\frac{\Omega_5}{\Omega_4^2-\Omega_5^2} \cos \Omega_4(t-t') \right] } \\ 
& {\displaystyle +\frac{F}{2 \omega_{{\bf k}} \omega_{{\bf k}+{\bf q}}} 
  (n_{{\bf q}-{\bf p}_i}-n_{{\bf p}_i}) \left[
\frac{\Omega_6}{\Omega_3^2-\Omega_6^2} \cos \Omega_3(t-t') -
\frac{\Omega_6}{\Omega_4^2-\Omega_6^2} \cos \Omega_4(t-t') \right] } \\ 
& {\displaystyle +\frac{F}{2 \omega_{{\bf k}} \omega_{{\bf k}+{\bf q}}} 
  (n_{{\bf q}-{\bf p}_i}+n_{{\bf p}_i}+1)
\left(\frac{\Omega_5}{\Omega_3^2-\Omega_5^2}-
\frac{\Omega_5}{\Omega_4^2-\Omega_5^2}\right)\cos \Omega_5 (t-t') } \\ 
& {\displaystyle 
-\frac{F}{2 \omega_{{\bf k}} \omega_{{\bf k}+{\bf q}}} (n_{{\bf q}-{\bf p}_i}
-n_{{\bf p}_i})
\left(\frac{\Omega_6}{\Omega_3^2-\Omega_6^2}-
\frac{\Omega_6}{\Omega_4^2-\Omega_6^2}\right)\cos \Omega_6 (t-t') } \ea \ee

\end{document}